%% file: Plasma_Talks.tex
\documentclass[letter, 12pt]{article}

\title{CfA Plasma Talks}
\author{Antoine Bret}
\date{ }

\usepackage{graphicx}
\usepackage{amsmath}
\usepackage{bm}
\usepackage{cancel}
\usepackage{amssymb}

\begin{document}

\maketitle

 Notes from a series of 13 one hour (or more) lectures on Plasma Physics given to Ramesh Narayan' research group at the Harvard-Smithsonian Center for Astrophysics, between January and July 2012.

 Lectures 1 to 5 cover various key Plasma Physics themes. Lectures 6 to 12 mainly go over the Review Paper on ``Multidimensional electron beam-plasma instabilities in the relativistic regime'' [\emph{Physics of Plasmas} \textbf{17}, 120501 (2010)]. Lectures 13 talks about the so-called Biermann battery and its ability to generate magnetic fields from scratch.

\tableofcontents

\include{PlasmaTalk01}\setcounter{equation}{0}\setcounter{footnote}{0}\setcounter{table}{0}

\include{PlasmaTalk02}\setcounter{equation}{0}\setcounter{footnote}{0}\setcounter{table}{0}
\include{PlasmaTalk03}\setcounter{equation}{0}\setcounter{footnote}{0}\setcounter{table}{0}

\include{PlasmaTalk04}\setcounter{equation}{0}\setcounter{footnote}{0}\setcounter{table}{0}
\include{PlasmaTalk05}\setcounter{equation}{0}\setcounter{footnote}{0}\setcounter{table}{0}

\include{PlasmaTalk06}\setcounter{equation}{0}\setcounter{footnote}{0}\setcounter{table}{0}

\include{PlasmaTalk07}\setcounter{equation}{0}\setcounter{footnote}{0}\setcounter{table}{0}

\include{PlasmaTalk08}\setcounter{equation}{0}\setcounter{footnote}{0}\setcounter{table}{0}

\include{PlasmaTalk09}\setcounter{equation}{0}\setcounter{footnote}{0}\setcounter{table}{0}

\include{PlasmaTalk10}\setcounter{equation}{0}\setcounter{footnote}{0}\setcounter{table}{0}

\include{PlasmaTalk11}\setcounter{equation}{0}\setcounter{footnote}{0}\setcounter{table}{0}

\include{PlasmaTalk12}\setcounter{equation}{0}\setcounter{footnote}{0}\setcounter{table}{0}
\include{PlasmaTalk13}

\end{document}

%% file: PlasmaTalk01.tex
\begin{center}
\section{Introduction}
\end{center}

\subsection*{When is a gas ionized?}
\begin{itemize}
 \item Ionization can come from the plasma itself, if hot enough. With $X$ = Ne/Nneutral, Saha equation gives,
\begin{equation}
 \frac{X^2}{1-X}=\frac{1}{nh^2}(2\pi m_e k_BT)^{3/2}e^{-I/k_BT},
\end{equation}
where $I$ is the Ionization energy. Comes from statistical physics inside atom + Maxwell distribution outside. $X \rightarrow 0$ for $k_BT <\ll I$, and $X \rightarrow 1$ for $k_BT \gg I$.

\begin{figure}[h]
\begin{center}
\includegraphics[width=0.5\textwidth]{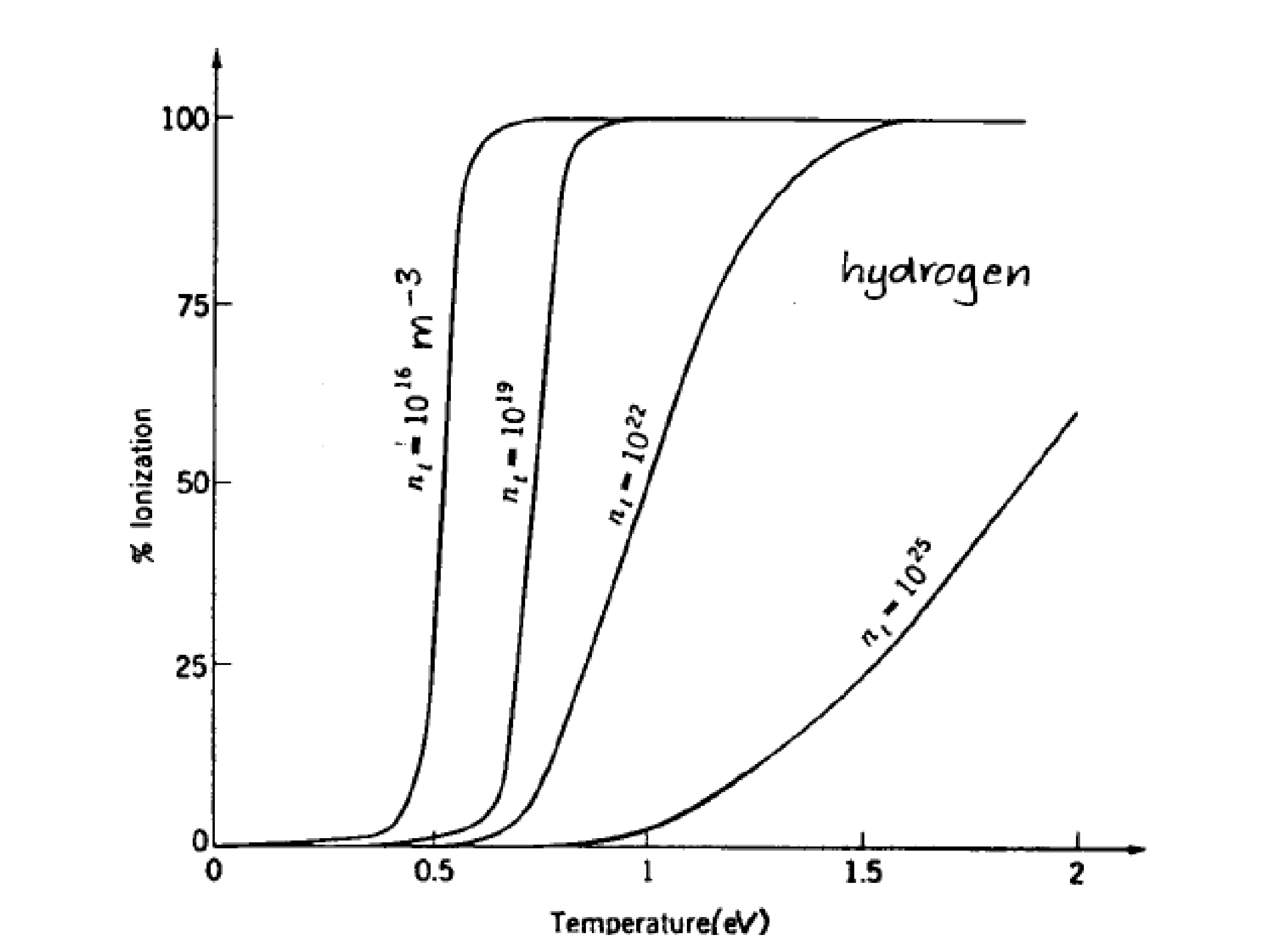}\label{saha}
\end{center}
\caption{Degree for Ionization for Hydrogen, with, I = 13.6 eV.}
\end{figure}

\item Ionization can come from external medium (Ionosphere ? T = say 1000 K).
\item Ionization can come from the proximity of atoms ? Share electrons : metal.

\end{itemize}

\subsection*{Classification}
Say temperature $T$, density $N$.

Typical distance between two electrons: $N^{-1/3}$.

Typical Coulomb energy: $e^2/N^{-1/3}$.

Typical kinetic energy in classical regime: $k_BT$.

More kinetic energy than Coulomb: $k_BT> e^2/N^{-1/3}$. Big frontier.

Classical relativistic: $k_BT > mc^2$.

Then come \textit{quantum} effects. When $T>$ Fermi temperature $T_F$, with $k_BT_F = \hbar^2(3 \pi^2 N)^{2/3}/2m_e$.

Thus, for $T < T_F$, energy increases with \textit{density}, not \textit{temperature}.

$k_BT_F < mc^2$, $k_B$ scales like $N^{2/3}$.

$k_BT_F > mc^2$, $k_B$ scales like $N^{1/3}$ (White Dwarfs to Neutron Stars?).

\subsection*{Important quantities}
Time it takes to neutralize charge in-balance: \textit{Plasma frequency}

\begin{equation}
 \omega_p^2 = \frac{4 \pi N e^2}{m_e} = 9000 \sqrt{N~[\mathrm{cm}^{-3}]}.
\end{equation}

That's why some waves bounce against the ionosphere.

Distance over which charge in-balance can exist: \textit{Debye length}
\begin{equation}
\lambda_D= \frac{V_{th}}{\omega_p} = \sqrt{\frac{k_BT}{4 \pi N e}} = 7.43~10^2 \sqrt{T~[\mathrm{K}]/N~[\mathrm{cm}^{-3}]}~[\mathrm{cm}] .
\end{equation}

\begin{figure}[h]
\begin{center}
\includegraphics[width=0.85\textwidth]{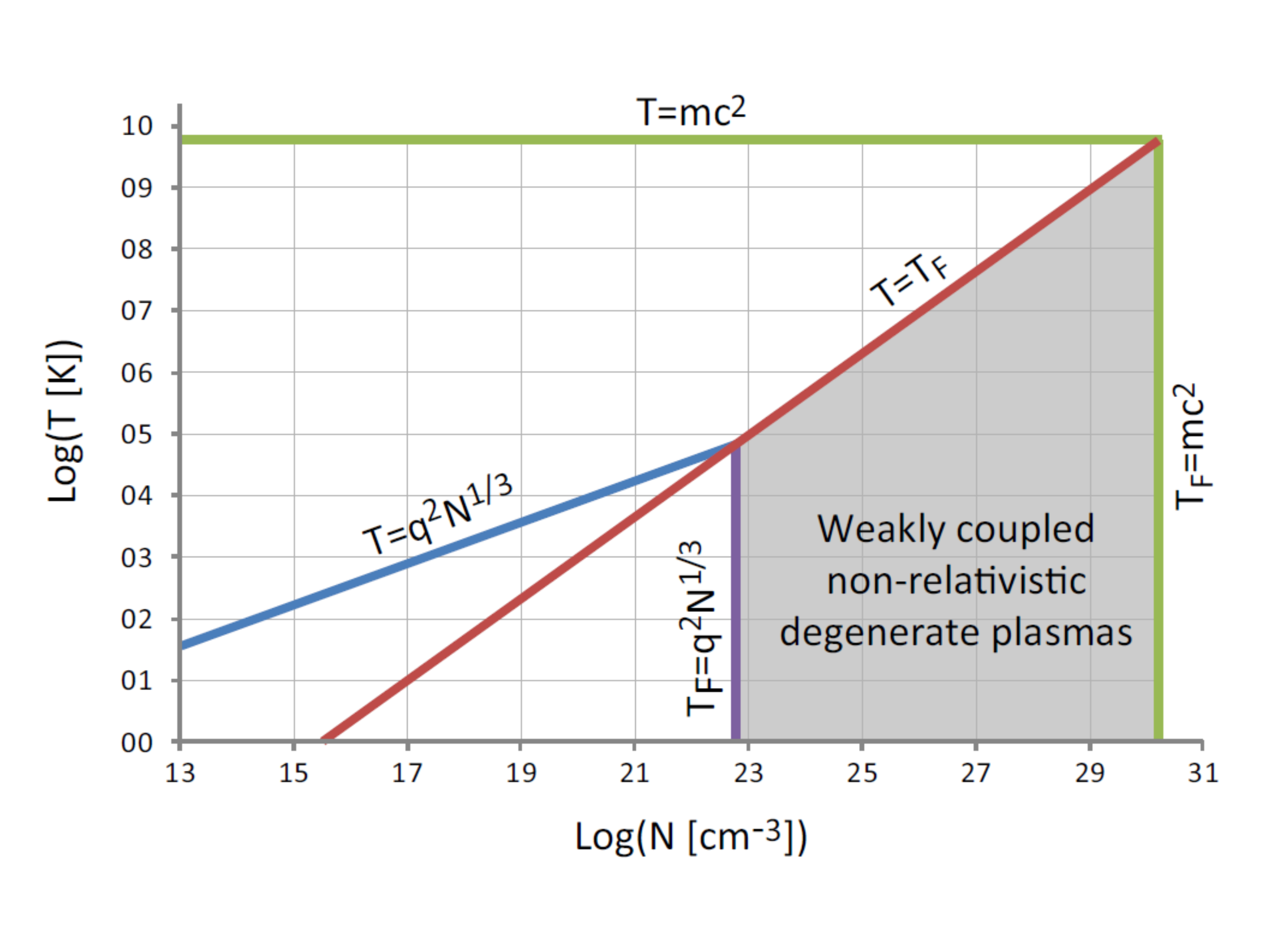}\label{class}
\end{center}
\caption{Classification of plasmas.}
\end{figure}

%% file: PlasmaTalk02.tex
\begin{center}
\section{Kinetic theory}\end{center}

\subsection*{Vlasov and Boltzmann equations}
Say only electrons + fixed positive background.
Most basic description level.
$G(\mathbf{r}, \mathbf{p}, t) d^3\mathbf{r} d^3\mathbf{p}$ = number of particles in $d^3\mathbf{r} d^3\mathbf{p}$ around $(\mathbf{r}, \mathbf{p})$ at time $t$. How does it evolve?

Say a particle has momentum $\mathbf{p}$ at position $\mathbf{r}$, at time $t$. Say a force $\mathbf{F}$ acts on it.
At time $t + dt$, it will have momentum $\mathbf{p} + \mathbf{F} dt$, and position $\mathbf{r} + \mathbf{p}/\gamma m~dt$ (\textit{non}-quantum treatment).\\
Therefore, ALL particles in  $d^3\mathbf{r} d^3\mathbf{p}$ around $(\mathbf{r}, \mathbf{p})$ at time $t$ MUST be in  $d^3\mathbf{r} d^3\mathbf{p}$ around $(r + \mathbf{p}/\gamma m~dt, \mathbf{p} + \mathbf{F} dt)$ at time $t + dt$. That means,
\begin{equation}
G\left( \mathbf{r} + \frac{\mathbf{p}}{\gamma m} dt, \mathbf{p} + \mathbf{F} dt, t + dt\right) =
G(\mathbf{r}, \mathbf{p}, t)  \end{equation}
The ``hyper'' volume element $d^3\mathbf{r} d^3\mathbf{p}$ does not change (Jacobian = 1 here). Just Taylor expand the left-hand-side to get the \textbf{Vlasov Equation},
\begin{eqnarray}
G(\mathbf{r}, \mathbf{p}, t)  +  \frac{\partial G}{\partial\mathbf{r}}\cdot\frac{\mathbf{p}}{\gamma m} dt
+ \frac{\partial G}{\partial\mathbf{p}}\cdot\mathbf{F} dt
+ \frac{\partial G}{\partial t} dt
&=&G(\mathbf{r}, \mathbf{p}, t), \nonumber \\
\Rightarrow\frac{\partial G}{\partial \mathbf{r}}\cdot \frac{\mathbf{p}}{\gamma m}
 + \frac{\partial G}{\partial\mathbf{p}}\cdot \mathbf{F} + \frac{\partial G}{\partial t} &=& 0.
\end{eqnarray}
Now, this result is NOT always right. Why?\\
We have assumed the force $\mathbf{F}$ does not change over $dt$. But $\mathbf{F}$ is an \textit{averaged} force, in the same way the function $G$ is averaged (IGM = $10^{-6}$ part/cm$^{-3}$. If not averaged, $d^3\mathbf{r}$ must be gigantic, not infinitesimal).
What if something ``un-smooth'' happened during $dt$ ?

\textit{Close collisions} are local, quasi-instantaneous processes, sending some particles OUT of  $d^3\mathbf{p}$ around  $\mathbf{p}$, and some other particles INSIDE $d^3\mathbf{p}$ around $\mathbf{p}$, during $dt$. (Think about billiard ball collisions: local and instantaneous). We'll then have,
\begin{eqnarray}
G(\mathbf{r}, \mathbf{p}, t)  &+&  \frac{\partial G}{\partial\mathbf{r}}\cdot\frac{\mathbf{p}}{\gamma m} dt
+ \frac{\partial G}{\partial\mathbf{p}}\cdot\mathbf{F} dt
+ \frac{\partial G}{\partial t} dt \nonumber \\
&=& G(\mathbf{r}, \mathbf{p}, t)
+[\mathrm{Collisions}(\mathbf{r}, t),~ \mathbf{q}\rightarrow \mathbf{p}, \forall \mathbf{q}]
-[\mathrm{Collisions}(\mathbf{r}, t),~ \mathbf{p}\rightarrow \mathbf{q}, \forall \mathbf{q}],
\end{eqnarray}
giving the \textbf{Boltzmann Equation},
\begin{equation}
\frac{\partial G}{\partial\mathbf{r}}\cdot\frac{\mathbf{p}}{\gamma m}
+ \frac{\partial G}{\partial\mathbf{p}}\cdot\mathbf{F}
+ \frac{\partial G}{\partial t}
= \int_{\mathbf{q}}[\mathrm{Collisions}(\mathbf{r}, t),~ \mathbf{q}\rightarrow \mathbf{p}]
-[\mathrm{Collisions}(\mathbf{r}, t),~ \mathbf{p}\rightarrow \mathbf{q}].
\end{equation}
The right-hand-side, referred to as the ``collision term'', is analitically untractable. Yet, that's the one driving the relaxation to a Maxwellian distribution $G_M\propto e^{-v^2}$. For practical purposes, alternative forms have been worked-out (Fokker-Planck/Landau, Balescu, Krook $\nu (G_M-G)$\ldots).

\subsection*{Vlasov or Boltzmann ?}
In a plasma, particles are influenced by,
\begin{itemize}
 \item Close collisions, changing $\mathbf{p}$ rapidly and appreciably (say $\theta > \pi/2$). Accounted for by the collision term in the kinetic equation.
 \item ``Distant'' collisions, which amount to the influence of the overall plasma ($\rho, \mathbf{J}\rightarrow \mathbf{E}, \mathbf{B}$). Accounted for by the Force term in the kinetic equation.
\end{itemize}
Define a ``close'' collision  by closest approach\footnote{Subscript $L$ stands (again) for $L$andau.} $<R_L$, such as $e^2/R_L=E_K$ where $E_K$ is the typical Kinetic energy ($k_BT$ or $k_BT_F$). Frequency for such collisions is roughly $\nu \sim n R_L^2 v_K$, with $mv_K^2=E_K$. Time scale for ``distant'' collisions if $\sim\omega_p^{-1}$. Vlasov's equation, with collision term = 0, is valid for $\nu\ll\omega_p$, i.e.,
\begin{equation}
n\left(\frac{e^2}{E_K} \right) ^2 \sqrt{\frac{E_K}{m}}  \ll \sqrt{\frac{4\pi n e^2}{m}}
\Leftrightarrow e^2n^{1/3} \ll E_K,
\end{equation}
which just defines \textit{weakly coupled plasmas}, where there is more kinetic energy than Coulomb potential energy, whether degenerate or not.
\subsection*{The Vlasov-Maxwell system}
For weakly coupled plasmas, the first equation needed is therefore Vlasov's with \textbf{F} = Lorentz,
\begin{equation}\label{vlasov02}
\frac{\partial G}{\partial t}
+ \frac{\mathbf{p}}{\gamma m}\cdot\frac{\partial G}{\partial \mathbf{r}}
+ q\left[\mathbf{E}(\mathbf{r},t)
+ \frac{\mathbf{v}}{c}\times\mathbf{B}(\mathbf{r},t)\right] \cdot \frac{\partial G}{\partial\mathbf{p}}  = 0.
\end{equation}
System is closed with Maxwell's equations, where charge and current densities are given by,
\begin{eqnarray}\label{Maxwell02}
\rho(\mathbf{r}, t)&=&\int G(\mathbf{r}, \mathbf{p}, t) d^3\mathbf{p},\nonumber\\
\mathbf{J}(\mathbf{r}, t)&=&\int q G(\mathbf{r}, \mathbf{p}, t)\mathbf{v} d^3\mathbf{p}.
\end{eqnarray}
Eqs. (\ref{vlasov02},\ref{Maxwell02}), together with Maxwell's, form the \textit{Vlasov-Maxwell} closed system of equations. In 1D along axis $x$, we just have for $G(x,p,t)$ and $E(x,t)$,
\begin{equation}\label{1D}
\frac{\partial G}{\partial t}
+ \frac{p}{\gamma m}\frac{\partial G}{\partial x}
+ q E \frac{\partial G}{\partial p}  = 0,~~~~~\frac{\partial E}{\partial x} = 4\pi q\int G(x,p,t)dp,
\end{equation}
with $q=-e$ for electrons. Landau damping comes from these 2, originally with $\gamma=1$.

%% file: PlasmaTalk03.tex
\begin{center}
\section{From Kinetic to Fluid to MHD Equations}
\end{center}

\subsection*{From Kinetic to Fluid}
Fluid equations can be deduced from the \textit{moments} of the kinetic equation\footnote{See the Appendix of Spitzer's \textit{Physics of Fully Ionized Gases} for details. Also, Chapter I of William L. Kruer, \textit{The Physics of Laser Plasma Interactions} (Previewed on \textit{Google Books}).}. The fluid macroscopic density $n$,  velocity $\mathbf{v}$ and pressure tensor $\mathbf{P}$ are defined through,
\begin{eqnarray}\label{FluidVar}
n(\mathbf{r},t)&=&\int F(\mathbf{r},\mathbf{u},t)d^3\mathbf{u},~~~~
\mathbf{P}(\mathbf{r},t)=\int m (\mathbf{u}-\mathbf{v})\otimes(\mathbf{u}-\mathbf{v})F(\mathbf{r},\mathbf{u},t)d^3\mathbf{u},
\nonumber\\
n(\mathbf{r},t)\mathbf{v}(\mathbf{r},t)&=&\int \mathbf{u}F(\mathbf{r},\mathbf{u},t)d^3\mathbf{u},~~
\end{eqnarray}
where $\otimes$ is ``dyadic'' product $\mathbf{u} \otimes \mathbf{v}=(u_iv_j)$. If our plasma is \textit{cold}, which \textit{kinetically} means $F(\mathbf{r},\mathbf{u},t)=\delta[\mathbf{u}-\mathbf{v}(\mathbf{r},t)]G(\mathbf{r},t)$, the density $n(\mathbf{r},t)$ and the velocity $\mathbf{v}(\mathbf{r},t)$ are what we would expect. Interestingly enough, the pressure tensor vanishes. \textit{Microsopic velocity spread translates to macroscopic pressure}. Consider now the non-relativistic Vlasov kinetic equation,
\begin{equation}
\frac{\partial F}{\partial t} + \mathbf{v}\cdot\frac{\partial F}{\partial\mathbf{r}}
+\frac{\mathbf{E}+\mathbf{v}\times\mathbf{B}/c}{m}\cdot \frac{\partial F}{\partial\mathbf{v}}= 0.
\end{equation}
The moments of the equation give,\footnote{Not straightforward. See Kruer for details. Note that $\partial/\partial \mathbf{r}$ is an alternative notation for $\nabla$.}
\begin{eqnarray}\label{FluidEq}
\int [\mathrm{Vlasov}]~~d^3\mathbf{p} &\Rightarrow&
\frac{\partial n}{\partial t}+\frac{\partial}{\partial \mathbf{r}}\cdot(n\mathbf{v})=0,\nonumber\\
\int m\mathbf{u}~[\mathrm{Vlasov}]~~d^3\mathbf{p} &\Rightarrow&
mn\left(\frac{\partial \mathbf{v}}{\partial t}+\mathbf{v}\cdot\frac{\partial \mathbf{v}}{\partial \mathbf{r}}\right) =
qn \left(\mathbf{E}+\frac{\mathbf{v}}{c}\times\mathbf{B}\right)-\frac{\partial}{\partial \mathbf{r}}\cdot \mathbf{P}.
\end{eqnarray}
For isotropic pressure\footnote{If the pressure tensor is anisotropic, with $\mathbf{P}=(p_{i,j})$,
\begin{equation*}
\frac{\partial}{\partial \mathbf{r}}\cdot \mathbf{P}=
\left(
\frac{\partial p_{xx}}{\partial x}+\frac{\partial p_{yx}}{\partial y}+\frac{\partial p_{zx}}{\partial z},~~
\frac{\partial p_{xy}}{\partial x}+\frac{\partial p_{yy}}{\partial y}+\frac{\partial p_{zy}}{\partial z},~~
\frac{\partial p_{xz}}{\partial x}+\frac{\partial p_{yz}}{\partial y}+\frac{\partial p_{zz}}{\partial z}
 \right) .
\end{equation*}} with $\mathbf{P}=p\mathbf{I}$, the last term is just the usual gradient $\partial p/\partial \mathbf{r} = \nabla p$.\\
The ``convective derivative'' term $(\partial_t+\mathbf{v}\cdot\nabla)$ simply \textit{follows} a fluid element.

At this stage, you can close the system introducing a relation between $n(\mathbf{r},t)$ and $p(\mathbf{r},t)$, that is, an \textit{equation of state}. Like for the first moment and the pressure, the Vlasov moment $\int u^n(~)d^3u$ always yields a macroscopic quantity $\propto\int u^{n+1}(~)d^3u$ from the $\mathbf{v}\cdot\partial F/\partial\mathbf{r}$ term.\\
Still regarding the micro/macro duality: a \textit{non-zero collision term} in the Vlasov equation is needed to recover \textit{viscosity} or \textit{friction} on the macro level.

\subsection*{From Fluid to MHD}
We have initially one distribution function $F_i(\mathbf{r},\mathbf{u},t)$ per species. The procedure above shows we eventually have one set of fluid equations per species. Assume we just have protons and electrons of densities $n_p(\mathbf{r},t)$ and  $n_e(\mathbf{r},t)$. If we want to describe fast phenomenon where electrons could be decoupled from protons (faster than $\omega_p^{-1}$, or smaller than $\lambda_D$), we need to keep two sets of equations. The so-called \textit{Braginskii Equations} might be the most elaborate version of this option.

What if we're interested in slow $\tau\ll\omega_p^{-1}$, and large scale $\gg\lambda_D$, effects? Electrons are expected to closely follow protons. The plasma is a electron/proton ``soup''. Electroneutrality on these scales gives $n_p(\mathbf{r},t)\sim n_e(\mathbf{r},t)$. In the same way we defined the fluid quantities (\ref{FluidVar}) and found they obey Eqs. (\ref{FluidEq}), we define the MHD variables,
\begin{eqnarray}\label{MHDVar}
\rho(\mathbf{r},t)&=&m_pn_p(\mathbf{r},t)+m_en_e(\mathbf{r},t),~~~~~~\mathbf{V}(\mathbf{r},t)= \frac{m_e\mathbf{v_e}+m_p\mathbf{v_p}}{m_e+m_p}\nonumber\\
\mathbf{J}(\mathbf{r},t)&=&q n_p(\mathbf{r},t)\mathbf{v}_p(\mathbf{r},t)
-q n_e(\mathbf{r},t)\mathbf{v}_e(\mathbf{r},t).
\end{eqnarray}
Combining the fluid equations for electrons and protons yields\footnote{Eqs. (\ref{FluidEq}) formally give a non-linear term $n_pm_p(\mathbf{v}_p\cdot\nabla)\mathbf{v}_p+n_em_e(\mathbf{v}_e\cdot\nabla)\mathbf{v}_e\neq \rho(\mathbf{V}\cdot\nabla)\mathbf{V}$. An ``='' is obtained neglecting the electron momentum, and considering $\mathbf{V} \sim \mathbf{v}_p$.},
\begin{eqnarray}
\frac{\partial \rho}{\partial t}+\frac{\partial (\rho\mathbf{V})}{\partial \mathbf{r}} &=&0, \label{MHD1}\\
\rho\left(\frac{\partial \mathbf{V}}{\partial t}+\mathbf{V}\cdot\frac{\partial \mathbf{V}}{\partial \mathbf{r}}\right) &=&
\frac{\mathbf{J}}{c}\times\mathbf{B} -\nabla (\overbrace{p_i+p_e}^P)+\rho \mathbf{g}\label{MHD2},
\end{eqnarray}
where $\rho \mathbf{E}$ is neglected with respect to the Lorentz force, as $n_e\sim n_p \Rightarrow \mathbf{E}\sim 0$. Also, a gravity term $\rho\mathbf{g}$ is added here. Its fluid counterpart in Eq. (\ref{FluidEq}) would obviously be $nm\mathbf{g}$. The system is closed through,
\begin{equation}\label{Maxwell03}
\frac{\partial \mathbf{B}}{\partial t} = -c\nabla\times \mathbf{E},~~~~~
\nabla\times \mathbf{B}=\frac{4\pi}{c}\mathbf{J}+\cancel{\frac{1}{c}\frac{\partial \mathbf{E}}{\partial t}}.
\end{equation}
Inserting $\mathbf{J}=c\nabla\times \mathbf{B}/4\pi$ into Eq. (\ref{MHD2}) gives the usual magnetic pressure and tension terms. The last equation used to close the system is Ohm's law, which simplest version reads
\begin{equation}\label{Ohm}
\mathbf{J} = \sigma\left(\mathbf{E}+\frac{\mathbf{V}}{c}\times\mathbf{B}\right),
\end{equation}
where $\sigma$ is the medium conductivity. This equation is just $\mathbf{J} = \sigma\mathbf{E}$ in the fluid-frame at velocity $\mathbf{V}$, transformed in the Lab. frame\footnote{J.D. Jackson, \textit{Classical Electrodynamics}, p. 472.}. \textit{Ideal} MHD sets $\sigma=\infty$, so that $\mathbf{E}=-\mathbf{V}\times\mathbf{B}/c$. Two concluding remarks:
\begin{itemize}
 \item Yes, we sometime consider $\mathbf{E}=0$, like in Eqs. (\ref{MHD2}) and (\ref{Maxwell03}-right), and sometime $\mathbf{E}\neq 0$ like in Ohm's law or (\ref{Maxwell03}-left). Kulsrud\footnote{R.M. Kulsrud, \textit{Plasma Physics for Astrophysics}, p. 44.} explains well how this proves reasonable.
\item We've cheated a little bit. We use the collision\textit{less} Vlasov's equation, and then talk about EOS or Ohm's law, which \textit{imply collisions}. It's just far simpler to forget about collisions at the \textit{kinetic/micro} level, derive the fluid equations, and then get collisions back into the game, kind of empirically, at the \textit{fluid/macro} level.
\end{itemize}

%% file: PlasmaTalk04.tex
\begin{center}
\section{Linear Landau damping - The Maths}
\end{center}

Just a piece of a vast problem: Energy exchange between waves and particles in a plasma. Simply put, in terms of the energy transfer direction:
\begin{itemize}
 \item Waves $\rightarrow$ Particles: Particle acceleration, wave damping.
 \item Particles $\rightarrow$ Waves: Wave instability.
\end{itemize}
The original paper is Ref. \cite{Landau04}. Landau damping is one of the most studied/debated problem in plasma physics. Nice Maths \textit{and} Physical derivation\footnote{See Kip Thorne's Caltech course ``Applications of Classical Physics'', Chapter 21 mostly for the Maths part at \textit{http://www.pma.caltech.edu/Courses/ph136/yr2004/}.}.

\subsection*{Calculation overview}
Since the calculation is quite subtle and long, it may be useful to get a general overview from the very beginning. Here are the steps we will follow:
\begin{enumerate}
 \item Derivation of the dispersion equation $\epsilon(k,\omega)=0$ from the 1D Vlasov-Poisson system.
 \item Landau contour, the continuity requirement and the Laplace transform.
 \item Resolution for small damping and any distribution function.
 \item Maxwellian distribution.
\end{enumerate}

\subsection*{Dispersion Equation}
Start from 1D non-relativistic equations\footnote{Easily generalized to 3D.} for $F(x,v=p/m,t)$ and field $E(x,t)$,
\begin{eqnarray}
0&=&\frac{\partial F}{\partial t}
+ v\frac{\partial F}{\partial x}
-e \frac{E}{m} \frac{\partial F}{\partial v},\label{1Da}\\
\frac{\partial E}{\partial x} &=& 4\pi e\left[n_0-\int F(x,v,t)dv\right] \label{1Db},
\end{eqnarray}
where $n_0=\int F_0 dv$ is the equilibrium density. Assume $F=F_0+F_1$, with $\mid F_1 \mid \ll \mid F_0 \mid$,  $F_0$ being an equilibrium solution. Same for $E$. The equilibrium electric field $E_0=0$. Linearizing Eqs. (\ref{1Da},\ref{1Db}), assuming $F_1, E_1 \propto \exp(ikx-i\omega t)$, gives
\begin{eqnarray}\label{1D-lin}
0&=&-i\omega F_1 + ikv F_1 -e \frac{E_1}{m} \frac{\partial F_0}{\partial v},\\
i k E_1 &=& -4\pi e\int F_1(x,v,t)dv.
\end{eqnarray}
Extract $F_1$ from the first equation, and plug it into the second,
\begin{eqnarray}
\epsilon(k,\omega)&=&0,~~~\mathrm{with},\nonumber\\
\epsilon(k,\omega)&=&1 - \frac{\omega_p^2}{k^2}\int \frac{f_0'}{v-\omega/k}dv,\label{disperb}
\end{eqnarray}
where $\omega_p^2=4\pi n_0 e^2/m$,  $f_0=F_0/n_0$ and $f_0'=\partial f_0/\partial v$. This dispersion relation was first obtained by Vlasov in 1925 \cite{Vlasov04}. It shows $\omega$ should be imaginary. Otherwise, we have a problem, unless $f_0'(\omega/k)=0$. The dielectric function $\epsilon(k,\omega)$ has therefore a real \textit{and} an imaginary part, which for all kind of systems, is related to \textit{dissipation}.

\begin{figure}[t]
\begin{center}
\includegraphics[width=0.5\textwidth]{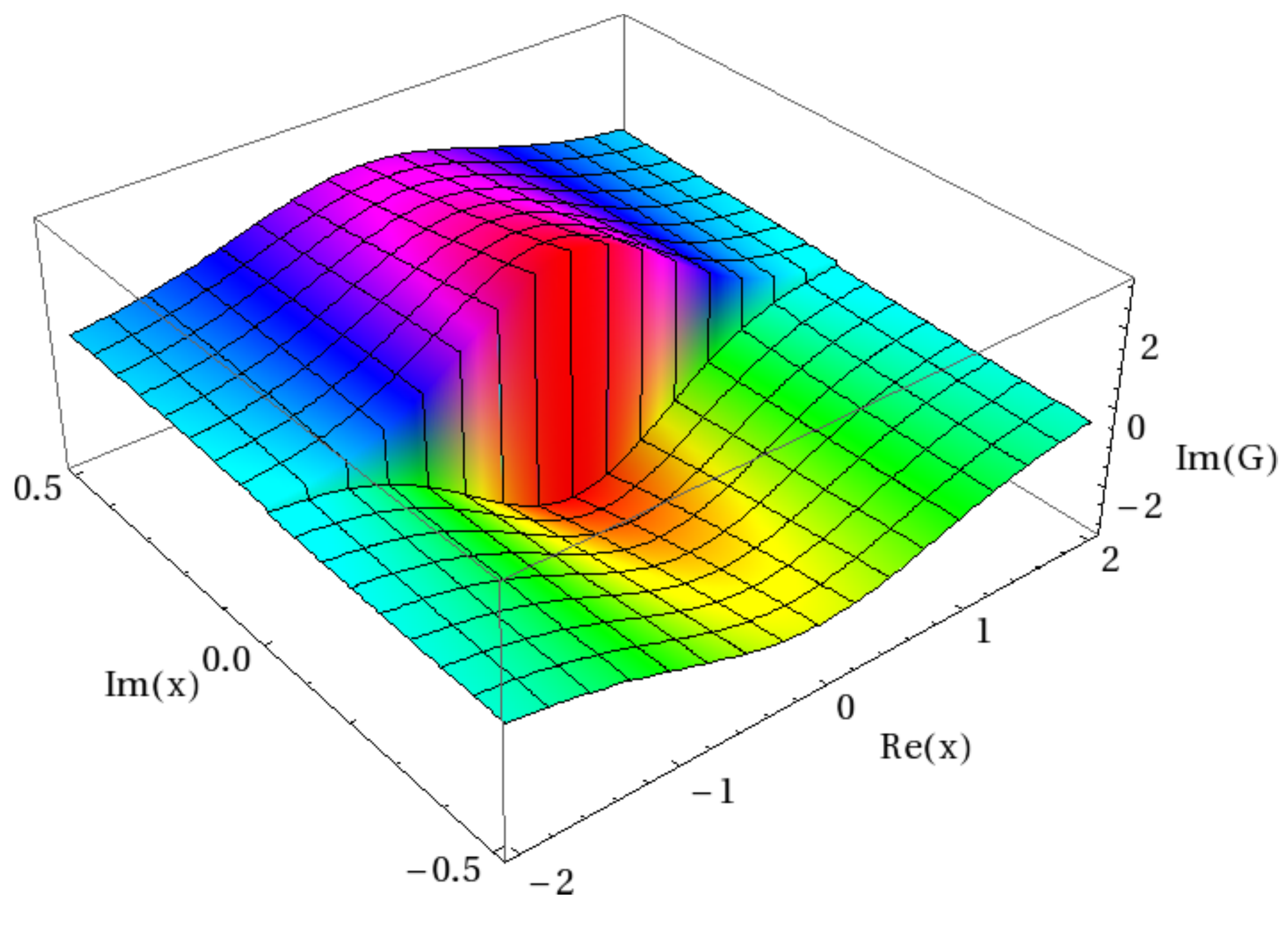}
\end{center}
\caption{Imaginary part of $G=\int e^{-u^2}du/u-x$, for $x\in \mathbb{C}$. The real axis is a discontinuity.} \label{fig:discon}
\end{figure}

We could just consider $\omega$ imaginary and take this quadrature as it is, integrating along the real axis. But there's a problem. The resulting function of $\omega$ is \textit{discontinuous}, precisely when crossing the real axis. As an illustration, Fig. \ref{fig:discon} displays the imaginary part of $G=\int e^{-u^2}du/u-x$, for $x\in \mathbb{C}$. The discontinuity is obvious around $\mathrm{Im}(x)=0$. One part of the plan has to be physically meaningful, and the other not. But which one? We could try both options, and check that damping comes only when choosing the upper one. But what if we didn't know in the first place that a Maxwellian is stable? We shall see that a Laplace analysis of the problem can fully answer the question, and will indeed tell us that the ``physical'' half-plane is the upper one.

Admitting for now the upper-plane is the physical one, what do we do with the lower one? The answer is that we have to ``analytically continuate'' the function we have on the upper-plane, to the lower one. This means finding a function on the lower plane which makes a continuous, ``analytical'' junction, with what we have on the upper one. In this respect, a uniqueness theorem from complex analysis helps: if somehow we find an expression in the lower plane matching what we have in the upper one, then this is the only one.
The Landau contour is going to do all of that for us: providing a contour of integration equivalent to an integration over the real axis for $\mathrm{Im} (\omega)>0$, and an analytical continuation of the later in the lower plane $\mathrm{Im} (\omega)<0$.

\subsection*{Landau contour, the continuity requirement and the Laplace transform}
\begin{figure}[t]
\begin{center}
\includegraphics[width=\textwidth]{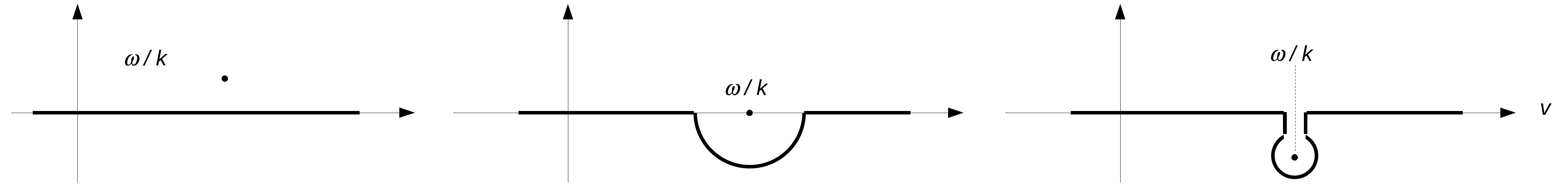}
\end{center}
\caption{The Landau integration contour. It is  not closed. It lies always on the same side of the pole. It lies below the pole.} \label{fig:contour}
\end{figure}

Let's first give the solution found by Landau, namely the famous ``Landau contour''. Figure \ref{fig:contour} shows this integration contour has 3 very distinctive features:
\begin{enumerate}
 \item The Landau contour is \textit{not} closed by the ``usual'' semi-circle in the lower or upper half-plane.
 \item The pole $\omega/k=(\omega_r+i\delta)/k$ must always lie on the \textit{same side} of the Landau contour.
 \item So, which side? The Landau contour goes \textit{below} the pole.
\end{enumerate}
These contour prescriptions are called the ``Landau prescriptions'', and the corresponding contour, the ``Landau contour''. We thus rewrite from now on Eq. (\ref{disperb}) as
\begin{equation}\label{disperL}
\epsilon(k,\omega)=1 - \frac{\omega_p^2}{k^2}\int_L \frac{f_0'}{v-\omega/k}dv,
\end{equation}
where $\int_L$ means integration along the Landau contour. Let's now find out about these 3 features.

\subsubsection*{The contour is not closed}\label{notclosed}
The contour just goes from $v=-\infty$ to $+\infty$, and is \textit{not} closed in the upper or lower half-plane, as ``usual'', because we have no guarantee $f_0'(v)$ behaves correctly there, so as to cancel the integration on the semi-circle at infinite radius. Indeed, considering a Maxwellian with $f_0'\propto e^{-v^2}$ and setting $v=R_ve^{i\theta_v}$ to parameterize the integration on a circle of radius $R_v$, we find $f_0'\propto e^{-R_v^2\cos 2\theta_v}$ which can hardly be considered a vanishing quantity at $R_v\rightarrow\infty$ for \textit{any} $\theta_v\in [0,\pi]$ (or $[0,-\pi]$, if you close in the lower half-plane).

\subsubsection*{Always on the same side}\label{side}
Assume $\omega=\omega_r+i\delta$, with $\delta>0$. As long as $\delta$ remains positive in Eq. (\ref{disperb}), the calculation does not pose any conceptual problem as the pole is not on the real axis, and continuity is guaranteed.

Now, what if $\delta$ approaches 0, and the pole $\omega/k$ even gets to cross the real axis? We would like $\epsilon(k,\omega)$ to be a \textit{continuous} function of $\omega$. Assume first we leave the integration contour unchanged (the real axis for $v$), and compare the quadrature for $\omega=\omega_r+i\delta$ and $\omega=\omega_r-i\delta$. The influence of the pole is mostly felt where the denominator is minimum at $v\sim\omega_r/k$, so let's locally get $f_0'$ out of the integral and compare,
\begin{equation}\label{compare}
I_1=\int \frac{dv}{v-(\omega_r+i\delta)/k}~~~~\mathrm{and}~~~~I_2=\int \frac{dv}{v-(\omega_r-i\delta)/k}.
\end{equation}
The difference $I_1-I_2$ is,
\begin{equation}\label{compare1}
I_1-I_2=2i\int \frac{\delta/k}{(v-\omega_r/k)^2+(\delta/k)^2}dv.
\end{equation}
The continuity of $\epsilon(k,\omega)$ demands the expression above vanishes when $\delta \rightarrow 0^+$. The problem is that it does \textit{not}. Instead, the quadrature tends to $\pi$ (see function $G_2$ in Appendix A), so that we indeed have a \textit{jump} of amplitude $2i\pi$ when crossing the real axis\footnote{It can also be said that for $I_1$, the integration path makes a \textit{counter}-clockwise half-turn around the pole, so that $I_1=i\pi$. But for $I_2$, the half-turn around the pole is clockwise, so that $I_2=-i\pi$ and $I_1-I_2=2i\pi$.}.

The only way to avoid this is to deform the integration contour in such a way that it always lies \textit{on the same side} of the pole $\omega/k$.

\subsubsection*{The contour goes below the pole}\label{below}
\begin{figure}[t]
\begin{center}
\includegraphics[width=\textwidth]{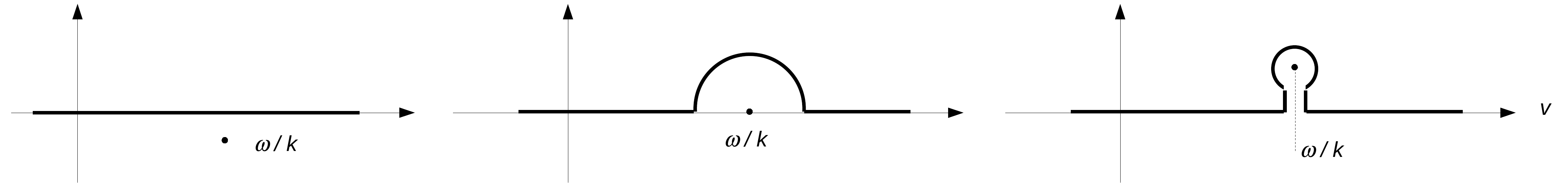}
\end{center}
\caption{Forbidden option for the contour. Continuity is preserved, but the contour lies \textit{above} the pole, in contradiction with the Laplace prescription.} \label{forbid}
\end{figure}

To understand why the contour goes \textit{below} the pole and not above like in Fig. \ref{forbid}, we need to follow Landau in rethinking the problem in terms of the time evolution of a perturbation applied at $t=0$. The Fourier technique is not well suited for that because it entails an integration from $t=-\infty$ to $+\infty$. By design, it does not single out any special moment in between. By contrast, the \textit{Laplace transform}  involves times only from \textit{zero} to $+\infty$. As shall be checked, the Laplace transform technique gives an unambiguous response about the location of the pole with respect to the integration contour.

\begin{figure}[t]
\begin{center}
\includegraphics[width=0.5\textwidth]{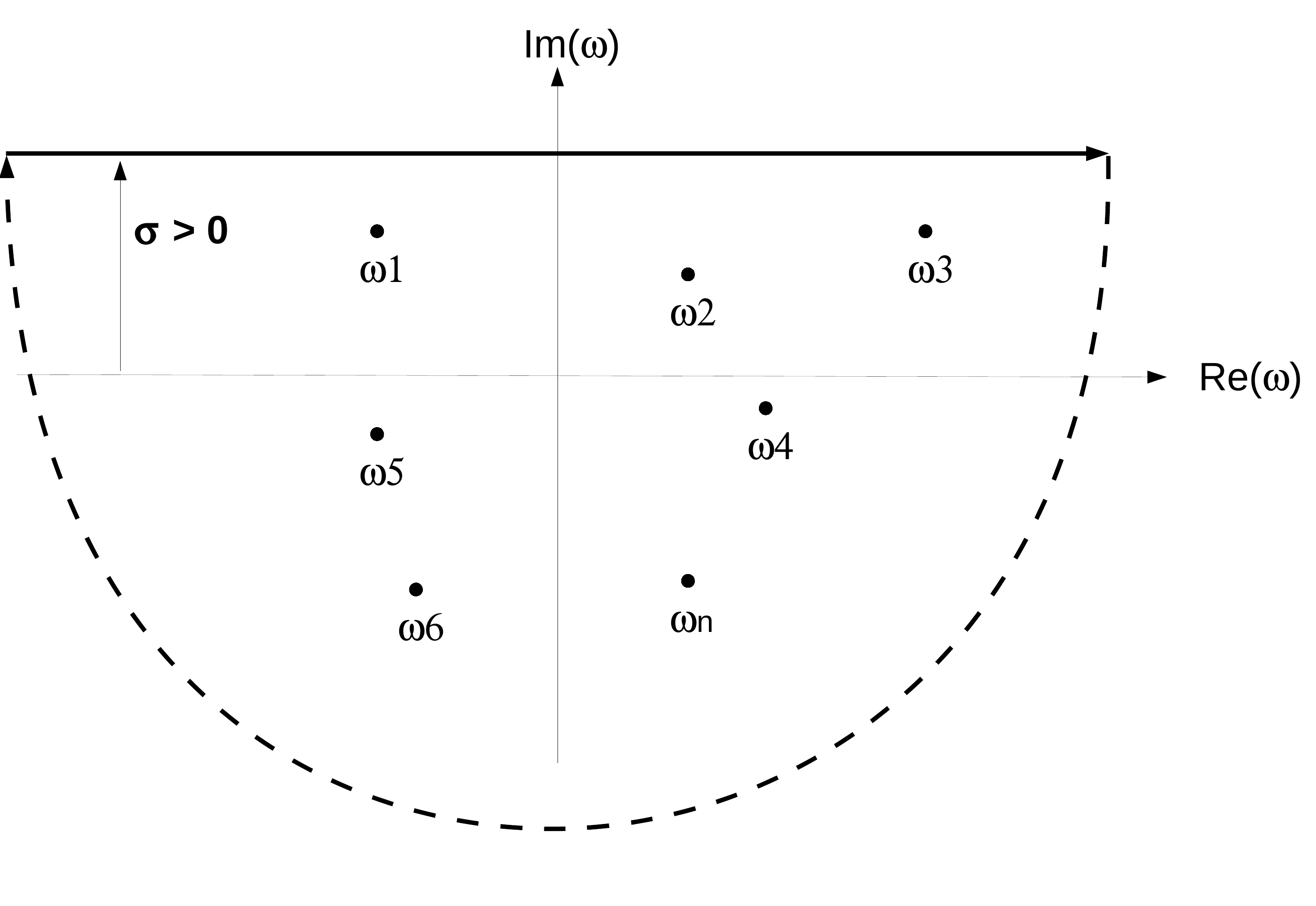}
\end{center}
\caption{Laplace integration contour. Goes from $\omega=-\infty +i\sigma$ to $+\infty +i\sigma$ with $\sigma>0$, and is closed in the lower half-plane. By design, $\sigma>0$ and such that every single poles $\omega_1,\ldots,\omega_n$ of the integrand lie \textit{inside} the contour.} \label{fig:contourL}
\end{figure}

Considering a function $h(t)$, its Laplace transform $\widehat{h}(\omega)$ and the inversion formula\footnote{I here follow Landau's book, \cite{LL04}, p. 139, in defining the Laplace transform this way. That's just the usual one,
\begin{equation}
g(p)=\int_0^\infty g(t)e^{-pt}dt,
\end{equation}
for $p=-i\omega$. It avoids having to rotate everything in the complex plane to relate the calculation to Eq. (\ref{disperb}).
}, read
\begin{eqnarray}\label{laplace}
\widehat{h}(\omega)&=&\int_0^\infty e^{i\omega t}h(t)dt,\label{laplacea}\\
h(t)&=&\int_{C_L}e^{-i\omega t}\widehat{h}(\omega)d\omega\label{laplaceb},
\end{eqnarray}
where the contour $C_L$ pictured on Fig. \ref{fig:contourL}, passes above \textit{all} the poles of $\widehat{h}(\omega)$ at height $\sigma>0$, and can be closed in the lower half-plane where $e^{-i\omega t}$ behaves conveniently as to cancel the integral at infinity there. Note that although the requirement $\sigma>0$ is emphasized in the book (p. 139), I still have to understand why being \textit{above} all the poles is not enough. And as we shall see very soon,  $\sigma>0$ is the key to the choice of the right part of the $\omega$ complex plane.

Let's compute from the Maxwell-Vlasov Eqs. (\ref{1Da},\ref{1Db}) the time evolution of the system considering,
\begin{eqnarray}\label{laplace1}
F(x,v,t)&=&n_0f_0(v) + F_1(v,t)e^{ikx},\\
E(x,t)&=&E_1(t)e^{ikx},
\end{eqnarray}
assuming $F_1,E_1$ are first order quantities, and $F_1(v,t=0)e^{ikx}$ is the perturbation initially applied. The linearized Vlasov equation reads,
\begin{equation}\label{VlasovL}
\frac{\partial F_1(v,t)}{\partial t}
+ ikvF_1(v,t)
-\frac{e n_0}{m}E_1(t) f_0'(v)  = 0.
\end{equation}
If we multiply by $e^{i\omega t}$ and take the integral from $t=0$ to $+\infty$, an integration by part on the time derivative term gives,
\begin{eqnarray}\label{ipp}
\int_0^\infty e^{i\omega t}\frac{\partial F_1(v,t)}{\partial t}dt
&=&\left[e^{i\omega t}F_1(v,t) \right]_0^\infty- i\omega\int_0^\infty e^{i\omega t}F_1(v,t)dt\nonumber\\
&=&-F_1(v,0) - i\omega\widehat{F}_1(v,\omega),
\end{eqnarray}
where $\lim_{t\rightarrow\infty}e^{i\omega t}F_1(v,t)=0$ has been assumed. On the one hand, the very existence of the Laplace transform of $F_1(v,\omega)=\int F_1(v,t)e^{i\omega t}dt$ implies it. On the other hand, a important conclusion of the paper is that for large times, $F_1(v,t)\propto e^{ikvt}$ (see \cite{Landau04} p. 452, and \textit{Plasma Talk 5}). This point is discussed neither in the book, nor in the original paper. Using Eqs. (\ref{ipp},\ref{VlasovL}) then gives,
\begin{equation}\label{VlasovLP}
(ikv-i\omega)\widehat{F}_1(v,\omega)
-\frac{e n_0}{m}\widehat{E}_1(\omega) f_0'(v)  = F_1(v,0),
\end{equation}
where $F_1(v,0)$ now acts like a ``source term'' at the right-hand-side. A few more manipulations exploiting Poisson's equation (\ref{1Db}) give,
\begin{equation}\label{VlasovLP1}
\widehat{E}_1(\omega) = \frac{1}{\epsilon(k,\omega)}\frac{4\pi e}{k^2}\int_{-\infty}^{\infty} \frac{F_1(v,0)dv}{v-\omega/k},
\end{equation}
where $\epsilon(k,\omega)$ is \textit{identical} to Eq. (\ref{disperb}). The time dependant electric field given by the inversion formula (\ref{laplaceb}) is,
\begin{equation}\label{LaplaceEt}
E_1(t) = \int_{C_L}e^{-i\omega t}\widehat{E}_1(\omega)d\omega
= \int_{C_L}\frac{e^{-i\omega t}}{\epsilon(k,\omega)}\left[\frac{4\pi e}{k^2}\int_{-\infty}^{\infty} \frac{F_1(v,0)dv}{v-\omega/k} \right] d\omega.
\end{equation}

In contradistinction with Eq. (\ref{disperb}) where the contour issue is puzzling, the Laplace technique used here is clear: The $v$-integration in $\epsilon(k,\omega)$ does go along the real axis, and the $\omega$-integration is performed at fixed $\mathrm{Im}(\omega)=\sigma>0$. It means that in Eq. (\ref{LaplaceEt}), which computes a physical quantity, the dielectric function $\epsilon(k,\omega)$ is calculated with $\omega$ \textit{above} the real $v$-axis.

That answers the question we had: the physically meaningful half-plane we were wondering about after Eq. (\ref{disperb}) is the \textit{upper} one. The kind of contour pictured on Fig. \ref{forbid} is thus ``forbidden''.

Incidentally, what are the poles of the integrand in Eq. (\ref{LaplaceEt})? For ``normal'', smooth initial excitations $F_1(v,0)$, the term between brackets won't have poles, so that our poles $\omega_1,\ldots,\omega_n$ \textit{are eventually the zeros of} $\epsilon(k,\omega)$. The $\omega$-integration of Eq. (\ref{LaplaceEt}) on the \textit{closed} contour $C_L$ will  thus give, with $\omega_j=\omega_{r,j}+i\delta_i$,
\begin{equation}\label{LaplaceEtF}
E_1(t) = 2i\pi\sum_{j=1}^n Res(j) \equiv \sum_{j=1}^n A_j\exp(-i\omega_jt)= \sum_{j=1}^n A_j\exp(-i\omega_{r,j}t)e^{\delta_j t},
\end{equation}
which for large times will be governed by the largest $\delta_j$. Therefore, the Laplace transform approach cannot spare us the resolution of $\epsilon(k,\omega)=0$, as these zeros are the building blocks of the temporal response of the system.

\subsection*{Resolution for small damping}
We suppose small damping, that is $|\delta| \ll |\omega_r|$, and Taylor expand Eq. (\ref{disperL}),
\begin{eqnarray}\label{epsi}
\epsilon(k,\omega_r+i\delta)&=&
\epsilon_r(k,\omega_r)+i\delta\left.\frac{\partial \epsilon_r}{\partial \omega_r}\right|_{\delta=0}
+ i\left[\epsilon_i(k,\omega_r)+i\delta\left.\frac{\partial \epsilon_i}{\partial \omega_r}\right|_{\delta=0} \right]
\nonumber\\
&=& \epsilon_r(k,\omega_r) + i\epsilon_i(k,\omega_r)
+ \delta\left[ i\frac{\partial \epsilon_r}{\partial \omega_r}-\frac{\partial \epsilon_i}{\partial \omega_r}
\right]_{\delta=0}\nonumber\\
&=& \epsilon(k,\omega_r)
+ i\delta\left. \frac{\partial \epsilon_r}{\partial \omega_r}\right|_{\delta=0}+o(\delta),
\end{eqnarray}
where the $o(\delta)$ (negligible with respect to $\delta$), comes from the fact that $\epsilon_i(\delta=0)=0$ (no damping, no dissipation, no imaginary dielectric function).

The first term $\epsilon(k,\omega_r)$ is given by Eq. (\ref{disperL}) setting $\omega=\omega_r$, or taking the limit of $\epsilon(k,\omega_r+i\delta)$ for $\delta \rightarrow 0^+$. The part of the integration along the real axis for $v\in [-\infty,\omega_r/k-\varepsilon]\cup [\omega_r/k+\varepsilon,+\infty]$ gives the so-called ``Cauchy Principal Part'', denoted \textbf{P} here. The part corresponding to the semi-circle (see Fig. \ref{fig:contour} middle) gives the semi-residue for $v=\omega_r/k$. An alternative way of deriving this result, considering the limit $\delta \rightarrow 0^+$, is reported in Appendix A. We thus get,
\begin{equation}\label{quadra0}
\epsilon(k,\omega_r) = 1-\frac{\omega_p^2}{k^2}\left[ \mathbf{P}\int \frac{f_0'}{v-\omega_r/k}dv
+i\pi f_0'(\omega_r/k)\right].
\end{equation}
This result allows to compute  $\partial \epsilon_r/\partial \omega_r$ in Eq. (\ref{epsi}), which eventually gives,
\begin{equation}\label{DisperFinal}
\epsilon(k,\omega)=1-\frac{\omega_p^2}{k^2}\mathbf{P}\int \frac{f_0'}{v-\omega_r/k}dv
-i\frac{\omega_p^2}{k^2}\left[
\pi f_0'(\omega_r/k)+
\delta\frac{\partial }{\partial \omega_r}
\mathbf{P}\int \frac{f_0'}{v-\omega_r/k}dv
\right].
\end{equation}
Equating the real part to zero yields,
\begin{equation}\label{DisperReal}
\frac{\omega_p^2}{k^2}\mathbf{P}\int \frac{f_0'}{v-\omega_r/k}dv=1,
\end{equation}
which was the result obtained by Vlasov in the first place. Canceling the imaginary part gives directly the damping rate,
\begin{equation}\label{DisperIm}
\delta = -\pi \frac{f_0'(\omega_r/k)}{\frac{\partial }{\partial \omega_r}
\mathbf{P}\int \frac{f_0'}{v-\omega_r/k}dv}.
\end{equation}
Eqs. (\ref{DisperReal}, \ref{DisperIm}) formally solve the problem in terms of the distribution function. A first order evaluation of \textbf{P}$\sim k^2/\omega_r^2$ (see Eq. (\ref{PPOK}) below), gives
\begin{eqnarray}\label{RealIm}
\omega_r &=& \omega_p,~~~~\mathrm{and}~\mathrm{then}\\
\frac{\delta}{\omega_p}&=&\frac{\pi}{2}\frac{\omega_p^2}{k^2}f_0'(\omega_p/k).\nonumber
\end{eqnarray}

The rate $\delta$ has the sign of $f_0'(\omega_p/k)$. That means that if $f_0$ decreases for $v=\omega_p/k$, the waves is damped because $\delta<0$. But if $f_0$ increases for $v=\omega_p/k$, we have $\delta>0$ and the wave can actually \textit{grow}.

One could argue we started initially assuming $\delta$ positive, and find it can be negative here. It is  not a problem for the following reason: Eq. (\ref{DisperFinal}) we found assuming $\delta>0$ is \textit{continuous} at $\delta=0$. It must therefore be identical to the integration \textit{on the Landau contour} on both sides of the real axis. We can therefore confidently solve it regardless of the sign of $\delta$. In other words, thanks to the Landau contour, we can compute the result as if $\delta$ was positive, and then don't care about the sign.

Historically, Vlasov first ran into Eq. (\ref{disperb}). He escaped the problem posed by the pole on the real axis by considering only the \textbf{P} of the quadrature. He did so apparently without much foundation, which Landau denounced without mercy in \cite{Landau04}. We understand from the analysis above that doing so, he missed the imaginary part which would have led to the ``Vlasov damping''.

\subsection*{Maxwellian distribution}
Let's finally consider a 1D Maxwellian distribution,
\begin{equation}
f_0(v)=\frac{1}{\sqrt{2\pi k_BT/m}}e^{-mv^2/2k_BT}.
\end{equation}
For phase velocities $\omega_r/k$ much larger than the thermal velocity $V_{th}=\sqrt{k_BT/m}$, we can expand the denominator in powers of $kv/\omega_r$, since that quantity is small where the numerator is relevant. We thus have,
\begin{eqnarray}\label{PPOK}
\mathbf{P}\int \frac{f_0'}{v-\omega_r/k}dv&=&
-\frac{k}{\omega_r}\int f_0' \left(1+\frac{kv}{\omega_r}+\frac{k^2v^2}{\omega_r^2}+\frac{k^3v^3}{\omega_r^3}+\cdots \right) dv\nonumber\\
&=&\frac{k^2}{\omega_r^2}+3\frac{k_BT}{m}\frac{k^4}{\omega_r^4}+\cdots
\end{eqnarray}
For small $k$, namely $kV_{th}/\omega_r\sim kV_{th}/\omega_p\ll 1$, Eq. (\ref{DisperReal}) now gives,
\begin{equation}\label{real}
\omega_r^2 = \omega_p^2(1+3k^2\lambda_D^2),~~~\mathrm{with}~~~\lambda_D=\frac{\sqrt{k_BT/m}}{\omega_p}.
\end{equation}
We finally (phew!) use Eq. (\ref{DisperIm}) to extract the damping rate. On the one hand, we compute the derivative of the \textbf{P} with respect to $\omega_r$ using Eq. (\ref{PPOK}), and then simply set $\omega_r=\omega_p$ in the result. On the other hand, we set $\omega_r=\omega_p$ in $f_0'$ to find\footnote{Some authors insert the full expression of $\omega_r$ from Eq. (\ref{real}), yielding $-1/2k^2\lambda_D^2-3/2$ in the argument of the exponential.}
\begin{equation}\label{Ima}
\delta = -\omega_p\frac{\sqrt{\pi/8}}{k^3\lambda_D^3}\exp\left(-\frac{1}{2k^2\lambda_D^2} \right) .
\end{equation}

Fluid theory just gives the real part of the frequency, namely Eq. (\ref{real}), so that Landau damping is a purely kinetic effect.

\subsection*{Appendix A}
Let's derive,
\begin{equation}\label{quadraa}
\int_L \frac{f_0'}{v-\omega_r/k}dv = \mathbf{P}\int \frac{f_0'}{v-\omega_r/k}dv +i\pi f_0'(\omega_r/k),
\end{equation}
used for Eq. (\ref{quadra0}), without using the residue theorem. For $\omega=\omega_r+i\delta$ with $\delta>0$, integration along the Landau contour is equivalent to an integration along the real axis. Let's thus assume $\delta>0$ and compute,
\begin{equation}
I=\lim_{\delta\rightarrow 0^+} \int_{-\infty}^\infty \frac{f_0'}{v-(\omega_r+i\delta)/k}dv.
\end{equation}

We multiply the numerator and the denominator of the integrand by $(v-\omega_r/k)+i\delta/k$, which is the complex conjugate of the denominator. We get an expression with a purely real, non singular denominator, and clearly separated real and imaginary parts,
\begin{equation}\label{quadra2}
I=\lim_{\delta\rightarrow 0^+}   \int_{-\infty}^\infty \underbrace{\frac{(v-\omega_r/k)^2 }{(v-\omega_r/k)^2+(\delta/k)^2}}_{G_1} \frac{f_0'}{(v-\omega_r/k)}dv
+ i  \int_{-\infty}^\infty \underbrace{\frac{\delta/k}{(v-\omega_r/k)^2+(\delta/k)^2}}_{G_2}f_0'dv.
\end{equation}

\begin{figure}[t]
\begin{center}
\includegraphics[width=0.45\textwidth]{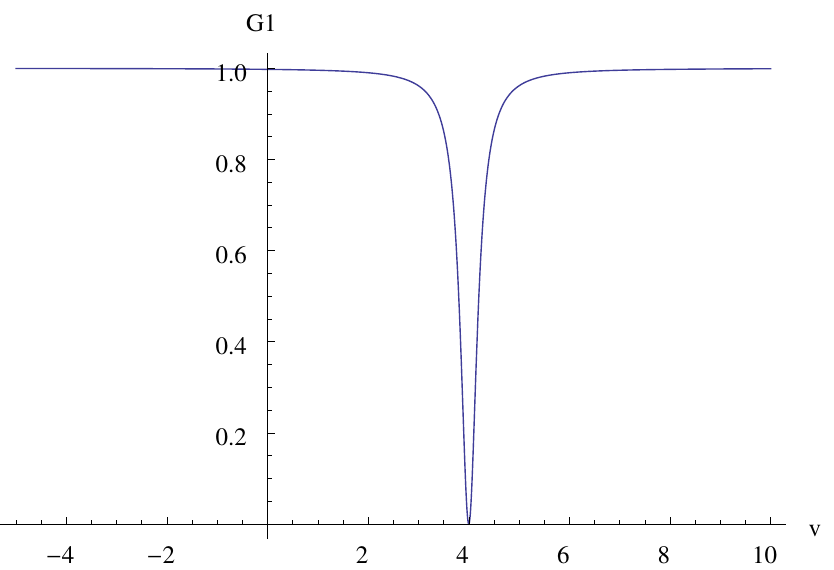}\includegraphics[width=0.45\textwidth]{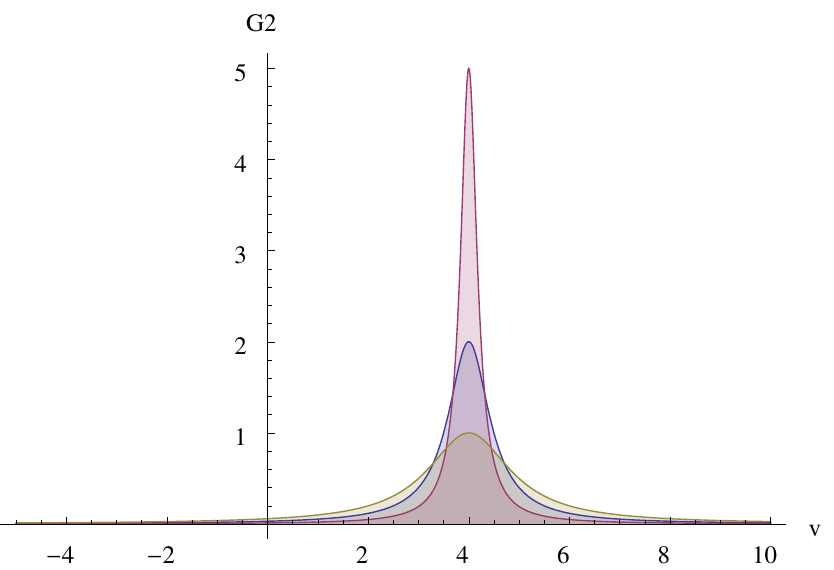}
\end{center}
\caption{Functions $G_1$ and $G_1$ involved in  Eq. (\ref{quadra2}). For small $\delta/k$, $G_1$ is almost 1 everywhere, except for $v=\omega_r/k$, where it is 0. $G_2$ peaks at $v=\omega_r/k$ and tends to 0 elsewhere, while its integral is always $\pi$, like a Dirac $\delta$ function. Parameters are $\omega_r/k=4$, $\delta/k=0.2$ for $G_1$, and $\delta/k=0.2, 0.5, 1$ for $G_2$.} \label{fig:Gs}
\end{figure}
Regarding the \emph{real part}, the factor $G_1$ of the integrand is 0 for $v=\omega_r/k$, and $\sim 1$ for $\delta/k\ll |v-\omega_r/k|$. It tends to the \textbf{P}$\int f_0'/(v-\omega_r/k)$ for small $\delta/k$ (see Fig.  \ref{fig:Gs}). The factor $G_2$ of the integrand of the \emph{imaginary part} departs from 0 only for $v\sim \omega_r/k$. But its integral is always $\pi$. For small $\delta/k$, the quadrature thus tends to $\pi f_0'(\omega_r/k)$, and we are back to (\ref{quadraa})\footnote{
The limit of $i\delta$ with $\delta\rightarrow 0^+$ is sometimes written ``$i0$''. The identity
\begin{equation*}
\lim_{\delta\rightarrow 0^+} \int_{-\infty}^\infty \frac{h(x)}{x-a-i\delta}dx
\equiv \int_{-\infty}^\infty \frac{h(x)}{x-a-i0}dx
=\mathbf{P}\int_{-\infty}^\infty \frac{h(x)}{x-a}dx + i\pi h(a),
\end{equation*}
can be referred to as the ``Plemelj Formula'' in the literature. For $\delta\rightarrow 0^-$, the imaginary part above is $-i\pi h(a)$.}.

This calculation is consistent with the Landau contour integration \textit{only} for $\delta\rightarrow 0^+$. This is because in such case, the real axis along which we perform the integration (\ref{quadra2}) coincide with the Laundau contour. If we were to compute Eq. (\ref{quadra2}) for $\delta\rightarrow 0^-$, we would find the opposite imaginary part, just because in this case, the real axis \textit{no longer} fits the Landau contour. The latter, instead, is deformed and keeps passing \textit{below} the pole, precisely to avoid the discontinuity.

%% file: PlasmaTalk05.tex
\begin{center}
\section{Landau damping - The Physics, Plasma Echo, and a (little) word about the non-linear problem}
\end{center}

While the original paper \cite{Landau} is purely mathematical, a clearer physical picture is provided in Landau's book (\cite{LL}, \S30 p. 126). Suppose we switch on at $t=0$ a 1D electrostatic wave  $\mathbf{E}=E_0\sin(kx-\omega t)\mathbf{x}$, traveling at $v_\phi=\omega/k$ along with a particle with velocity $v_0$ at $t=0$. For $v_0$ slightly \textit{larger} than $v_\phi$, the particle is trapped in the wave potential, where it is going to oscillate. Doing so, it ends up with an average velocity close to the wave velocity $v_\phi$. It should thus loose energy, and the energy goes to the wave. Situation is reversed for particles initially slightly \textit{slower} than the wave. They end up gaining energy from the wave.

If slower particles are more numerous than the faster ones, the wave looses more than it gains, which means it is damped. Let's now ``Fermi-calculate'' this, not following Fermi but Jackson \cite{Jackson} and Spitzer \cite{Spitzer} (who follows Jackson). Landau uses a slightly different approach, still implying a calculation with some small parameter eventually tending to zero. I chose Jackson\footnote{The J.D. Jackson  who wrote \textit{Classical Electrodynamics}.} precisely because there's no such trick in his strategy. The reasoning is \textit{non}-relativistic.

To start with, which particles can enter the game? If their velocity is too high \textit{relatively} to the wave, they will flow from one potential crest to another, without much net energy exchange. The ones for which energy exchange is possible, are the ones which will be trapped by the potential. The wave potential height is,
\begin{equation}\label{high}
\Delta \varphi=\frac{q E_0}{k}.
\end{equation}
The maximum particle velocity $\Delta v$ in the wave-frame at $v_\phi$ must then satisfy,
\begin{equation}\label{trapped}
\frac{1}{2}m(\Delta v)^2=\Delta \varphi~~\Rightarrow~~(\Delta v)^2=\frac{2q E_0}{mk}.
\end{equation}

\begin{figure}[t]
\begin{center}
\includegraphics[width=0.5\textwidth]{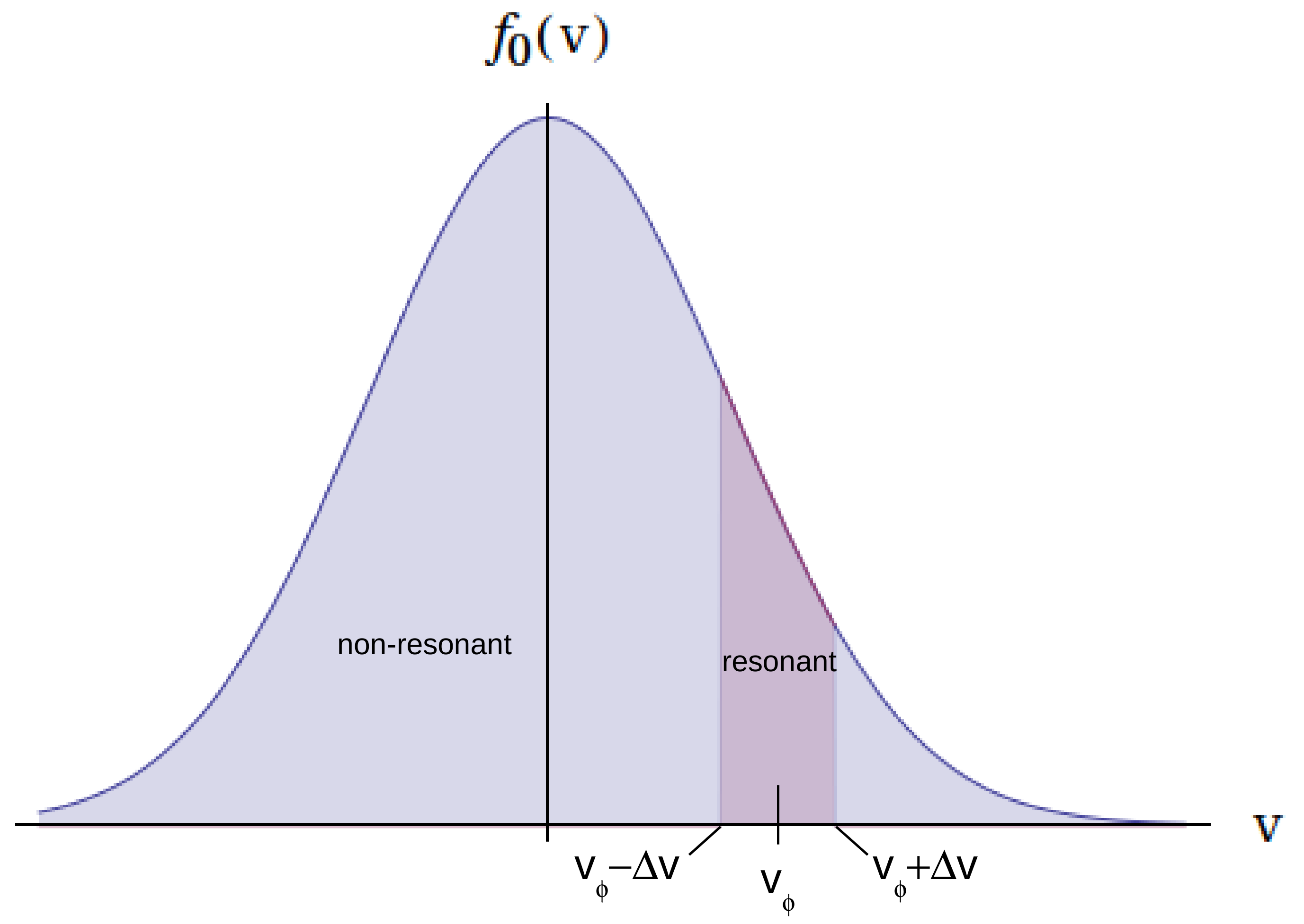}
\end{center}
\caption{Division of the plasma between non-resonant and resonant (trapped) particles. Only resonant particles contribute to the calculation. After \cite{Dawson}.} \label{fig:trap}
\end{figure}

Thus, only particles with velocity $v\in[v_\phi-\Delta v,v_\phi+\Delta v]$ in the lab-frame can be caught by the wave. This is pictured on Fig. \ref{fig:trap}, taken from another great work on Landau damping by Dawson \cite{Dawson}. For a particle near the center of this interval, we take $\sin x\sim x$ for the field, and the equation of motion in the wave-frame reads
\begin{equation}\label{motion}
m\frac{d^2x}{dt^2}+qE_0kx=0~~\Rightarrow~~\frac{d^2x}{dt^2}+\omega_{osc}^2~x=0,~~\mathrm{where}~~\omega_{osc}^2=\frac{qE_0k}{m},
\end{equation}
showing it oscillates in the wave potential with frequency $\omega_{osc}$.

Before it was trapped, the particle energy in the lab-frame was just $W_1=mv^2/2$. After the trapping, the energy is
\begin{equation}\label{nrj}
W_2=\frac{1}{2}m\left(v_\phi^2 + (v-v_\phi)^2\right),
\end{equation}
where $\frac{1}{2}mv_\phi^2$ is the end translational kinetic energy, and $\frac{1}{2}m(v-v_\phi)^2$ can be viewed as an internal energy of oscillation\footnote{The energy of a mass oscillating is a potential is split between its kinetic energy and its potential energy. At the bottom of the potential well, all the energy is kinetic. The term $\propto (v-v_\phi)^2$ in Eq. (\ref{nrj}) is the kinetic energy, in the wave frame, of the particle at the bottom of the well.}. If there are $n_0f_0(v)dv$ particles with such velocities, with $\int f_0=1$, the energy shift is
\begin{equation}
dW=n_0f_0(v)dv(W_2-W_1),
\end{equation}
which we integrate over all particles capable of such exchange,
\begin{equation}\label{nrjtot}
\Delta W=\int_{v_\phi-\Delta v}^{v_\phi+\Delta v}n_0f_0(v)dv
\frac{1}{2}m\left(v_\phi^2 + (v-v_\phi)^2-v^2\right).
\end{equation}
Expanding $f_0(v)=f_0(v_\phi)+(v-v_\phi)f_0'(v_\phi)+\cdots$, the term corresponding to $f_0(v_\phi)$ in Eq. (\ref{nrjtot}) vanishes\footnote{It does \textit{not} vanish if you omit the ``internal energy'' term in Eq. (\ref{nrj}).}, and we find
\begin{equation}\label{nrjtot1}
\Delta W=-\frac{2}{3}m n_0 v_\phi \Delta v^3 f_0'(v_\phi).
\end{equation}
If, like on Fig. \ref{fig:trap}, particles slower than $v_\phi$ are more numerous than faster ones, $f_0'(v_\phi)<0$ and $\Delta W>0$, which means particles \textit{gain} energy at the expense of the wave. The wave is \textit{damped}. We now just have to write that this energy leaves the field \textbf{E} over a time scale $\omega_{osc}^{-1}$,
\begin{equation}\label{field}
\frac{d (E_0^2/8\pi)}{dt} = -\omega_{osc}\Delta W = \omega_{osc}\frac{2}{3}m n_0 v_\phi \Delta v^3 f_0'(v_\phi).
\end{equation}
Plugging here the expressions for $\Delta v$ and $\omega_{osc}$ from Eqs. (\ref{trapped},\ref{motion}) we find,
\begin{eqnarray}\label{field1}
\frac{d (E_0^2/8\pi)}{dt} &=&
\frac{8\sqrt{2}}{3}\omega \frac{\omega_p^2}{k^2}f_0'(v_\phi)\left( \frac{E_0^2}{8\pi}\right),\nonumber\\
&\equiv& 2\delta\left( \frac{E_0^2}{8\pi}\right) .
\end{eqnarray}
As the field energy $\propto E_0^2$ is damped at $2\delta$, the field itself is damped at $\delta$. Setting finally $\omega=\omega_p$, we have
\begin{equation}\label{deltafield}
\frac{\delta}{\omega_p} = \frac{4\sqrt{2}}{3}\frac{\omega_p^2}{k^2}f_0'(\omega_p/k),
\end{equation}
identical to Eq. (25) of \textit{Plasma Talk 4}, up to a numerical pre-factor close to 1 ($\pi/2=1.57$ and $4\sqrt{2}/3=1.88$). A discussion of the \textit{non}-Galilean invariance of Eq. (\ref{deltafield}) is available in \cite{Jackson} (p. 180).

\subsection*{A word on Landau damping and gravitation}
According to Ref. \cite{Gayer}, Landau Damping of Gravitational Waves would \textit{not} be possible. Much has been done with respect of Landau Damping of more mundane ``gravity waves''. \textit{The stability and vibrations of a gas of stars}, by Lynden-Bell, seems to be a quite influential paper \cite{Lynden}. The abstract concludes stating ``Landau Damping occurs for wave-length smaller than the critical one [Jean's]''.

\subsection*{Plasma Echo}
Fascinating consequence of the fact that the \textit{density} relaxes whereas the \textit{distribution function} does \textit{not} (many functions have the same integral). Original idea by Gould \textit{et al.} \cite{Gould}.

Suppose we produce an initial electric field perturbation $\propto e^{-ik_1x}$ in the plasma. The Laplace analysis \cite{Landau} of the \textit{distribution function} temporal evolution (not the field, nor the density) shows it \textit{indefinitely oscillates} with $F=f_0+f_1(v)\exp(ik_1vt-ik_1x)$. For large times, any velocity integral  ``phase''-vanishes,
\begin{equation}\label{phase}
\lim_{t\rightarrow\infty}\int f_1(v)e^{ik_1vt-ik_1x}dv=0,
\end{equation}
which is how we recover zero field and density perturbations. The density perturbation and the field associated with $f_1$ die out, but $f_1$ \textit{doesn't}. This is how you reconcile the necessary reversibility of the Vlasov-Maxwell system, with the apparent irreversibility of Landau Damping. There only seems to be a \textit{macroscopic} irreversibility, but the evolution in \textit{microscopically} reversible.

Is it possible to detect this ever oscillating $f_1(v)$ at later times ? Yes. Assume we wait for a time $\tau$, and send another perturbation in the plasma $\propto e^{ik_2x}$. The second perturbation is going to modulate both $f_0$ and $f_1$ according to $e^{ik_2v(t-\tau)-ik_2x}$. Regarding $f_1$, we will recover something varying like
\begin{equation}\label{phase1}
e^{ik_1vt-ik_1x}e^{ik_2x-ik_2v(t-\tau)}=e^{i(k_2-k_1)x+ik_2v\tau-i(k_2-k_1)vt }.
\end{equation}
The key-point here is that contrary to Eq. (\ref{phase}), where $k_1t\neq 0$ implies the velocity integral vanishes at large times, together with the first order density and field, the coefficient of $v$ in the exponential above is exactly canceled at time,
\begin{equation}\label{phase2}
t=\frac{k_2}{k_2-k_1}\tau.
\end{equation}
At this time, the velocity integral will \textit{not} vanish, and an electric field should reappear in the plasma. So you perturb a plasma. You wait until everything apparently calmed down. Then you send another perturbation, and at the time prescribed by Eq. (\ref{phase2}), an electric field will suddenly pop-up ``out of nowhere'', related to the perturbation you first sent. That is the ``plasma echo''.

The idea was experimentally tested soon after the theory came, and the echo was found \cite{EchoXt}. Mouhot \& Villani put it this ways: ``A plasma which is apparently back to equilibrium after an initial disturbance, will react to a second disturbance in a way that shows that it has not forgotten the first one'' (\cite{Villani}, p. 40).

Regarding gravitational systems, Lynden-Bell wrote ``A system whose density has achieved a steady state will have information about its birth still stored in the peculiar velocities of its stars'' (\cite{Lynden}, p. 295).

\subsection*{\textit{Nonlinear} Landau damping}
We found linear waves are damped. Landau Damping  has been \textit{experimentally} confirmed \cite{Malmberg}. Here are a few landmarks for \textit{large} amplitude waves (1D, non-relativistic)\footnote{Thanks to Giovanni Manfredi for the summary!}:
\begin{itemize}
 \item Isichenko 1997 \cite{Isichenko}: Landau damping valid $\forall$ amplitude (Theory).
 \item Manfredi 1997 \cite{Manfredi}: Some large amplitude waves \textit{do not} decay until $t=\infty$ (Numerical).
 \item Lancellotti \& Dorning 1998 \cite{Lancellotti}: Existence of ``critical initial states'' for which   $\lim_{t\rightarrow\infty}E\neq 0$ (Theory).
 \item Caglioti \& Maffei 1998 \cite{Caglioti}: Mathematical proof of the existence of \textit{some} damped solutions (Theory).
 \item Medvedev \textit{et al.} 1998 \cite{Medvedev}: Damping of waves of finite amplitude and arbitrary
shape according to $e^{\delta t}$, with $\lim_{t\rightarrow\infty}\delta =0$ (Theory).
\end{itemize}

Mouhot \& Villani 2010 \cite{Villani,VillaniShort}: End of the controversy. Nonlinear Landau damping for general interactions, including Coulomb \textit{and} Newton (therefore also including the case of galactic dynamics).

For any potential $V(\mathbf{r})$ such that $|V(\mathbf{k})|=O\left( |k|^{-2-\varepsilon}\right) $, with $\varepsilon>0$, and any \textit{linearly} stable distribution function $f_0(x,v)$, large amplitude perturbations relax  in such a way that all observables (density, field\ldots),
\begin{equation}
\Psi(t)=\int f(t,x,v) \psi(x,v) dx dv,
\end{equation}
relax exponentially with time. The distribution function itself does \textit{not} relax to its value at $t=0$. For small perturbations, $f(t,x,v)$ converges to something that is close to $f_0(x,v)$. For larger perturbations, the distribution function converges to something that is far from $f_0$, or it does not converge
 at all. The large time behavior of a strongly disturbed solution
is still an open mystery.

See \cite{Villani} for the full report, and a great history of the problem, or \cite{VillaniShort} for a shorter version. Villani was awarded the \textit{2010 Fields Medal} for this.

%% file: PlasmaTalk06.tex
\begin{center}
\section{Beam Plasma Instabilities - Introduction}
\end{center}

\subsection*{Miscellaneous}
From now on, and for a number of Lectures, I'll just go through the Review Paper, ``Multidimensional electron beam-plasma instabilities in the relativistic regime'', \textit{Physics of Plasmas}, \textbf{17}, 120501 (2010).

Counter-streaming flows, possibly relativistic. Lots of them. Basic system: counter-streaming \textit{electron beams} with $n_{b0}, n_{p0}, \mathbf{v}_{b0}, \mathbf{v}_{p0}$ over a background of \textit{fixed} protons $n_i$. Main hypothesis:

\begin{itemize}
 \item Collisionless, Vlasov-Maxwell plasmas (i.e. weakly coupled, see \textit{Plasma Talk 2}),
 \item Homogenous, no boundaries (system size $\gg c/\omega_p$),
\item Initially current and charge neutral, $n_{b0}v_{b0}=n_{p0}v_{p0}$ and $n_{b0}+n_{p0}=n_i$,
\item No $\mathbf{B}_0$, to start with\ldots
\end{itemize}

\textit{Motivations}: simplest system + Fast Ignition Scenario for Inertial Fusion + Shock Acceleration physics (SNR's, GRB's). See Fig. 2 of Review Paper.

Particle-In-Cell Simulations: great tool for testing/guiding - See Fig. 3 of Review Paper.

\subsection*{A \textit{multidimensional} unstable spectrum}
\begin{itemize}
 \item 1948: some perturbations with \textbf{k} $\parallel$ to the flow are unstable - \textit{Two-stream} modes.
 \item 1959: some perturbations with \textbf{k} $\perp$ to the flow are unstable - \textit{Filamentation} modes.\\
 Still 1959: collisionless plasma with $T_x>T_y$, unstable for $\mathbf{k}\parallel \mathbf{y}$. \textit{Weibel}.\\
Difference between them discussed in Sec. III. F of Review.
\item 1960: some perturbations with \textbf{k} arbitrarily oriented are unstable - \textit{Oblique} modes.
\end{itemize}

\textit{Bottom line} here: Which one will Nature choose? The fastest. Need to tackle the problem globally.

First: look at flow aligned, then flow-perp and the oblique modes. Second: which one grows faster?

%% file: PlasmaTalk07.tex
\begin{center}
\section{Two-stream Instability}
\end{center}

\subsection*{Two-stream (flow-aligned) modes}
Interesting starting with a cold fluid 1D model. Equivalent to Vlasov with $f_0(v)\propto\delta(v-v_0).$ \textit{Non}-relativistic. General case shows it's still relevant for the 3D case.\\
Linearize conservation and Euler equations. One set for each electron species, and I omit subscripts for clarity. Consider first orders quantities $n_{1p},n_{1b}, E_1\propto e^{ikx-i\omega t}$.
Conservation and Euler equations read,
\begin{eqnarray}
\frac{\partial n}{\partial t} + \frac{\partial (nv)}{\partial x}&=&0,\\
m\frac{\partial v}{\partial t} + m v\frac{\partial v}{\partial x}&=&q E.
\end{eqnarray}
Once linearized, they respectively give
\begin{eqnarray}
n_{1}=n_{0}\frac{kv_{1}}{\omega-kv_{0}},\\
v_{1}=i\frac{qE_1/m}{\omega-kv_{0}},
\end{eqnarray}
so that,
\begin{equation}\label{specie}
n_{1} = \frac{qn_{0}}{m}\frac{ikE1}{(\omega-kv_{0})^2}.
\end{equation}
Then, from Poisson's equation\footnote{Poisson's equation brings a \textit{vectorial} equation down to a \textit{scalar} one. We thus loose information, unless $\mathbf{k}\cdot \mathbf{E}=kE$. The full 3D analysis shows modes with $k_\perp=0$ are precisely like this.}
\begin{equation}
ikE_1=4\pi q(n_{1b}+n_{1p}),
\end{equation}
we get,
\begin{equation}\label{disper}
1 = \frac{\omega_{pb}^2}{(\omega-kv_{0b})^2}+\frac{\omega_{pp}^2}{(\omega-kv_{0p})^2},
~~~\mathrm{with}~~~\omega_{p,bp}^2 = \frac{4\pi  q^2n_{0,bp}}{m}.
\end{equation}

The frequency $\omega-kv$ is the Doppler shifted frequency. Can't help but thinking it looks like an energy conservation equation. Without drifts, $v_{0,bp}=0$ and we just have
\begin{equation}
(\hbar\omega)^2 = (\hbar\omega_{pb})^2+(\hbar\omega_{pp})^2.
\end{equation}

Any ideas?

Until Eq. (\ref{specie}), species are disconnected from each other in the calculation. It is Poisson's equation which puts them together, summing the contribution of each species. Assume an infinite amount of these, each beamlet going at velocity $v$, with density $n_0f_0(v)dv$, $\int f_0=1$. The extension of Eq. (\ref{disper}) reads,
\begin{equation}
1 = \int \frac{4 \pi q^2 n_0f_0(v)dv}{m(\omega-kv)^2}=\frac{\omega_{p0}^2}{k^2}\int \frac{f_0(v)dv}{(v-\omega/k)^2},
~~~~~~\omega_{p0}^2 = \frac{4\pi  q^2n_0}{m}.
\end{equation}
identical with the one encountered in \textit{Plasma Talk 4}, up to an integration by part.

So if we ``Fermi understand'' Eq. (\ref{disper}), we have everything.

Introducing the dimensionless variables,
\begin{equation}
x = \frac{\omega}{\omega_{pp}},~~Z=\frac{k v_{0b}}{\omega_{pp}}, ~~\alpha=\frac{n_{0b}}{n_{0p}},
\end{equation}
Eq. (\ref{disper}) reads,
\begin{equation}\label{disperLess}
1 = \frac{\alpha}{(x-Z)^2}+\frac{1}{(x+\alpha Z)^2}.
\end{equation}

\subsubsection*{Diluted beam, $\alpha\ll 1$}
There are techniques\footnote{See S. A. Bludman, K. M. Watson, and M. N. Rosenbluth, \textit{Phys.
Fluids} \textbf{3}, 747 (1960).} to solve Eq. (\ref{disperLess}) in this regime, always approximately, for all $Z$. I just show here how to find \textit{the} mode growing the most.

\begin{figure}[t]
\begin{center}
\includegraphics[width=0.45\textwidth]{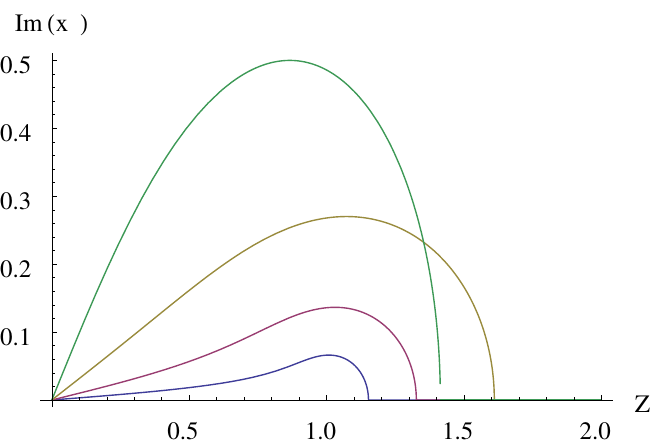}\includegraphics[width=0.45\textwidth]{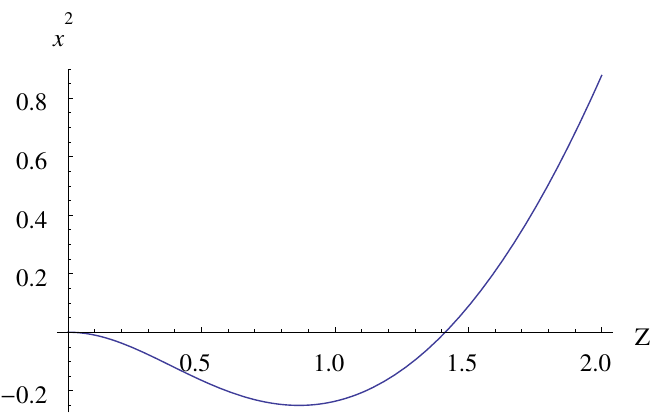}
\end{center}
\caption{
\textbf{Left}: Plot of Im$(x)$ in terms of $Z$ for $\alpha=10^{-3},10^{-2},10^{-1}$ and 1.
\textbf{Right}: Plot of Eq. (\ref{sym0}), $x^2 = 1+Z^2-\sqrt{1+4 Z^2}$. The system is unstable, $x^2<0$, for $Z<\sqrt{2}$.
} \label{fig:1}
\end{figure}

The beam is just a \textit{perturbation} to the plasma. The modes of the system should be close to the modes of the plasma alone. We thus look for solutions at $\omega\sim\omega_{pp}$, i.e. $x=1+\epsilon$.
We also know that the fastest growing mode should efficiently exchange energy with the beam. It should thus have have $\omega/k\sim v_{0b}$. With $\omega\sim\omega_p$, that means $Z\sim 1$. Eq (\ref{disperLess}) now reads,
\begin{equation}
1 = \frac{\alpha}{(1+\epsilon-Z)^2}+\frac{1}{(1+\epsilon+\alpha Z)^2}.
\end{equation}
As we'll checked, $ \mid 1-Z \mid \ll \epsilon$ and $\alpha Z\sim\alpha \ll \epsilon$, which gives
\begin{equation}
1 = \frac{\alpha}{\epsilon^2}+\frac{1}{(1+\epsilon)^2}~~~~\Rightarrow~~~~
1 = \frac{\alpha}{\epsilon^2}+1-2\epsilon~~~~\Rightarrow~~~~\epsilon^3 = \frac{\alpha}{2}.
\end{equation}
By setting $\epsilon=\rho e^{i\theta}$, we find
\begin{equation}
\rho=\left( \frac{\alpha}{2}\right)^{1/3},~~~~
\theta=-\frac{2\pi}{3},~0,~\frac{2\pi}{3}.
\end{equation}
With $e^{\pm i2\pi/3}=-1/2\pm i\sqrt{3}/2$, we obtain 3 modes
\begin{eqnarray}
x=\frac{\omega}{\omega_{pp}}&=&1-\frac{\alpha^{1/3}}{2^{4/3}} - i\frac{\sqrt{3}}{2^{4/3}}\alpha^{1/3},\\
&=&1-\frac{\alpha^{1/3}}{2^{4/3}},\\
&=&1-\frac{\alpha^{1/3}}{2^{4/3}} + i\frac{\sqrt{3}}{2^{4/3}}\alpha^{1/3},~~~~Unstable.\label{root}
\end{eqnarray}
As evidenced on Fig. \ref{fig:1}, the most unstable mode has $Z\sim 1$, that is $k/\omega_{pp}\sim v_{0b}$. An electron from the beam always sees the same electric field.

\textit{How to compute these results in a Fermi-like way?
}

\subsubsection*{Symmetric beams, $\alpha =1$}
Eq. (\ref{disperLess}) now reads,
\begin{equation}\label{SymNR}
1 = \frac{1}{(x-Z)^2}+\frac{1}{(x+Z)^2},
\end{equation}
which can be solved \textit{exactly} for all $Z$, giving
\begin{equation}\label{sym0}
x^2 = 1+Z^2\pm\sqrt{1+4 Z^2}.
\end{equation}
For $Z<\sqrt{2}$, the solution with a minus sign is unstable (see Fig. \ref{fig:1}), with a most unstable wave-vector $Z_m$ and its frequency $x_m$ given by
\begin{equation}\label{sym07}
Z_m=\frac{\sqrt{3}}{2},~~~~x_m = 0 + i\frac{1}{2}.
\end{equation}

For the diluted beam regime, unstable modes are plasma Langmuir waves at $\omega\sim\omega_p$, traveling with the beam. Things are not so clear here. The beam is no longer a perturbation. The waves have Re$(\omega)=0$, and are the modes of the full counter-streaming system ``beam+plasma'', each of equal density.

To wrap-up the most unstable mode characteristics in terms of $\alpha\in [0,1]$:
\begin{itemize}
 \item Growth-rate: Im$(\omega/\omega_{pp})=\frac{\sqrt{3}}{2^{4/3}}\alpha^{1/3} \longrightarrow \frac{1}{2}$ (Note that $\frac{\sqrt{3}}{2^{4/3}}\sim 0.68 > \frac{1}{2}$).
 \item Frequency: Re$(\omega/\omega_{pp})=1-\frac{\sqrt{3}}{2^{4/3}}\alpha^{1/3} \longrightarrow 0$.
 \item Most unstable wave-vector: $Z=1 \longrightarrow \frac{\sqrt{3}}{2}$.

\end{itemize}

\subsection*{Relativistic effects}
Maxwell's and conservation equations are the same. Euler is now (subscripts omitted),
\begin{equation}
m\frac{\partial (\gamma v)}{\partial t} + m v\frac{ \partial (\gamma v)}{\partial x}=q E.
\end{equation}
It turns out that when linearizing ``$\gamma v$'' instead of ``$v$'', one finds,
\begin{equation}
\gamma v = \gamma_0 v_0 + v_1\gamma_0^3+\cdots
\end{equation}
As a result, Eq. (\ref{disperLess}) is replaced by,
\begin{equation}\label{SymR}
1 = \frac{\alpha}{(x-Z)^2\gamma_b^3}+\frac{1}{(x+\alpha Z)^2\gamma_p^3}.
\end{equation}

Intuitively, where does these $1/\gamma^3$ come from ?  If a particle oscillates along its main direction of motion, its mass gets a $\gamma^3$ relativistic boost. Changing $m$ to $m\gamma^3$ is Eq. (\ref{disper}) gives the result above.

\subsubsection*{Diluted beam, $\alpha\ll 1$}
Here, $\gamma_p\sim 1$, so that we can recycle the non-relativistic results for diluted beam, formally replacing $\alpha\rightarrow \alpha/\gamma_b^3$, i.e. $n_b\rightarrow n_b/\gamma_b^3$. The unstable modes given by Eq. (\ref{root}) now reads,
\begin{equation}
x=\frac{\omega}{\omega_{pp}}=1-\frac{1}{2^{4/3}}\frac{\alpha^{1/3}}{\gamma_b} + i\frac{\sqrt{3}}{2^{4/3}}\frac{\alpha^{1/3}}{\gamma_b}.
\end{equation}

\subsubsection*{Symmetric beams, $\alpha =1$}
With two symmetric beams, the Lorentz factors are the same $\gamma_p=\gamma_b\equiv\gamma$. Equation (\ref{SymR}) now reads,
\begin{equation}
1 = \frac{1}{(x-Z)^2\gamma^3}+\frac{1}{(x+Z)^2\gamma^3}.
\end{equation}
Here again, we just replace $x\rightarrow x\gamma^{3/2}$ and $Z\rightarrow Z\gamma^{3/2}$, and we're formally back to the non-relativistic case. Equation (\ref{sym07}) then gives
\begin{equation}
Z_m=\frac{\sqrt{3}}{2\gamma^{3/2}},~~~~x_m = 0 + i\frac{1}{2\gamma^{3/2}}.
\end{equation}

%% file: PlasmaTalk08.tex
\begin{center}
\section{Filamentation Instability - Part 1}
\end{center}

We still consider the same counter-streaming system, but look now at perturbations with $\mathbf{k} \perp$ to the flow. With respect to the Two-stream instability ($\mathbf{k}\parallel$ to the flow), the situation is reversed: The physics is simple, but the full maths are involved. Let's start with the physics.

\subsection*{Physical picture}
\begin{figure}[t]
\begin{center}
\includegraphics[width=0.8\textwidth]{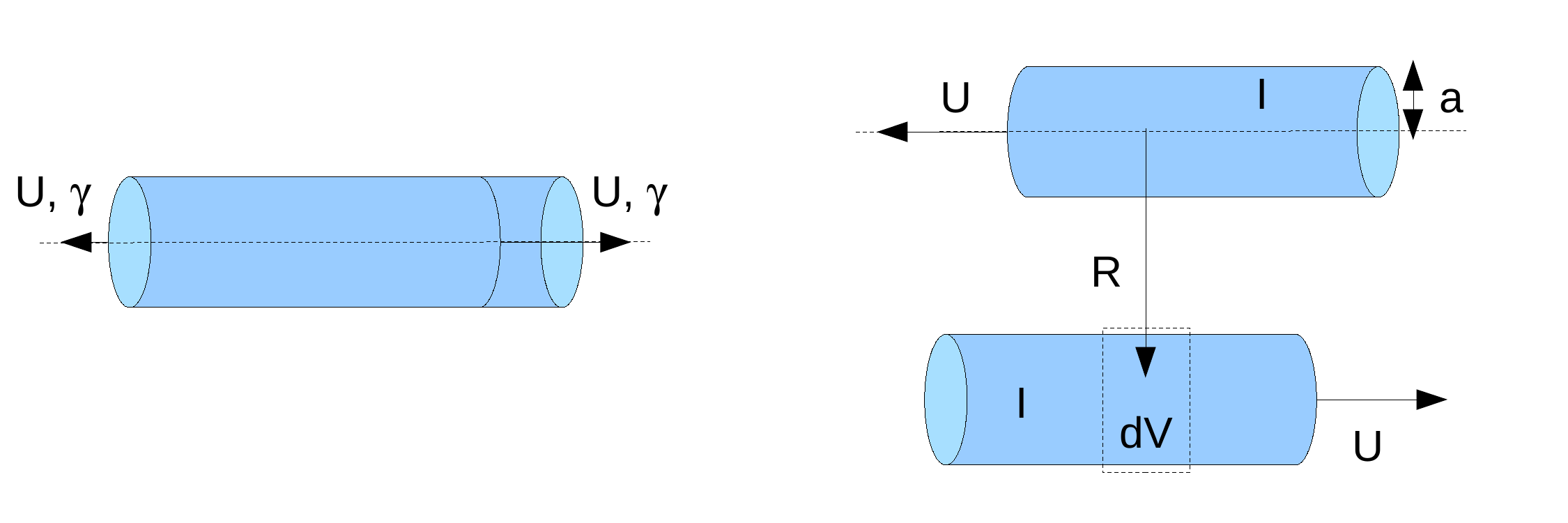}
\end{center}
\caption{System unstable to the filamentation instability.} \label{fig:fila1}
\end{figure}

Suppose two particle currents of same radius $a$ and density $n$ but opposite velocities $u$, perfectly overlap (Fig. \ref{fig:fila1}, left). The system is charge and current neutral, in equilibrium. We now set them apart by a distance $R$ (Fig. \ref{fig:fila1}, right). The first current generates a \textbf{B} field at the level on the second one. The field is such that the Lorentz force \textbf{F} produced \textit{repels} the other current even more. \textit{Unstable} system. We can write,
\begin{equation}\label{eq:1}
 F = dM \frac{d^2R}{dt^2},
\end{equation}
where $dM$ is the mass of the volume element. The force reads,
\begin{equation}
 F = dq\frac{u}{c}B,
\end{equation}
where $dq$ is the charge of the volume element. With a density $n$, and particles of charge $q$ and rest mass $m$, the charge $dq$ and the mass $dM$ of the volume $dV$ read respectively,
\begin{equation}
 dq = qndV,~~\mathrm{and}~~dM = \gamma mndV,
\end{equation}
where $\gamma m$ is the relativistic mass boost for transverse motion. Equation (\ref{eq:1}) now reads,
\begin{equation}
 qndV\frac{u}{c}B = \gamma mndV\frac{d^2R}{dt^2},~~~~\mathrm{i.e,}~~~~ q\frac{u}{c}B = \gamma m\frac{d^2R}{dt^2}.
\end{equation}
B is the field created by the current, so that
\begin{equation}
 B = \frac{2 I}{cR},~~\mathrm{where}~~I = n q u \pi a^2.
\end{equation}
Replacing the current $I$ by its expression, we find
\begin{equation}\label{eq:GRPhys}
 \frac{d^2\xi}{dt^2}=\frac{\delta^2}{\xi},
~~\mathrm{with}~~\delta=\omega_p\frac{\beta}{\sqrt{2\gamma}},
~~\mathrm{and}~~\xi=\frac{R}{a},~~\beta=\frac{u}{c},
\end{equation}
with $\omega_p^2=4\pi n q^2/m$. Although this equation won't give $\xi\propto e^{\delta t}$, it does tell the system does not relax to its initial state, on a time scale $\propto\delta^{-1}$, which fits \textit{exactly} the result of the linear theory\footnote{Up to a factor of order unity, as usual.}. Maybe an exponential grow would be obtained starting from opposite current \emph{partially} overlapping.

\subsection*{The Maths}
\subsubsection*{The Dispersion Equation: Calculation Pattern}
The dispersion equation for the filamentation instability is not easier to derive than the one for arbitrarily oriented $\mathbf{k}$'s. I will thus go over the general case $k_\parallel k_\perp\neq 0$, and then focus on $k_\parallel = 0$.
For the Two-stream instability, the flow of the calculation was\footnote{See \textit{Plasma Talks 7}.}:
\bigskip
\begin{center}
\begin{tabular}{c}
 Euler + Conservation eqs. (or Vlasov) for each species  \\
$\downarrow$  \\
First order densities $n_1$'s in terms of $\mathbf{E}_1$ \\
$\downarrow$  \\
Merge info for all species through Poisson's equation \\
$\downarrow$  \\
\textit{Dispersion Equation}
\end{tabular} \end{center}

We could use Poisson's equation for modes with $k \parallel$ to the flow because we know\footnote{We'll soon find out it is true.} they have $\mathbf{k} \parallel \mathbf{E}$. In plasma jargon, we say these modes are \textit{longitudinal}, or \textit{electrostatic}. For the, Poisson equation, which convert a \textit{vectorial} into a \textit{scalar} identity, doesn't result in a loss of information, precisely because  $\mathbf{k} \parallel \mathbf{E}$.

For filamentation modes, we don't know about the respective orientation of $\mathbf{k}$ and $\mathbf{E}$. The divergence of the electric field would introduce the cosine of the $\widehat{\mathbf{k},\mathbf{E}}$ angle, which is unknown. Poisson's $\nabla\cdot\mathbf{E}=4\pi\rho$ gives $k_xE_x+k_yE_y+k_zE_z=4\pi\rho$, which cannot be used as a dispersion equation because it yields only \textit{one} equation for \textit{three} components of the field. We need a 3D ``merging species'' equation which does not result in information loss. This is Maxwell-Amp\`{e}re, which merges the \textit{currents} instead of the \textit{charges}.

The general pattern of the calculation is indeed quite similar:
\bigskip
\begin{center}
\begin{tabular}{c}
 Euler + Conservation eqs. (or Vlasov) for each species  \\
$\downarrow$  \\
First order \textit{currents} $\mathbf{J}_1$'s in terms of $\mathbf{E}_1$\\
$\downarrow$  \\
Merge info for all species through Maxwell-Amp\`{e}re equations \\
$\downarrow$  \\
\textit{Dispersion Equation}
\end{tabular} \end{center}

Let's see this more in details, reasoning again from the fluid equations. Every equilibrium quantities are now slightly perturbed with terms $\propto \exp(i\mathbf{k}\cdot \mathbf{r}-i\omega t)$. The linearized conservation equations give for each species:
\begin{equation}\label{n1}
n_{1}=n_{0}\frac{\mathbf{k}\cdot\mathbf{v}_1}{\omega-\mathbf{k}\cdot\mathbf{v}_0}.
\end{equation}
 The linearized non-relativistic (so far) Euler equation give, still for each species:
\begin{equation}\label{v1}
\mathbf{v}_1=\frac{-iq/m}{\omega-\mathbf{k}\cdot\mathbf{v}_0}
\left(\mathbf{E}_1+\frac{\mathbf{v}_0\times\mathbf{B}_1}{c} \right) .
\end{equation}
It is easy to eliminate $\mathbf{B}_1$ through Maxwell-Faraday equation,
\begin{equation}\label{Faraday}
\mathbf{B}_1 = \frac{c}{\omega}\mathbf{k}\times \mathbf{E}_1,
\end{equation}
so that we see how Eqs. (\ref{n1},\ref{v1}) eventually give $n_1$ and $\mathbf{v}_1$ in terms of $\mathbf{E}_1$ alone, for each species,
\begin{eqnarray}\label{speciesOK}
\mathbf{v}_1&=&\frac{-iq/m}{\omega-\mathbf{k}\cdot\mathbf{v}_0}
\left(\mathbf{E}_1+\frac{\mathbf{v}_0\times(\mathbf{k}\times \mathbf{E}_1)}{\omega} \right),\nonumber \\
n_{1}&=&\frac{-iqn_{0}}{m}\frac{\mathbf{k}}{(\omega-\mathbf{k}\cdot\mathbf{v}_0)^2}\cdot
\left(\mathbf{E}_1+\frac{\mathbf{v}_0\times(\mathbf{k}\times \mathbf{E}_1)}{\omega} \right).
\end{eqnarray}

We may now write Maxwell-Amp\`{e}re equation, to merge the information from all the species into one single equation depending of $\mathbf{E}_1$ only,
\begin{equation}
i\mathbf{k}\times\mathbf{B}_1 = \frac{-i\omega}{c}\mathbf{E}_1+\frac{4\pi}{c}\mathbf{J}_1,
\end{equation}
and eliminate $\mathbf{B}_1$ from Maxwell-Faraday Eq. (\ref{Faraday}) to obtain,
\begin{equation}\label{MaxOK}
\frac{c^2}{\omega^2}\mathbf{k}\times(\mathbf{k}\times \mathbf{E}_1)
+\mathbf{E}_1 + \frac{4i\pi}{\omega}\mathbf{J}_1=0.
\end{equation}
The first order current is finally expressed through,
\begin{equation}\label{J1}
\mathbf{J}_1 =
\underbrace{n_{0,b}\mathbf{v}_{1,b}+n_{1,b}\mathbf{v}_{0,b}}_{\mathrm{Beam ~part}}
+\underbrace{n_{0,p}\mathbf{v}_{1,p}+n_{1,p}\mathbf{v}_{0,p}}_{\mathrm{Plasma ~part}}.
\end{equation}
Although the end result is not really ``user friendly'', we can see how Eqs. (\ref{speciesOK},\ref{MaxOK},\ref{J1}) eventually yield a \textit{tensorial} equation of the form
\begin{equation}\label{tensor08}
\mathbf{T}\cdot\mathbf{E}_1 = 0.
\end{equation}

When starting from the \textit{Vlasov} equation, linearization gives the first order distribution function for each species,
\begin{equation}\label{vlasov}
f_1(\mathbf{k},\mathbf{v},\omega) = \frac{iq/m}{\omega-\mathbf{k}\cdot\mathbf{v}}
\left(\mathbf{E}_1+\frac{\mathbf{v}\times\mathbf{B}_1}{c} \right)\cdot\frac{\partial f_0}{\partial \mathbf{v}}.
\end{equation}
Here again, Maxwell-Faraday Eq. (\ref{Faraday}) together with $n_1=\int f_1 dv$ and $\mathbf{v}_1=\int f_1 \mathbf{v}dv$, allow to reach the dispersion equation.

%% file: PlasmaTalk09.tex
\begin{center}
\section{Filamentation Instability - Part 2}
\end{center}

\subsection*{Dispersion Equation Analysis}
The tensorial equation $\mathbf{T}\cdot\mathbf{E}_1$ at the end of \textit{Plasma Talk 8} has the obvious solution $\mathbf{E}_1 = 0$. Now, the proper modes of our system are precisely the \textit{non-trivial solutions} $\mathbf{T}\cdot\mathbf{E}_1 = 0$, with $\mathbf{E}_1 \neq 0$\footnote{We could also say we look for the eigen-vectors associated with the eigen-value $\lambda=0$.}.

That tells us two things:

\begin{itemize}
 \item If $(\exists~\mathbf{E}_1 \neq 0~/~\mathbf{T}\cdot\mathbf{E}_1 = 0) \Rightarrow \det\mathbf{T} = 0$. That's the \textit{dispersion equation}, yielding $\omega$ in terms of $\mathbf{k}$.\\
Assume we pick up one wave vector $\mathbf{k}$. The dispersion equation
\begin{equation}\label{disper09}
\det\mathbf{T}(\mathbf{k},\omega) = 0,
\end{equation}
gives one or more $\omega$'s,  $(\omega_{1,\mathbf{k}},\ldots,\omega_{N,\mathbf{k}})\in \mathbb{C}^N$. Each couple $(\mathbf{k},\omega_{j,\mathbf{k}})$ defines a proper mode of the system. Unstable modes have Im$(\omega)<0$.\\
The fluid model usually gives a polynomial dispersion equation. Each new ingredient to the model (mobile ions, magnetic field,\ldots), adds waves. Polynomial of degree larger than 10 are common.

\item The proper modes of the system $\mathbf{E}_1(\mathbf{k},\omega)$ are in the \textit{Kernel} of \textbf{T}, which is precisely the set of non-zero $\mathbf{E}_1$'s fulfilling $\mathbf{T}\cdot\mathbf{E}_1 = 0$.\\
Assume again we picked up one wave vector $\mathbf{k}$. The dispersion equation gives a series of frequencies $(\omega_{1,\mathbf{k}},\ldots,\omega_{N,\mathbf{k}})$. We thus have $N$ tensors with vanishing determinants. Each of these $N$ tensors has a Kernel of dimension 1 or 2 (a Kernel of dimension 3 would imply \textbf{T}=\textbf{0}).
\begin{eqnarray}
\mathbf{T}( \mathbf{k},\omega_{1,\mathbf{k}})~~&\Rightarrow&
~~\left\lbrace \mathbf{E}_{1,i}\left( \mathbf{k},\omega_{1,\mathbf{k}}\right)\right\rbrace_{i=1~\mathrm{or}~2} ,\nonumber\\
&\vdots&\nonumber\\
\mathbf{T}( \mathbf{k},\omega_{N,\mathbf{k}})~~&\Rightarrow&
~~\left\lbrace \mathbf{E}_{1,i}( \mathbf{k},\omega_{N,\mathbf{k}})\right\rbrace_{i=1~\mathrm{or}~2} ,\nonumber
\end{eqnarray}
So, for one couple $(\mathbf{k},\omega_\mathbf{k})$, the formalism tells how is the $\mathbf{E}_1$ field. It lies either along a given direction, or in a plane. In particular, the formalism tells us about the $\widehat{\mathbf{k},\mathbf{E}}$ angle. We don't have to assume waves are longitudinal\footnote{Also referred to as ``electrostatic''.}  ($\mathbf{k}\parallel\mathbf{E}$), or transverse ($\mathbf{k}\perp\mathbf{E}$). The formalism \textit{decides for us}.
\end{itemize}

\begin{figure}[t]
\begin{center}
\includegraphics[width=0.5\textwidth]{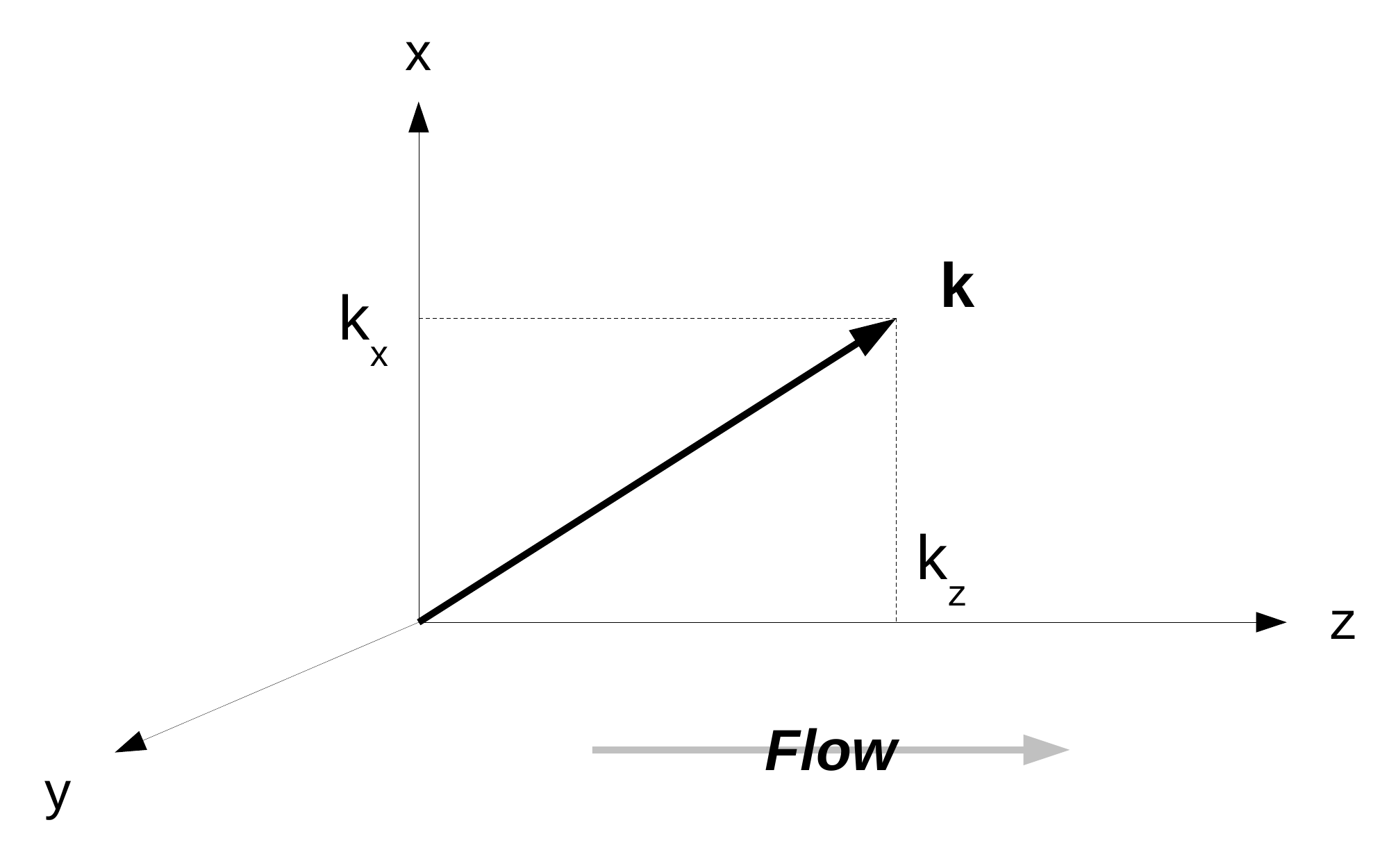}
\end{center}
\caption{Axis conventions.} \label{axis09}
\end{figure}

For a flow $\parallel \mathbf{z}$, and $\mathbf{k}=(k_x,0,k_z)$ as pictured on Fig. \ref{axis09}, the final form of the tensor \textbf{T} is given by,
\begin{equation}\label{eq:T09}
\mathbf{T}=\left|
\begin{array}{ccc}
\eta ^{2}\varepsilon _{xx}-k_{z}^{2} & 0 & \eta ^{2}\varepsilon
_{xz}+k_{z}k_{x} \\
0 & \eta ^{2}\varepsilon _{yy}-k^{2} & 0 \\
\eta ^{2}\varepsilon _{xz}+k_{x}k_{z} & 0 & \eta ^{2}\varepsilon
_{zz}-k_{x}^{2}
\end{array}
\right|,
\end{equation}
where $\eta=\omega/c$ and $\varepsilon _{\alpha\beta}$ is given by Eq. (8) of the Review Paper.

\subsection*{Two-stream Check}
Let's check our assumption from \textit{Plasma Talk 7}, that for $\mathbf{k}\parallel$ flow, i.e. $k_x=0$, there are longitudinal modes with $\mathbf{k}\parallel\mathbf{E}$. Setting $k_x=0$ in Eq. (\ref{eq:T09}) gives,
\begin{equation}\label{eq:TTS}
\mathbf{T}(k_z,k_x=0)=\left|
\begin{array}{ccc}
\eta ^{2}\varepsilon _{xx}-k^{2} & 0 & \eta ^{2}\varepsilon
_{xz} \\
0 & \eta ^{2}\varepsilon _{yy}-k^{2} & 0 \\
\eta ^{2}\varepsilon _{xz} & 0 & \eta ^{2}\varepsilon
_{zz}
\end{array}
\right|.
\end{equation}
For such wave vectors, the system is perfectly symmetric around the flow axis $z$. We thus have $\varepsilon_{xx}=\varepsilon_{yy}\equiv\varepsilon_\perp$, and\footnote{Less obvious, but true.} $\varepsilon _{xz}=0$, so that
\begin{equation}\label{eq:TTS1}
\mathbf{T}(k_z,k_x=0)=\left|
\begin{array}{ccc}
\eta ^{2}\varepsilon _\perp-k^{2} & 0 & 0 \\
0 & \eta ^{2}\varepsilon_\perp-k^{2} & 0 \\
0 & 0 & \eta ^{2}\varepsilon
_{zz}
\end{array}
\right|.
\end{equation}
The equation $\mathbf{T}\cdot \mathbf{E}_1=0$ defines two kinds of waves:

\begin{itemize}
 \item Assume $(\mathbf{k},\omega)$ fulfills,
\begin{equation}
\varepsilon_\perp=k^2c^2/\omega^2,
\end{equation}
then, $\varepsilon_{zz}$ will in general \textit{not} vanish for the \textit{same} $(\mathbf{k},\omega)$. For these $(\mathbf{k},\omega)$, the tensor will thus have the form,
\begin{equation}
\mathbf{T}=\left|
\begin{array}{ccc}
0 & 0 & 0 \\
0 & 0 & 0 \\
0 & 0 & \eta ^{2}\varepsilon
_{zz}\neq 0
\end{array}
\right|,
\end{equation}
and waves with with $\mathbf{E}_1 \in (x,y)$ satisfy $\mathbf{T}\cdot \mathbf{E}_1=0$. Since $\mathbf{k}=(0,0,k_z)$, these are \textit{transverse} modes, $\mathbf{k}\perp \mathbf{E}_1$. In general, they are stable.
 \item If we consider now $(\mathbf{k},\omega)$ fulfilling
\begin{equation}\varepsilon_{zz}=0,
\end{equation}
we find non-zero solutions of $\mathbf{T}\cdot \mathbf{E}_1=0$ are waves with $\mathbf{E}_1 \in (z)$, as the tensor now takes the form,
\begin{equation}
\mathbf{T}=\left|
\begin{array}{ccc}
\eta ^{2}\varepsilon _\perp-k^{2}\neq 0 & 0 & 0 \\
0 & \eta ^{2}\varepsilon_\perp-k^{2}\neq 0 & 0 \\
0 & 0 & 0
\end{array}
\right|.
\end{equation}
Since $\mathbf{k}=(0,0,k_z)$, these are \textit{longitudinal} modes, $\mathbf{k}\parallel \mathbf{E}_1$, with dispersion equation,

which indeed are our two-stream modes. It is thus checked that the modes we investigated in \textit{Plasma Talk 7} do exist.
\end{itemize}

\subsection*{The Filamentation Instability}
\subsubsection*{About the Dispersion Equation}
Let's now consider $k_z=0$ in Eq. (\ref{eq:T09}). We find,
\begin{equation}\label{eq:TFila}
\mathbf{T}=\left|
\begin{array}{ccc}
\eta^2\varepsilon _{xx} & 0 & \eta^2\varepsilon
_{xz} \\
0 & \eta^2\varepsilon _{yy}-k^2 & 0 \\
\eta^2\varepsilon _{xz} & 0 & \eta^2\varepsilon
_{zz}-k^2
\end{array}
\right|,
\end{equation}
where $\mathbf{T}\cdot \mathbf{E}_1=0$ again defines two kinds of modes:
\begin{itemize}
 \item Modes with $\mathbf{E}_1 \in (y)$, therefore transverse since $\mathbf{k}\parallel \mathbf{x}$, with dispersion equation,
\begin{equation}
\varepsilon_{yy}=k^2c^2/\omega^2.
\end{equation}
 \item The \textit{Filamentation modes} (at last), with $\mathbf{E}_1 \in (x,z)$ and dispersion equation,
\begin{equation}\label{fila_OK}
\varepsilon_{xx}(\varepsilon_{zz}-k^2c^2/\omega^2)=\varepsilon_{xz}.
\end{equation}
\end{itemize}
Of course, we would like to have $\varepsilon_{xz}=0$, which would ease our life and give a simpler, two branches dispersion equation,
\begin{eqnarray}
\varepsilon_{xx}&=&0,\nonumber\\
\varepsilon_{zz}&=&k^2c^2/\omega^2.\label{fila_trans}
\end{eqnarray}
Eq. (\ref{fila_trans}) has been frequently used in the literature to study the Filamentation instability\footnote{See Bret \textit{et al.}, Phys. Plasmas, \textbf{14}, 032103 (2007).}. It defines purely transverse waves with $\mathbf{E}_1 \in (z)$, that is, $\parallel$ to the flow. The problem is that these papers never say they assume $\varepsilon_{xz}=0$. In general, they are wrong.

I wrote ``in general'', because on rare occasions, they study settings for which truly, $\varepsilon_{xz}=0$. Which are they? Remember that even if we now focus on $\mathbf{k}=(k_x,0,0)$, this tensor element still depends on the beam and plasma distribution functions. A detailed study\footnote{\textit{Ibid}.} shows $\varepsilon_{xz}$ strictly vanishes \textit{only if} our counter streaming species are \textit{perfectly symmetric}.

So, unless our density ratio is 1, and we have the same temperatures on the beam and the plasma, the same Lorentz factors, the same\ldots~ everything, the correct dispersion equation is Eq. (\ref{fila_OK}), \textit{not} (\ref{fila_trans}).

\subsubsection*{Cold Analysis - Relativistic effects}
What we've said is so far non-relativistic. Still in the fluid model, the main relativistic effect is displayed when linearizing the Euler equation. The relativistic Euler equation reads,
\begin{equation}
\frac{\partial \mathbf{p}}{\partial t} + (\mathbf{v}\cdot\nabla)\mathbf{p}=
q\left(\mathbf{E}+\frac{\mathbf{v}\times\mathbf{B}}{c} \right),~~\mathbf{p}=\gamma m \mathbf{v}.
\end{equation}
Its two linearized versions are,
\begin{eqnarray}
im(\mathbf{k}\cdot\mathbf{v}_0-\omega)\mathbf{v}_1 &=&
q\left(\mathbf{E}_1+\frac{\mathbf{v}_0\times\mathbf{B}_1}{c} \right),~~~non-relativistic ,\nonumber\\
im(\mathbf{k}\cdot\mathbf{v}_0-\omega)
\left(\gamma_0\mathbf{v}_1+\gamma_0^3\frac{\mathbf{v}_1\cdot\mathbf{v}_0}{c^2}\mathbf{v}_0 \right) &=&
q\left(\mathbf{E}_1+\frac{\mathbf{v}_0\times\mathbf{B}_1}{c} \right),~~~relativistic.
\end{eqnarray}
Everything is in the anisotropic linearization of $\gamma_0 \mathbf{v}$ around $\mathbf{v}_0$. We see above that for a small motion along the flow, the relativistic mass increase goes like $\gamma_0^3$. But for small motion normal to the flow, $\mathbf{v}_1\cdot\mathbf{v}_0=0$ and the mass increase only goes with $\gamma$. This of course, adds a level of complexity to the general calculation, as Eq. (10) from \textit{Plasma Talk 8} for $\mathbf{v}_1$ is even more involved.

For the filamentation instability, we have $\mathbf{v}_1\cdot\mathbf{v}_0=0$, and we find we can just formally replace $m\rightarrow\gamma m$. Assuming a cold beam with density $n_b$, Lorentz factor $\gamma_b$, and cold plasma electrons with density $n_p$ and Lorentz factor $\gamma_p$, the tensor elements are\footnote{\textit{Ibid}.},
\begin{eqnarray}\label{eq:tensorfluide}
  \varepsilon_{xx} &=& 1-\frac{\alpha}{x^2\gamma_b}-\frac{1}{x^2\gamma_p},\nonumber\\
  \varepsilon_{yy} &=& 1-\frac{\alpha}{x^2\gamma_b}-\frac{1}{x^2\gamma_p}, \nonumber\\
  \varepsilon_{zz} &=& 1-\frac{\alpha}{x^2\gamma_b}-\frac{\alpha Z^2}{x^4\gamma_b}-\frac{1}{x^2\gamma_p}-\frac{\alpha^2Z^2}{x^4\gamma_p}, \nonumber\\
  \varepsilon_{xz} &=& \frac{\alpha  Z}{x^3\gamma_p}\left(\frac{1}{\gamma_p} -\frac{1}{\gamma_b}\right) ,
\end{eqnarray}
with again,
\begin{equation}\label{dimless09}
x=\frac{\omega}{\omega_{pp}},~~~~Z=\frac{k v_b}{\omega_{pp}},~~~\alpha=\frac{n_b}{n_p}.
\end{equation}

\begin{figure}[t]
\begin{center}
\includegraphics[width=0.6\textwidth]{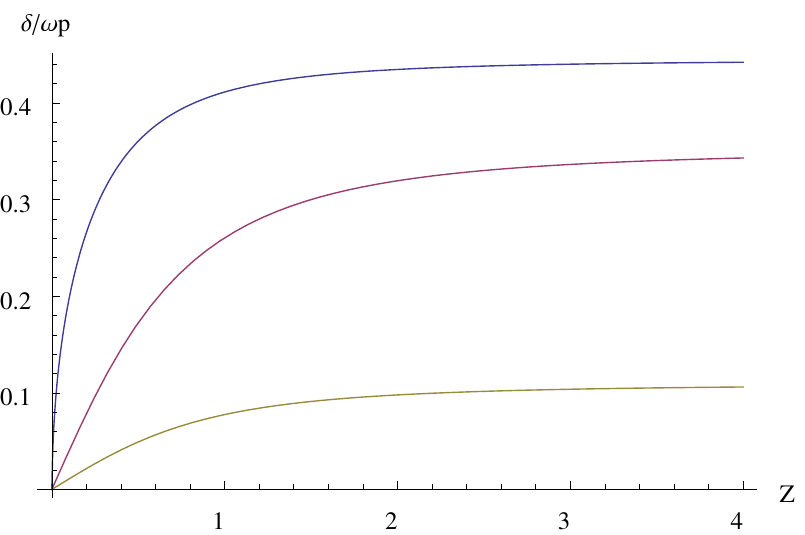}
\end{center}
\caption{Filamentation instability growth-rate for density ratios $\alpha=1$, 0.5 and 0.1, from higher to lower curves respectively. The beam Lorentz factor is $\gamma_b=10$.} \label{fig:fila}
\end{figure}

The numerical resolution of Eq. (\ref{fila_OK}), when plugging the tensor elements above, yields the growth-rate curves pictured on Fig. \ref{fig:fila}. As evidenced, the growth-rate just saturates for large $Z$. A trick to recover the large $Z$ growth-rate, consists in extracting the coefficient $a_n$ of $Z^n$ in the polynomial dispersion equation, as $a_n=0$ is the asymptotic dispersion equation for $Z\rightarrow\infty$. Doing so, one finds a zero real frequency and
\begin{eqnarray}
\lim_{Z\rightarrow\infty} \frac{\delta}{\omega_{pp}}&=&\frac{v_b}{c}\sqrt{\frac{\alpha}{\gamma_b}},~~~\alpha\ll 1,\\
&=&\frac{v_b}{c}\sqrt{\frac{2}{\gamma_b}},~~~\alpha = 1,\label{linearOK}
\end{eqnarray}
where the agreement with Eq. (6) of \textit{Plasma Talk 8} can be checked. Note that for $\alpha=1$, the tensor elements (\ref{eq:tensorfluide}) simplify substantially. Equation (\ref{fila_trans}) for unstable modes is valid and reads,
\begin{equation}
x^2-\frac{2}{\gamma_b^3}-\frac{2 Z_x^2}{x^2 \gamma_b}=\frac{Z_x^2}{\beta^2},
\end{equation}
which can be solved exactly.

We can follow the same line of reasoning for the ``wrong'' transverse dispersion equation (\ref{fila_trans}), in order to check its inaccuracy. The exact result is for any $\alpha$,
\begin{eqnarray}\label{asymgr_w}
\lim_{Z\rightarrow\infty} \frac{\delta_T}{\omega_{pp}}=
\beta\sqrt{\frac{\alpha(\alpha \gamma_b+\gamma_p)}{\gamma_b \gamma_p}}
&=&\beta\sqrt{\frac{\alpha(\alpha \gamma_b+1)}{\gamma_b}},~~~\alpha\ll 1,\\
&=&\beta\sqrt{\frac{2}{\gamma_b}},~~~\alpha= 1.
\end{eqnarray}
As expected, the result for the symmetric case $\alpha=1$ is the same. But for the diluted beam regime $\alpha\ll 1$, the ``transverse'' growth-rate $\delta_T$ differs from the exact one by,
\begin{equation}
\delta_T=\delta\sqrt{1+\alpha\gamma_b},
\end{equation}
so that the transverse calculation overestimates the growth-rate by a factor which can be arbitrarily large.

%% file: PlasmaTalk10.tex
\begin{center}
\section{Oblique Modes}
\end{center}

\begin{figure}[h]
\begin{center}
\includegraphics[width=0.5\textwidth]{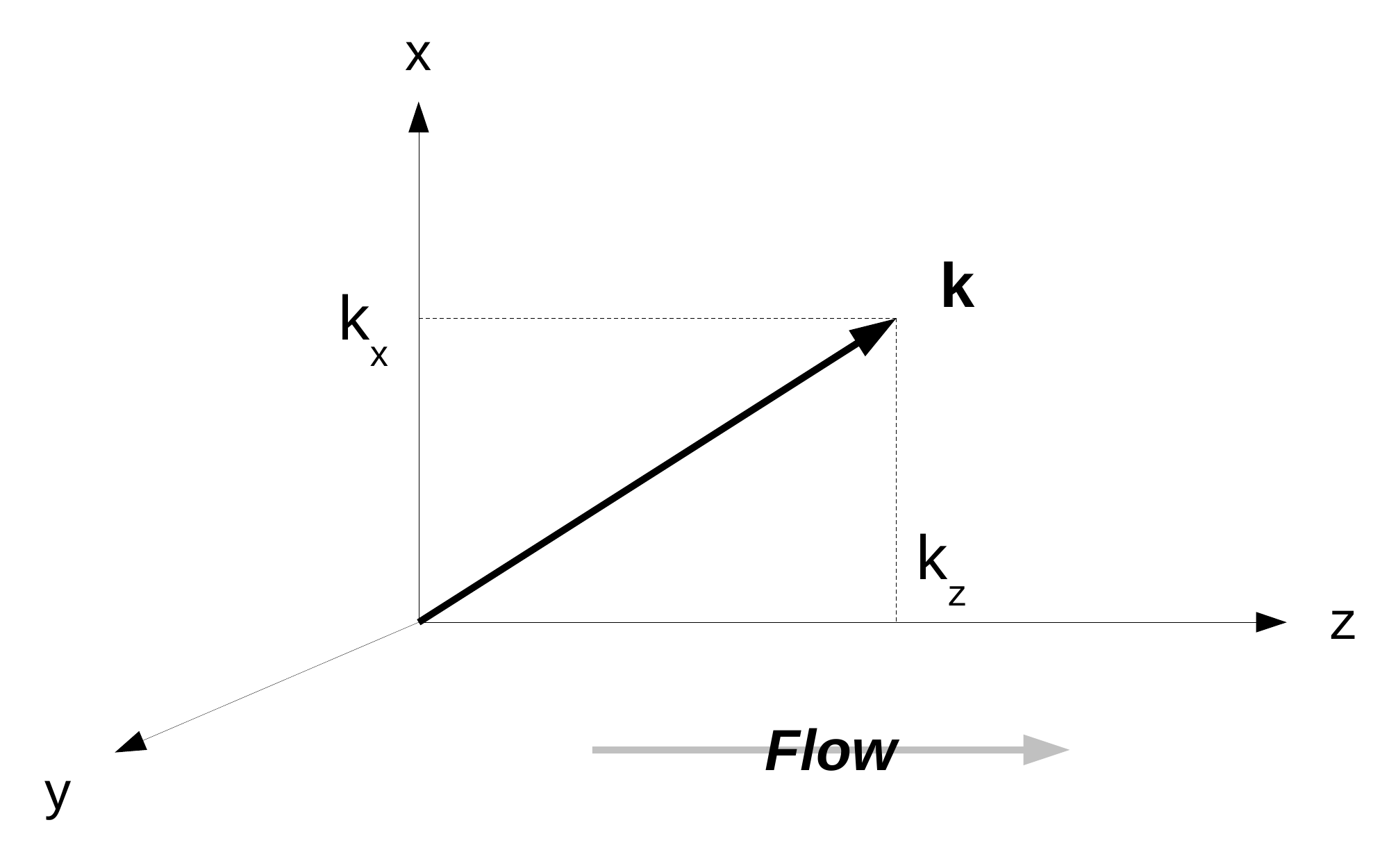}
\end{center}
\caption{Axis conventions and setup.} \label{axis10}
\end{figure}

\subsection*{The electrostatic approximation}
We now come to these fast growing ``oblique'' modes in the relativistic regime. They are found for both $k_\parallel\neq 0$ and $k_\perp\neq 0$, thus their name.

Let's remind the general dispersion equation for a set-up like to one pictured on Fig. \ref{axis10}. From Eqs. (1,2) of \textit{Plasma Talk 9}, we have the dispersion equation\footnote{There are some general symmetry requirements on the distribution function. See Review Paper.}
\begin{equation}\label{disper10}
\det\mathbf{T}(\mathbf{k},\omega) = 0,
\end{equation}
with,
\begin{equation}\label{eq:T10}
\mathbf{T}=\left|
\begin{array}{ccc}
\eta ^{2}\varepsilon _{xx}-k_{z}^{2} & 0 & \eta ^{2}\varepsilon
_{xz}+k_{z}k_{x} \\
0 & \eta ^{2}\varepsilon _{yy}-k^{2} & 0 \\
\eta ^{2}\varepsilon _{xz}+k_{x}k_{z} & 0 & \eta ^{2}\varepsilon
_{zz}-k_{x}^{2}
\end{array}
\right|,
\end{equation}
where $\eta=\omega/c$ and $\varepsilon _{\alpha\beta}$ is given by Eq. (8) of the Review Paper. We could just go on with this expression, plug some distribution functions for the beam and the plasma, and solve the dispersion equation. Doing so, we would realize something important for these oblique modes: unless we're really close the $k_\parallel=0$, these modes have $\mathbf{k}\times \mathbf{E}\sim \mathbf{0}$. That had already been noticed long ago in the first papers on the topic\footnote{
Bludman \textit{et al.}, \textit{Phys. Fluids} \textbf{3}, 741 \& 747 (1960).
Fainberg \textit{et al.}, \textit{Sov. Phys. JETP} \textbf{30}, 528 (1970).
}, for the cold case.
That's been confirmed recently for the hot case with various distribution functions\footnote{Bret \textit{et al.}, \textit{Phys. Rev. E} \textbf{70}, 046401 (2004) \& \textit{Phys. Rev. E} \textbf{81}, 036402 (2010).}, but to my knowledge, it hasn't been proved from the formalism.

It is then fruitful to assume $\mathbf{k}\times \mathbf{E}=\mathbf{0}$. Although this approximation breaks down for \textbf{k} near the normal direction, it has so far been found valid for the fastest growing oblique mode. The approximation is called the ``electrostatic'' or ``longitudinal'' approximation.

Poisson's equation can still deliver a dispersion equation, but in a slighter intricate way because of relativistic effects (a derivation from Amp\`{e}re's equation, similar to the filamentation one, is exposed in the Appendix). We simply go through the calculations of \textit{Plasma Talk 8 \& 9}, assuming at each steps $\mathbf{k}\times \mathbf{E}_1= \mathbf{0}$, implying $\mathbf{B}_1= \mathbf{0}$ as well.

The relativistic linearized conservation and Euler equations give for each species,
\begin{eqnarray}\label{line}
n_1&=&n_0\frac{\mathbf{k}\cdot\mathbf{v}_1}{\omega-\mathbf{k}\cdot\mathbf{v}_0}\nonumber\\
m\left(\gamma_0\mathbf{v}_1+\gamma_0^3\frac{\mathbf{v}_1\cdot\mathbf{v}_0}{c^2}\mathbf{v}_0 \right) &=&
q\left(\mathbf{E}_1+\cancel{\frac{\mathbf{v}_0\times\mathbf{B}_1}{c}} \right).
\end{eqnarray}
Solving these two equations gives for the perturbed density,
\begin{equation}\label{dens}
n_1=i\frac{k_zE_{1z}+\gamma^2k_xE_{1x}}{(\omega-\mathbf{k}\cdot\mathbf{v}_0)^2}.
\end{equation}
Inputs from each species are then merged through Poisson's equation,
\begin{equation}
\mathbf{k}\cdot\mathbf{E}_1=\omega_{pb}^2\frac{k_zE_{1z}+\gamma_b^2k_xE_{1x}}{(\omega-\mathbf{k}\cdot\mathbf{v}_{0b})^2}
+\omega_{pp}^2\frac{k_zE_{1z}+\gamma_p^2k_xE_{1x}}{(\omega-\mathbf{k}\cdot\mathbf{v}_{0p})^2},
\end{equation}
which may be put under the form $\mathbf{W}\cdot\mathbf{E}_1=0$, where the \textit{vector} $\mathbf{W}$ reads,
\begin{equation}
\mathbf{W}=
\left(
\begin{array}{c}
k_x \\
0 \\
k_z
\end{array}
\right)-\frac{\omega_{pb}^2}{(\omega-\mathbf{k}\cdot\mathbf{v}_{0b})^2}
\left(
\begin{array}{c}
\gamma_b^2k_x \\
0 \\
k_z
\end{array}
\right)-\frac{\omega_{pp}^2}{(\omega-\mathbf{k}\cdot\mathbf{v}_{0p})^2}
\left(
\begin{array}{c}
\gamma_p^2k_x \\
0 \\
k_z
\end{array}
\right).
\end{equation}
Now,  $\mathbf{k}\parallel\mathbf{E}_1$ and $\mathbf{W}\cdot\mathbf{E}_1=0$, implies $\mathbf{W}\cdot\mathbf{k}=0$, which gives,
\begin{equation}\label{eq:disper}
1
-\frac{k_z^2+k_x^2 \gamma_b^2}{k_z^2+k_x^2}\frac{\omega_{pb}^2/\gamma_b^3}{ (\omega-k_zv_b)^2}
-\frac{k_z^2+k_x^2 \gamma_p^2}{k_z^2+k_x^2}\frac{\omega_{pp}^2/\gamma_p^3}{ (\omega-k_zv_p)^2}=0.
\end{equation}

Note that in this longitudinal approximation, there are \textit{no} oblique effects for $\gamma_b=\gamma_p=1$. A generalization of the result to the kinetic level is not as obvious as in the 1D theory for the two-stream instability, precisely because we are not 1D. The kinetic equation reads\footnote{The quantity equal to 0 here is called the dispersion ``function''. See S. Ichimaru, \textit{Basic Principles of Plasma Physics: A Statistical Approach}, Chapter 3.},
\begin{equation}\label{kinetic}
0=1+\frac{4\pi q^2}{k^2}
\int \frac{\mathbf{k}\cdot \partial f_0(\mathbf{p})/\partial \mathbf{p}}{\omega - \mathbf{k}\cdot\mathbf{v}}d^3p,
~~~~\mathbf{p}\frac{m \mathbf{v}}{\sqrt{1-v^2/c^2}}.
\end{equation}
It should not be very difficult to derive intuitively Eq. (\ref{kinetic}) from Eq. (\ref{eq:disper}) summing the beamlets contributions, as we did for the 1D case. In terms of the usual dimensionless variables,
\begin{equation}
x=\frac{\omega}{\omega_{pp}},~~~~Z=\frac{k v_b}{\omega_{pp}},~~~\alpha=\frac{n_b}{n_p},
\end{equation}
and using $v_p=-\alpha v_b$, Eq. (\ref{eq:disper}) reads,
\begin{equation}\label{dimless10}
0=1-
\underbrace{\frac{Z_z^2+\gamma_b^2 Z_x^2}{Z_z^2+ Z_x^2}\frac{\alpha /\gamma_b^3}{(x-Z_z)^2}}_{beam}
-
\underbrace{\frac{Z_z^2+\gamma_p^2 Z_x^2}{Z_z^2+ Z_x^2}\frac{1/\gamma_p^3}{(x+Z_z \alpha )^2}}_{plasma}
.
\end{equation}

\subsection*{Diluted beam}
For $\alpha \ll 1$, $\gamma_p\sim 1$ and Eq. (\ref{dimless10}) reads,
\begin{equation}\label{dimless_dil}
0=1-
\frac{Z_z^2+\gamma_b^2 Z_x^2}{Z_z^2+ Z_x^2}\frac{\alpha/\gamma_b^3}{(x-Z_z)^2}
-\frac{1}{(x+Z_z \alpha )^2}
.
\end{equation}
This equation is very similar to the one we found for the diluted two-stream case (non-relativistic). We formally deal with a diluted beam of equivalent density ratio,
\begin{equation}
\alpha'=\frac{\alpha}{\gamma_b^3}~\frac{Z_z^2+\gamma_b^2 Z_x^2}{Z_z^2+ Z_x^2}.
\end{equation}
The maximum growth rate will be found for $Z_z\sim 1$, and the frequency of the unstable mode reads,
\begin{eqnarray}\label{taux_dilue}
\Im(x)&=&\frac{\sqrt{3}}{2^{4/3}}\frac{\alpha^{1/3}}{\gamma_b}\left(\frac{1+\gamma_b^2 Z_x^2}{1+ Z_x^2}\right)^{1/3}, \\
\Re(x)&=&1-\frac{1}{2^{4/3}}\frac{\alpha^{1/3}}{\gamma_b}\left(\frac{1+\gamma_b^2 Z_x^2}{1+ Z_x^2}\right)^{1/3}.\nonumber
\end{eqnarray}
The growth rate (\ref{taux_dilue}) displays THE oblique effect: For perp components of the wave vector such that $\gamma_b^2 Z_x^2\gg 1$, we switch from a $\gamma_b^{-1}$ to a $\gamma_b^{-1/3}$ scaling.

\begin{figure}[t]
\begin{center}
\includegraphics[width=\textwidth]{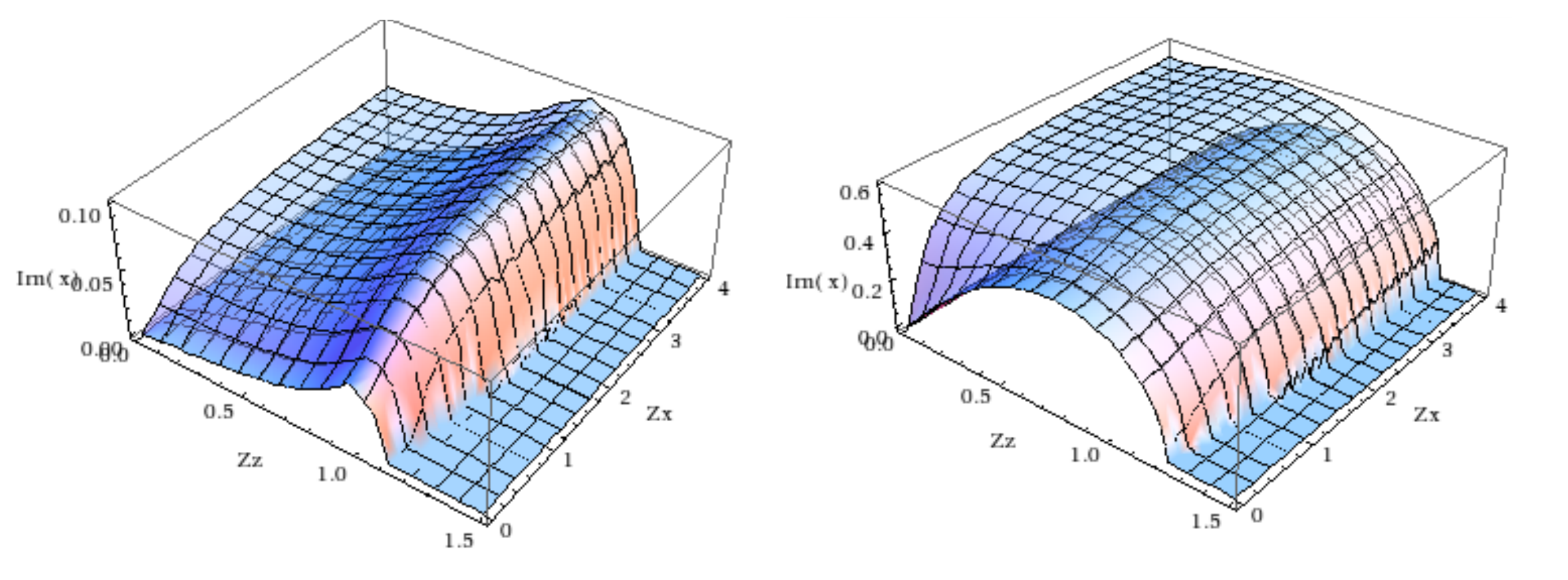}
\end{center}
\caption{LEFT: Exact growth rate from dispersion equation (\ref{eq:exact}) (transparent) vs. longitudinal for $\alpha=10^{-2}$ and $\gamma_b=3$. RIGHT: Same for $\alpha=1$ and $\gamma_b=1.1$.} \label{fig:2ddiluted}
\end{figure}

\begin{figure}[t]
\begin{center}
\includegraphics[width=0.5\textwidth]{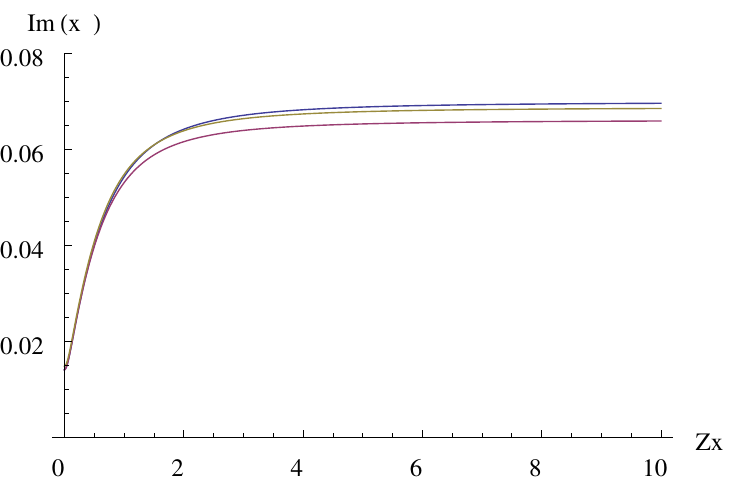}
\end{center}
\caption{Comparison of the exact growth rate (top, blue curve), the longitudinal result (lower, purple curve) and Eq. (\ref{taux_dilue}) [middle, yellow curve] along the perp $Z_x$ direction at fixed $Z_z=1$. The density ratio is $\alpha=0.01$ and the beam Lorentz factor $\gamma_b=10$.} \label{fig:compa}
\end{figure}

The validity of the longitudinal approximation is tested on Fig. \ref{fig:2ddiluted} where it is compared to the exact solution. As expected, it breaks down for small $Z_z$ while the exact calculation renders the filamentation growth rate as well.

Fixing $Z_z=1$, we can compare the exact solution, the longitudinal result and Eq. (\ref{taux_dilue}) along the perp $Z_x$ direction. The result is displayed on Fig. \ref{fig:compa}.

\subsection*{Symmetric beams}
For $\alpha = 1$, $\gamma_b=\gamma_p\equiv\gamma$ and Eq. (\ref{dimless10}) reads,
\begin{equation}\label{sym}
0=1-\frac{1}{(x-Z_z)^2\Gamma^3}-\frac{1}{(x+Z_z )^2\Gamma^3},
~~~~\Gamma=\gamma\left( \frac{Z_z^2+ Z_x^2}{Z_z^2+\gamma^2 Z_x^2}\right)^{1/3}.
\end{equation}
With an equation formally equivalent to the one studied for the two-stream symmetric case in \textit{Plasma Talk 7}. Although the equation above can be exactly solved for $x$, studying the fastest growing $Z_z$ for any given $Z_x$ is difficult because both are eventually inside $\Gamma$. Once the equation is solved, we can however look at the large $Z_x$ limit of the growth rate which reads,
\begin{equation}
\delta^2_{Z_x\rightarrow\infty}=\frac{1+Z_z^2 \gamma_b-\sqrt{1+4 Z_z^2 \gamma_b}}{\gamma_b},
\end{equation}
reaching the extremum,
\begin{equation}\label{symGR}
\delta_{m,Z_x\rightarrow\infty}=\frac{1}{2\sqrt{\gamma_b}},~~~\mathrm{for}~~~~Z_{z,m}=\frac{\sqrt{3}}{2\sqrt{\gamma_b}}.
\end{equation}

\subsection*{Exact dispersion equation}
Without the longitudinal approximation, and for arbitrarily oriented $\mathbf{k}$'s, we are back to the determinant of the tensor (\ref{eq:T10}) for the dispersion equation. It has two branches corresponding to the two factors of the determinant,
\begin{eqnarray}
\varepsilon _{yy}=k^{2}c^2/\omega^2&,&~~~\mathbf{E}_1\in (y),\\
(\eta^2\varepsilon _{xx}-k_z^2)(\eta^2\varepsilon _{zz}-k_x^2)=(\eta^2\varepsilon _{xz}-k_xk_z)^2&,&~~~\mathbf{E}_1\in (x,z).\label{eq:exact}
\end{eqnarray}
The second branch therefore holds the two-stream, the oblique and the filamentation instabilities. As evidenced by the exact plot on Figs. \ref{fig:2ddiluted}, there is a continuous transition from two-stream to filamentation modes, probably linked to a common underlying physics. \textit{Any ideas} ?

\subsection*{Cold hierarchy}
We may finally establish the hierarchy of modes for the cold regime in the $(\alpha,\gamma_b)$ phase space. The competing modes, with their variation from $\alpha\ll 1$ to 1, are
\begin{eqnarray}\label{compet}
\mathrm{Two-stream},~~\Im(x)&=&\frac{\sqrt{3}}{2^{4/3}}\frac{\alpha^{1/3}}{\gamma_b}\rightarrow \frac{1}{2\gamma_b^{3/2}}, \nonumber\\
\mathrm{Oblique},~~\Im(x)&=&\frac{\sqrt{3}}{2^{4/3}}\left( \frac{\alpha}{\gamma_b}\right)^{1/3}\rightarrow\frac{\sqrt{3}}{2\gamma_b^{1/2}}, \nonumber\\
\mathrm{Filamentation},~~\Im(x)&=&\frac{v_b}{c}\sqrt{\frac{\alpha}{\gamma_b}}\rightarrow\frac{v_b}{c}\sqrt{\frac{2}{\gamma_b}}.
\end{eqnarray}
The Two-stream case is quickly settled: it is always slower than the oblique unless $\gamma_b=1$. In the cold regime, the two-stream instability \textit{never} governs the unstable spectrum\footnote{We'll see later that temperature effects change this. This is why the two-stream instability can be observed in some real systems.}.

We are thus left comparing oblique and filamentation modes. For the diluted regime, the $\gamma_b$ scaling clearly favors the oblique. Situation is more involved near the symmetric regime. As evidenced on Fig. \ref{fig:2ddiluted} RIGHT, the longitudinal approximation gives the good order of magnitude for the growth rate, but is not enough to render the ``fine structure'' of the problem.

\begin{figure}[t]
\begin{center}
\includegraphics[width=0.45\textwidth]{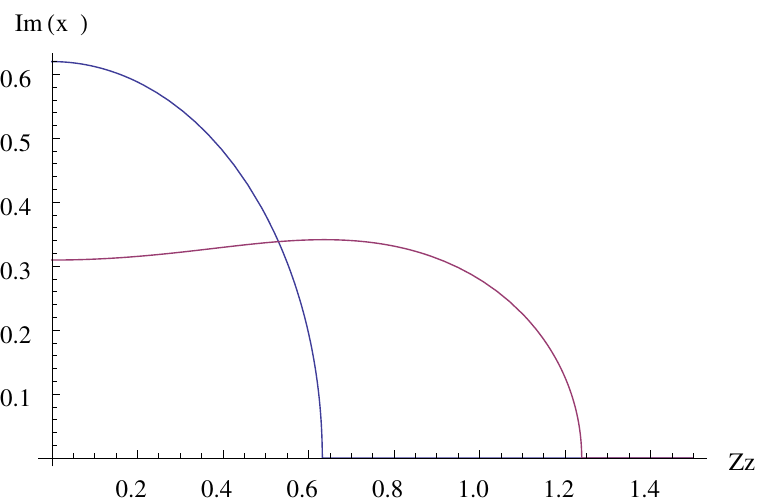}\includegraphics[width=0.5\textwidth]{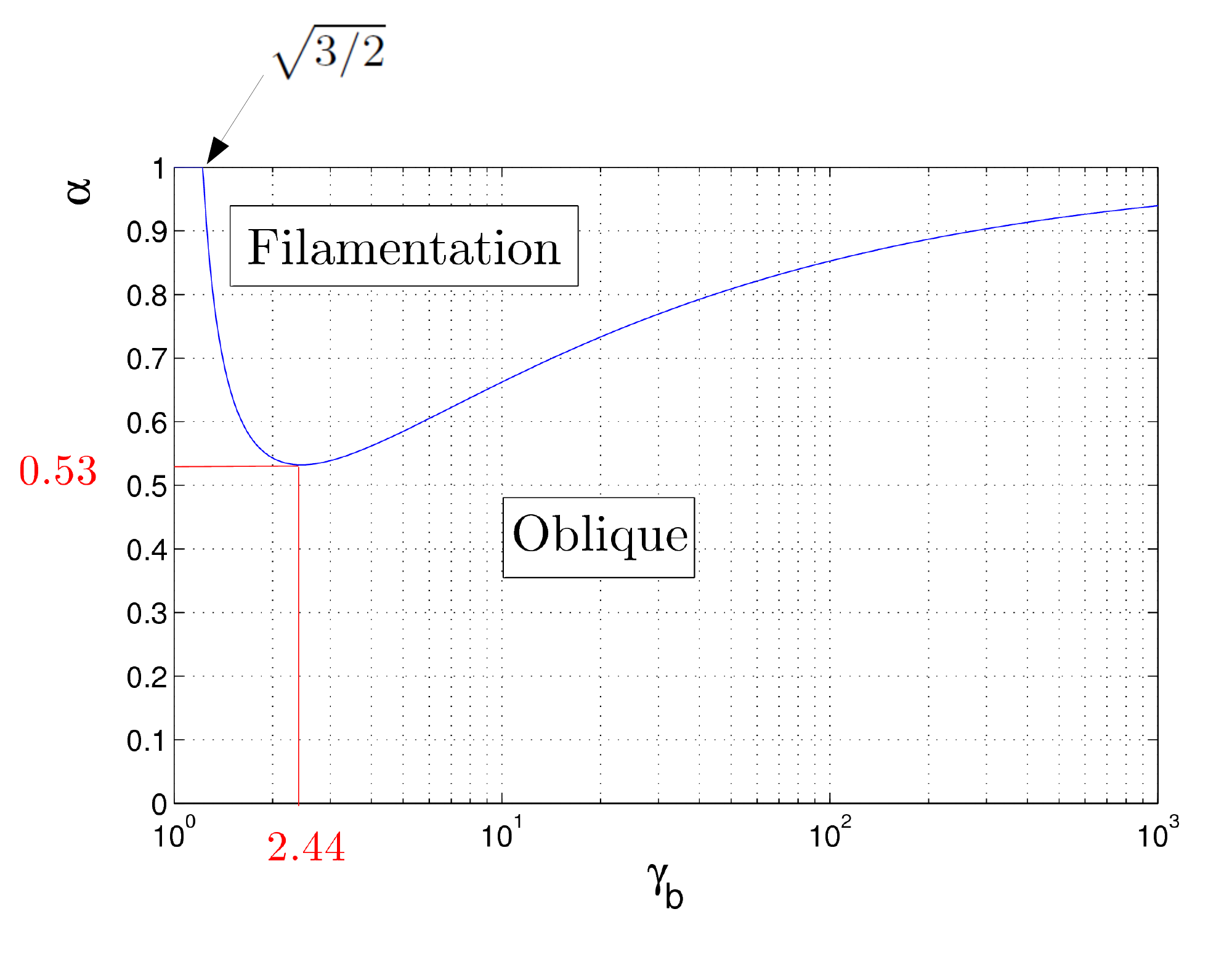}
\end{center}
\caption{LEFT: Growth rate $Z_x\rightarrow\infty$ in terms of the parallel wave vector $Z_z$, for $\gamma_b=5$, and $\alpha=0.3$ and 1 (lower and upper curves respectively). The local oblique extremum vanishes when approaching the symmetric case. RIGHT: Dominant mode in terms of $(\gamma_b,\alpha)$.} \label{GdZx}
\end{figure}

What is this ``fine structure''? In the diluted beam regime, we clearly have a local extremum for oblique vectors corresponding to oblique modes. When approaching the symmetric regime, this may no longer be the case. To evidence this, we've plotted on Fig. \ref{GdZx} LEFT the growth rate at $Z_x\rightarrow\infty$ for different parameters. For $\alpha=0.3$, we clearly find an oblique extremum, which turns to be the dominant mode. But in the symmetric case $\alpha=1$, the local extremum disappears, giving rise to a monotonous behavior and a system governed by filamentation.

As a consequence, the oblique/filamentation frontier has to be determined numerically for large density ratios. The resulting hierarchy plot can be found on Fig. \ref{GdZx}, RIGHT. The frontier position for $\alpha=1$ can be determined exactly from the dispersion equation at $Z_x\rightarrow\infty$. At $\alpha=1$, the equation can be solved, and the local oblique extremum vanishes when the second derivative of the growth rate at $Z_z=0$ vanishes. For $\alpha\neq 1$, this second derivative vanishes on the blue line. Note the frontier gets closer to $\alpha=1$ for large $\gamma_b$'s, with a convergence numerically found like $\gamma_b^{-0.395}$.

More detailed in the Review Paper, Section IV-A.

\subsection*{Appendix}
We here derive the dispersion equation (\ref{eq:disper}) from Maxwell-Amp\`{e}re equation. The equation we used to merge information from each species, namely Eq. (12) from \textit{Plasma Talk 8} simplifies in the longitudinal approximation,
\begin{equation}
\cancel{\frac{c^2}{\omega^2}\mathbf{k}\times(\mathbf{k}\times \mathbf{E}_1)}
+\mathbf{E}_1 + \frac{4i\pi}{\omega}\mathbf{J}_1=0.
\end{equation}
From Eqs. (\ref{line},\ref{dens}), the first order current is expressed  in terms of $\mathbf{E}_1$, leading again to a tensorial equation of the form,
\begin{equation}\label{eq:tensor}
\mathbf{T}\cdot \mathbf{E}_1=\mathbf{0}.
\end{equation}
If we have but two counter-streaming species (\textit{b}eam + \textit{p}lasma), the tensor reads\footnote{Interestingly, it is not symmetric. I don't understand why. I did check you don't find the correct result if you artificially add the missing element [1,3] to make it symmetric.},
\begin{equation}\label{tensor10}
\mathbf{T}=\left(
\begin{array}{ccc}
   \omega +\frac{\omega_{pb}^2}{\gamma_b (k_z v_b -\omega) }+\frac{\omega_{pp}^2}{\gamma_p (k_z v_p -\omega) } & 0 & 0 \\
 0 & 0 & 0 \\
 -k_x   \left(\frac{v_b \omega_{pb}^2}{\gamma_b (k_z v_b-\omega )^2}+\frac{v_p \omega_{pp}^2}{\gamma_p (k_z v_p-\omega )^2}\right) & 0 & \omega \left(1-\frac{\omega_{pb}^2}{\gamma_b^3 (k_z v_b-\omega )^2}-\frac{\omega_{pp}^2}{\gamma_p^3 (k_z v_p-\omega )^2}\right)
\end{array}
\right).
\end{equation}
The summing of elements from each species is here obvious again. Note that when $\mathbf{k}$ is aligned with the main axis, that is $k_\perp=0$ or $k_\parallel =0$, the respective orientation of $\mathbf{k}$ and $\mathbf{E}_1$ is easily determined, because  $\mathbf{E}_1$ is also found along the very same main axis. Things are here different because the orientation of $\mathbf{k}$ is arbitrary while the ``easy'' axis of our tensor are still the main ones.

Assume we have $\mathbf{E}_1$ fulfilling Eq. (\ref{eq:tensor}) and $\mathbf{E}_1\parallel\mathbf{k}$. Because $\mathbf{T}$ is a linear operator, that implies $\mathbf{T}\cdot \mathbf{k}$ is also the null vector:
\begin{equation}
\mathbf{T}\cdot \mathbf{k}=\mathbf{0}.
\end{equation}
The scalar product $\mathbf{k}\cdot(\mathbf{T}\cdot \mathbf{k})$ must therefore also vanish,
\begin{equation}\label{eq:disper_Long}
\mathbf{k}\cdot(\mathbf{T}\cdot \mathbf{k})=0.
\end{equation}
The advantage is that the left-hand-side of Eq. (\ref{eq:disper_Long}) is now a \textit{scalar}, giving us the dispersion equation for longitudinal waves with arbitrarily oriented $\mathbf{k}$'s. That quantity can be calculated from (\ref{tensor10}), and gives the dispersion equation (\ref{eq:disper}).

%% file: PlasmaTalk11.tex
\begin{center}
\section{Temperature Effects}
\end{center}

I will quickly go through the main temperature, i.e. energy spread effects, on our instabilities. Let's first start finding out about the limits of the cold regime.

\subsection*{When are we no longer ``cold''?}
The instability process is a matter of wave-particle interaction\footnote{Though there could be some issues here. See the end of the two-stream section.}. Assume a mode $\mathbf{k}$ is exchanging energy with a group of particles. If during one growth period, all particles remain in phase with the wave, the interaction is virtually cold. The wave grows as if there was no thermal spread at all. Writing that after one growth period, the velocity spread along \textbf{k} produces a spatial spread \textit{smaller} than the wavelength, we find the condition for the validity of the cold model\footnote{
Fainberg \textit{et al.}, \textit{Sov. Phys. JETP} \textbf{30}, 528 (1970).
},
\begin{equation}\label{cold}
\Delta v_\mathbf{k}~\delta^{-1}\ll k^{-1},
\end{equation}
where $\delta$ is the growth rate. Note worthily, the condition is \textit{not} homogenous throughout the $\mathbf{k}$ space. The spread $\Delta v_\mathbf{k}$ and the growth rate both depend on $\mathbf{k}$. A given system may be virtually cold for the two-stream instability, and hot for the filamentation.

The same physical picture allows to understand the main effect of temperature. Thermal spread reduces the growth rates, precisely because if condition (\ref{cold}) is \textit{not} fulfilled, the wave can exchange energy only with a fraction of the particles involved in the cold regime.

See Section III.C of the Review.

\subsection*{Two-stream modes}
\begin{figure}[t]
\begin{center}
\includegraphics[width=0.5\textwidth]{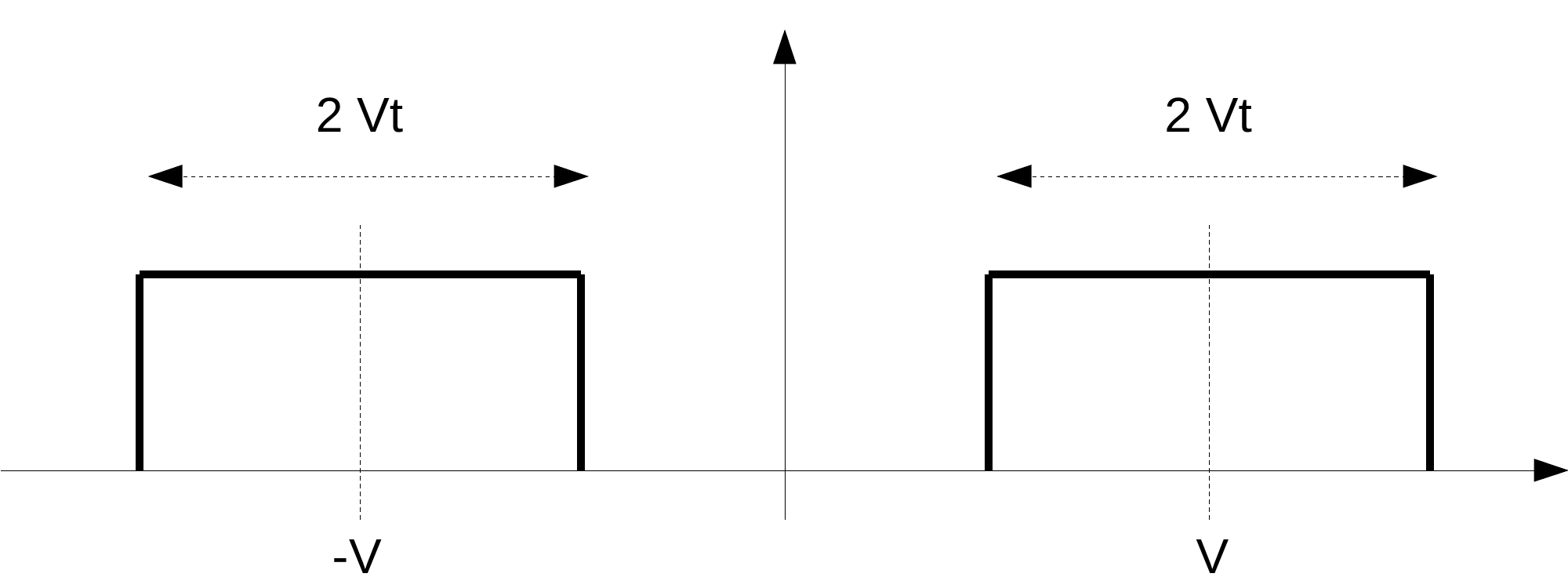}~~\includegraphics[width=0.45\textwidth]{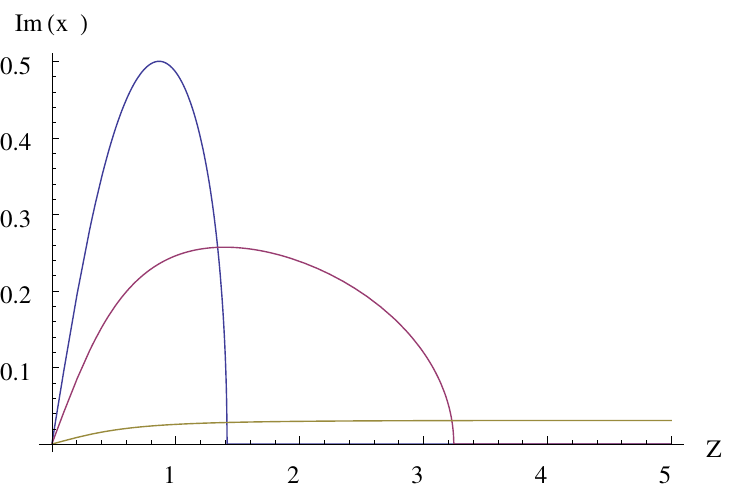}
\end{center}
\caption{LEFT: Simple toy model for the stabilization of the 1D two-stream instability. RIGHT: Growth rate in terms of $Z=kV/\omega_p$ for $\rho=0$, 0.9 and 0.999 from higher to lower curves respectively. The system is stable for $\rho>1$.} \label{fig:TS}
\end{figure}

Assume a 1D beam/plasma system with a velocity distribution such as the one pictured on Fig. \ref{fig:TS} LEFT, and define the temperature parameter,
\begin{equation}
\rho=\frac{V_{t}}{V}.
\end{equation}
For $\rho=0$, we have two counter-streaming symmetric beams. But it is obvious that if $\rho=1$, the two distributions make contact, and we end up with a total distribution equivalent to an homogenous \textit{stable} plasma at rest, with velocity spread equal to $\pm 2V$.

Indeed, the dispersion equation is easily computed, and reads in terms of the usual dimensionless parameters,
\begin{equation}
1-\frac{1}{(x+Z)^2-(\rho Z)^2}-\frac{1}{(x-Z)^2-(\rho Z)^2}=0.
\end{equation}
A plot of the growth rate is pictured on Fig. \ref{fig:TS} RIGHT, evidencing the progressive stabilization of the system for $\rho$ approaching unity.

The same pattern holds for more realistic distribution functions. The so-called ``Penrose Criterion''\footnote{\textit{Oliver} Penrose, brother of \textit{Roger} Penrose, \textit{Phys. Fluids} \textbf{3}, 258 (1960).} states that distribution functions are unstable if they have more than one local extremum. Bottom line: for hot enough beam and/or plasma, two-stream can be stabilized, relativistic or not\footnote{Buschauer, \textit{MNRAS}, \textbf{137} 99 (1977).}.

Something interesting: It is tempting to relate the former criterion to the formula for Landau Damping giving a growth rate $\propto f'_0(\omega/k)$. Nevertheless, we find here unstable waves with a distribution function which derivative is almost always zero\footnote{To be more accurate, the derivative of the distribution function pictured in Fig. \ref{fig:TS} LEFT, goes like a Dirac' $\delta$ for $v=\pm V \pm V_t$.}! In addition, when the system is unstable for $\rho<1$, the real part of the unstable modes is found at $\omega=0$, so that $f'_0(\omega/k)=0$ in our case, while there are \textit{no} particles at $v=0$. This is \textit{not} an artifact of our distribution functions, because $\omega=0$ also with two counter-streaming symmetric Maxwellian species.

 I have never seen this kind of issues discussed, except in one single paper\footnote{Phys. Rev. B \textbf{43}, 14009 (1991), where the problem is pointed out, but not solved.}. There are things left to understand\ldots

\subsection*{Filamentation modes}
\begin{figure}[t]
\begin{center}
\includegraphics[width=\textwidth]{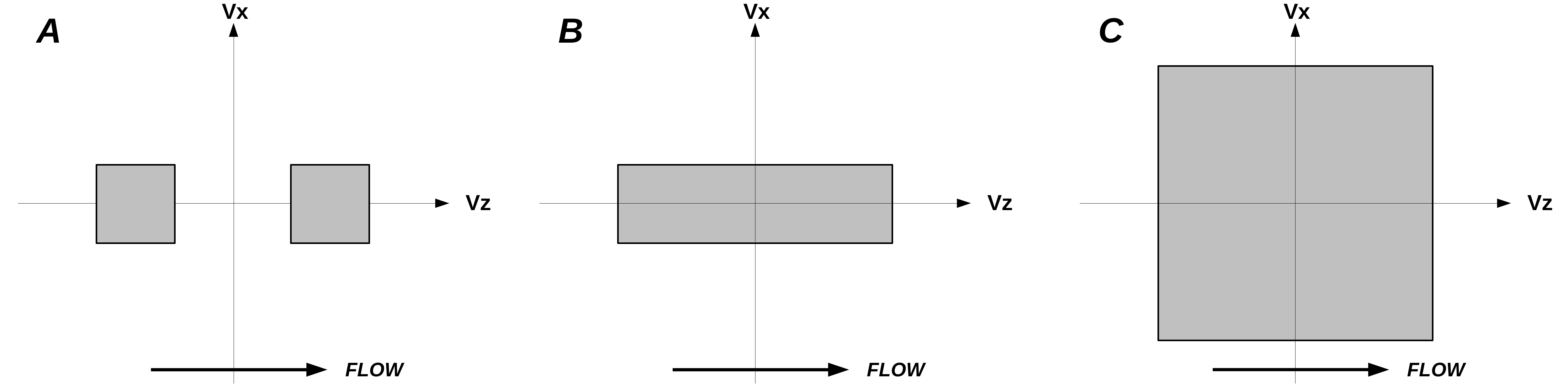}
\end{center}
\caption{Distribution functions with different stability properties.} \label{Toy2D}
\end{figure}

Let's extend our toy model consisting in  distributions flat up to a certain velocity (``waterbag''). Consider the 3 distribution functions pictured on Fig. \ref{Toy2D}. The shaded areas are uniformly filled with particles in velocity space.
\begin{itemize}
 \item \textbf{A} is a counter-streaming system. Unstable to both two-stream and filamentation instabilities.
 \item In \textbf{B}, we just extend the parallel spread until the distributions come in contact. According to the previous paragraph, the result is two-stream stable. But the result is also anisotropic. Weibel found\footnote{Weibel, \textit{Phys. Rev. Lett.}, \textbf{2}, 83 (1959).} it is unstable to perturbations with $\mathbf{k}$ normal to the highest thermal spread: that's filamentation here. \textbf{B} is therefore two-stream stable \textit{and} filamentation unstable.
 \item \textbf{C} is built from \textbf{B}, equating the spread in every directions. The result is stable.
\end{itemize}
We thus find we can ``play'' with temperature parameters in order to stabilize some parts of the spectrum, or all of it.
The bottom line here is that filamentation can be stabilized.

\bigskip

Note that in Fig. \ref{Toy2D}, we carefully tailor the spreads of every species to cancel the instability. What if we just play with the beam, that is, the component around $v_z>0$ on Fig.  \ref{Toy2D}-A? The end result depends on the distribution:
\begin{itemize}
 \item For waterbag distributions, filamentation is stabilized beyond a certain amount of \textit{beam transverse spread}. That can be understood as follows\footnote{Silva \textit{et al.}, \textit{Physics of Plasmas}, \textbf{9}, 2458 (2002).}:
Assume a current filament of radius $1/k$ and density $n_b$. The current is $I\sim qn_bk^{-2}v_b$. It creates at the surface of the filament a field $B=2I/ck^{-1}$. A charge at the surface is therefore pulled \textit{in} by the Lorentz force
\begin{equation}
F_B=q\frac{v_b B}{c}=\left(\frac{v_b}{c} \right)^2q^2n_bk^{-1}.
\end{equation}
If there is no temperature, nothing prevents the filament from further pinching, which is why the instability extends up to $k=\infty$ in the cold regime. But if we're hot, kinetic pressure opposes the pinching: A little piece of filament near the surface, with volume $dV$ and surface $dS$, is pulled in by $F_Bn_bdV$, and pushed out by $n_bk_BTdS$. Pinching is prevented if,
\begin{equation}
n_bk_BTdS>F_Bn_bdV~~\Rightarrow~~k> q\frac{v_b}{c}\sqrt{\frac{n_b}{k_BT}}~\equiv~k_{m\perp}\propto \frac{\sqrt{n_b}}{v_{tb}},
\end{equation}
which is the scaling found from the theory with waterbag distributions\footnote{Bret \textit{at al.}, \textit{Physical Review E}, \textbf{72}, 016403 (2005).} (we have used $dV\sim k^{-1}dS$, and $v_{tb}$ is the thermal beam spread.).

\item For relativistic Maxwellians, it has been proved that filamentation never vanishes completely\footnote{Gremillet, \textit{Unpublished}, Bret \textit{et al.}, \textit{Physical Review E}, \textbf{81}, 036402 (2010),}. If $T_b$ is the beam temperature, the maximum filamentation growth-rate scales like $T_b^{-3/2}$. In addition, this result still holds for \textit{any} plasma temperature.
\end{itemize}

\subsection*{Oblique modes - general thermal ``rules''}
Two-stream modes are unstable up to a finite $k_{m\parallel}$, and filamentation up to yet another finite $k_{m\perp}$. How do we close the unstable domain? Two different answers according to the distribution. With waterbag, we close from, and to infinity. See Fig. 10a of the Review. With a Maxwellian, these large $k$ oblique modes are stabilized, and we close ``normally'', as pictured for example on Fig. 14 of the Review.

Oblique modes' temperature sensitivity is intermediate between two-stream and filamentation. As evidenced in \textit{Plasma Talk 10}, they tend to be interesting only in the relativistic regime, due to the $\gamma_b$ scaling which favors them. Two thermal ``rules'' are useful to grasp the influence of the \textit{beam} spread over the full spectrum:
\begin{itemize}
 \item Parallel spread \textit{hardly matters}. Why? Because in the relativistic regime, it takes a huge parallel \textit{energy} spread to get a reasonable parallel \textit{velocity} spread. All velocities are squeezed against $c$. See Fig. 13 of the Review.
\item Two-stream does not care about the transverse spread, because particles differing only by their \textit{transverse} velocity will equally stay tuned with a plane wave at $\mathbf{k} \parallel flow$. For the same reason, filamentation do care.\\
As a result, modes are all the more affected by beam temperature than they are oblique.
\end{itemize}
We could rank, from best to worst, the 3 kinds of modes in terms of the way they ``resist'' the various effects:

\begin{description}
 \item[Relativistic:]   Oblique $\rightarrow$ Filamentation $\rightarrow$ two-stream.
 \item[Density ratio:]  Filamentation $\rightarrow$ Oblique/two-stream.
 \item[Beam temperature:]  Two-stream $\rightarrow$ Oblique $\rightarrow$ Filamentation.
 \end{description}

All this ends up with the \textit{mode hierarchy} pictured on Fig. 20 of the Review Paper.

\subsection*{The phase velocity diagram - Fig. 17 of the Review}
A great tool to understand the physics. Plot, on the \textit{very same graph}, the distribution functions and the phase velocities of the unstable modes. On can straightforwardly check which species are in resonance with which kind of modes.

%% file: PlasmaTalk12.tex
\begin{center}
\section{Non-Linear Regime}
\end{center}

\begin{figure}[h]
\begin{center}
\includegraphics[width=0.6\textwidth]{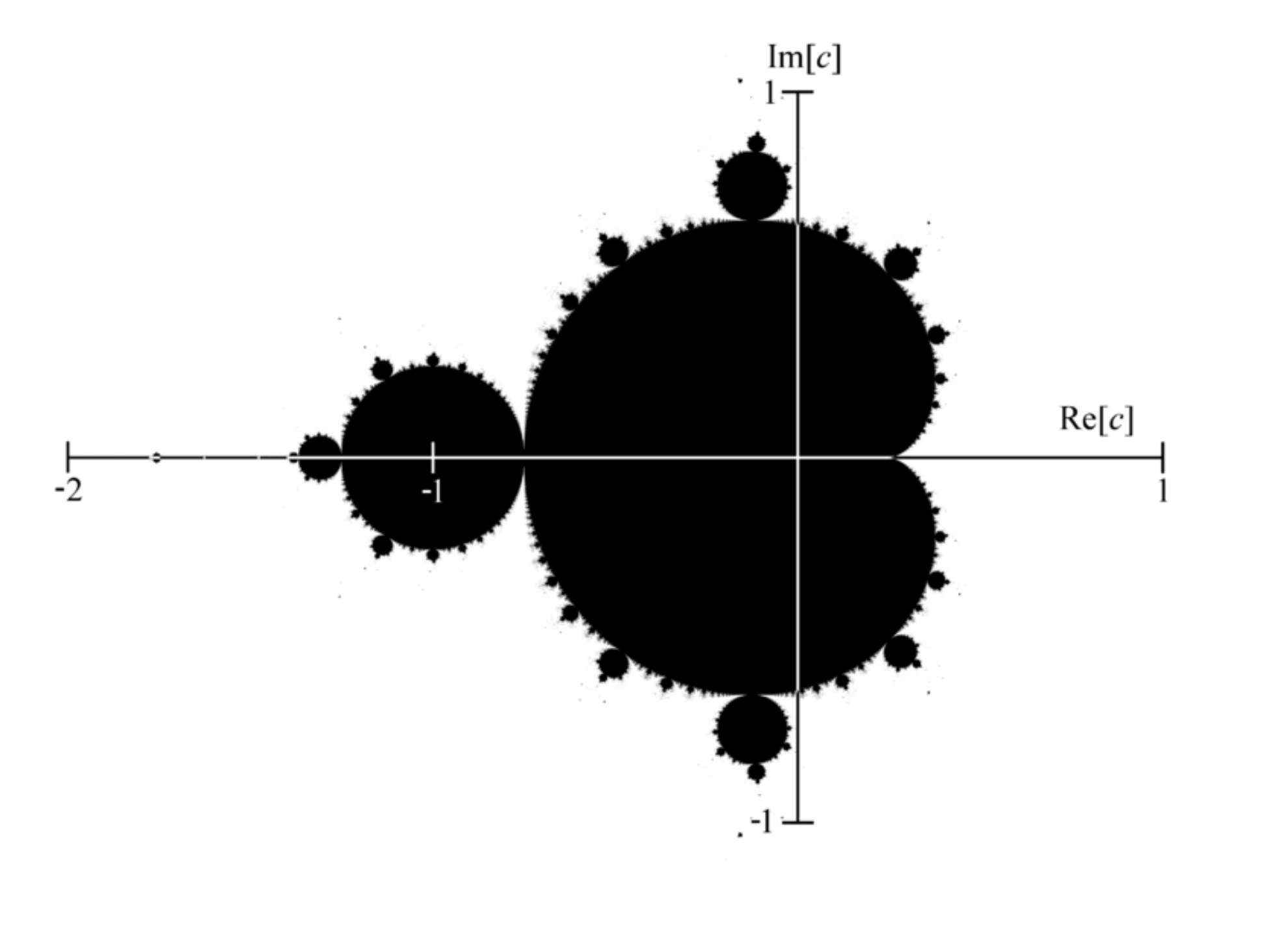}\includegraphics[width=0.4\textwidth]{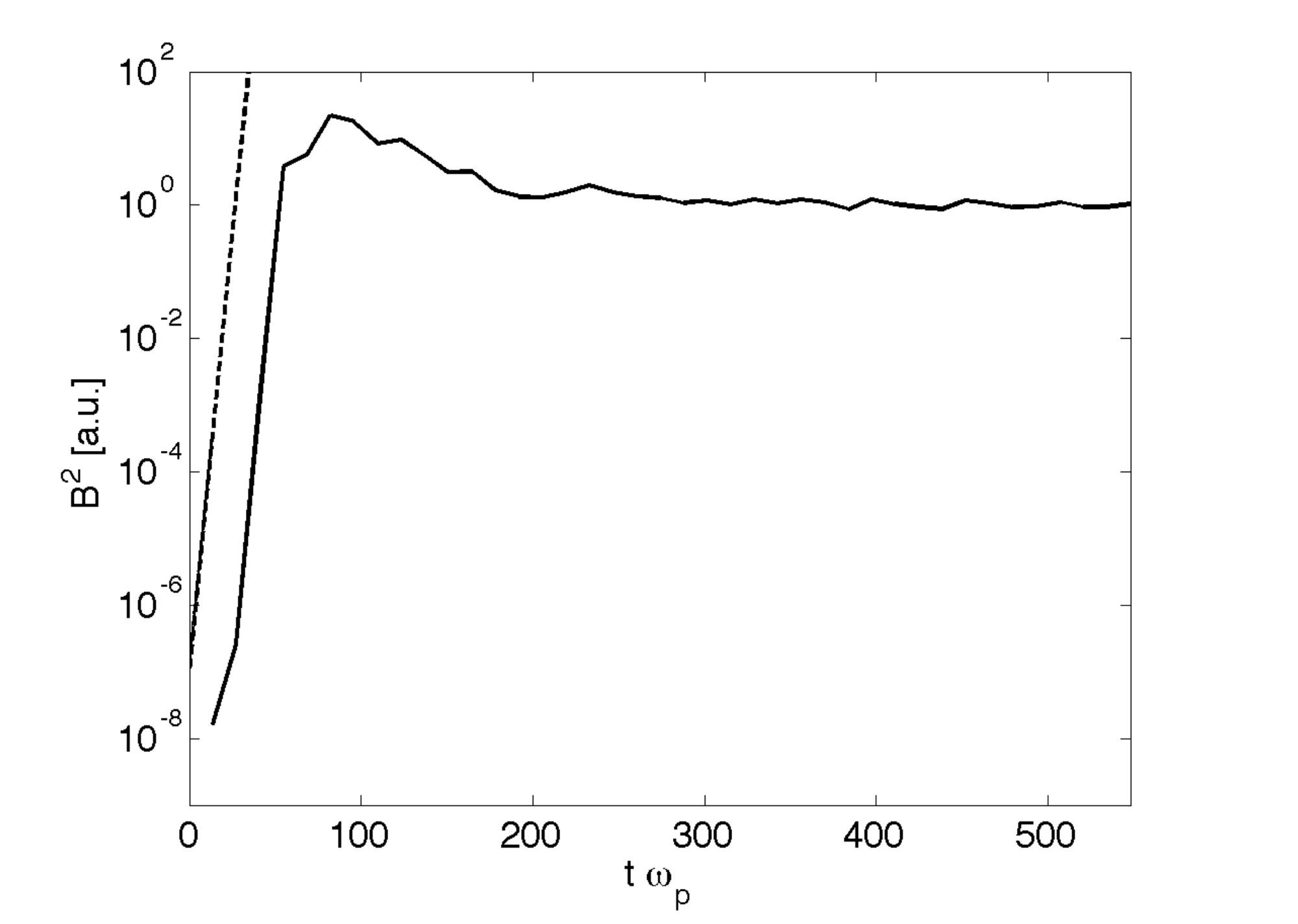}
\end{center}
\caption{\textit{Left}: The ``Mandelbrot set'', defined by Eqs. (\ref{eq:mand1},\ref{eq:mand2}). \textit{Right}: Typical field growth extracted from a PIC simulation. Growth roughly stops when the linear exponential growth stops. The line shows the expected growth from theory.} \label{fig:mand}
\end{figure}

Non-linear problems may be complicated with just 1 or 2 degrees of freedom. Think about this 2D example: take the very simple sequence\footnote{Non-linear in the sense that if two sequences fulfill $z_{n+1}=z_n^2 + c$, their linear combinations do not.}
\begin{eqnarray}\label{eq:mand1}
z_{n+1}&=&z_n^2 + c,\nonumber\\
z_0&=&c,~~c \in \mathbb{C},
\end{eqnarray}
and just define the set
\begin{equation}\label{eq:mand2}
M=\left\{ c \in \mathbb{C}~/~\lim_{n\rightarrow\infty} \mid z_n \mid \neq \infty \right\}.
\end{equation}
You have the famous, and incredibly complicated, fractal and everything, ``Mandelbrot set'' pictured on Fig. \ref{fig:mand}. People were stunned when they realized something as trivial as Eq. (\ref{eq:mand1}) could generate such amount of complexity\footnote{A Math-guy friend of mine once told me people would laugh at Mandelbrot, as ``the guy who works on polynomial of degree 2''!}.

A plasma has $\infty$ number of degrees of freedom. Yet, to my knowledge, something as beautiful as Fig. \ref{fig:mand} is still lacking in plasma physics. Maybe because it's too complicated\ldots

At any rate, there's no hope of analytically finding out about the long term evolution of our beam plasma systems in the general case. Remember it took a Fields Medal to prove non-linear Landau damping. Even for the cold case, things are not easy.

I'll go through some results on the saturation of the various instabilities, always assuming the fastest growing mode is the \textit{only one} excited, and that everything is \textit{cold} at $t=0$. The former is quite reasonable, as the most unstable mode grows \textit{exponentially} faster than the rest. Relativistic effects help as they drive a sharper unstable spectrum, where growth rates vary rapidly from one mode to another. The later is a limitation.

The idea is to find out \textit{when} the linear theory should break down, and to claim that growth \textit{stops} at that point. Granted, \textit{exponential} growth should stop there. But other kind of growth could keep on. Yet the observed field growth in PIC simulations is always like to one pictured on Fig. \ref{fig:mand}. Why?

See Section V of the \textit{Review Paper}, and references therein.

\subsection*{Two-stream and oblique instabilities}
\subsubsection*{Two-stream, non-relativistic}
We assume a \emph{diluted beam}. There, two-stream is resonant with the beam. Say the wave $Ee^{ikx-i\omega t}$ is growing, traveling along with the beam electrons. Electrons will start oscillating in the field at the ``bouncing'' frequency\footnote{See \textit{Plasma Talk 5} on Landau damping.},
\begin{equation}\label{eq:Ebounce}
\omega_b^2=\frac{q E k}{m}.
\end{equation}
The linear assumption that all electrons have $v=v_b$ during one growth period $\delta^{-1}$ breaks down when,
\begin{equation}
\omega_b=\delta~~\Rightarrow~~E_s=\frac{\delta^2}{\omega_p}\frac{mv_b}{q},
\end{equation}
giving the value of the field at saturation (I've set here $k\sim\omega_p/v_b$). A great by-product of $E_s$ is the beam energy loss $\Delta W_b$. Since the field energy can only come from the beam energy, we can write,
\begin{eqnarray}\label{eq:DeltaEe}
\frac{\Delta W_b}{\frac{1}{2}n_b m v_b^2}&=&\frac{E_s^2/8\pi}{\frac{1}{2}n_b m v_b^2}\\
&=&\frac{n_p}{n_b}\left( \frac{\delta}{\omega_p}\right)^4
\sim \left( \frac{n_b}{n_p}\right)^{1/3},\nonumber
\end{eqnarray}
where I just replaced the growth-rate by its cold value $\frac{\sqrt{3}}{2^{4/3}}(n_b/n_p)^{1/3}\omega_p$. We could even write the energy lost is shared between the plasma and the field\footnote{Lorenzo Sironi told me you see this in the PICs. But why?}. In such case, $\Delta W_b/W_b$ is half the result (\ref{eq:DeltaEe}).

\begin{figure}
\begin{center}
\includegraphics[width=0.8\textwidth]{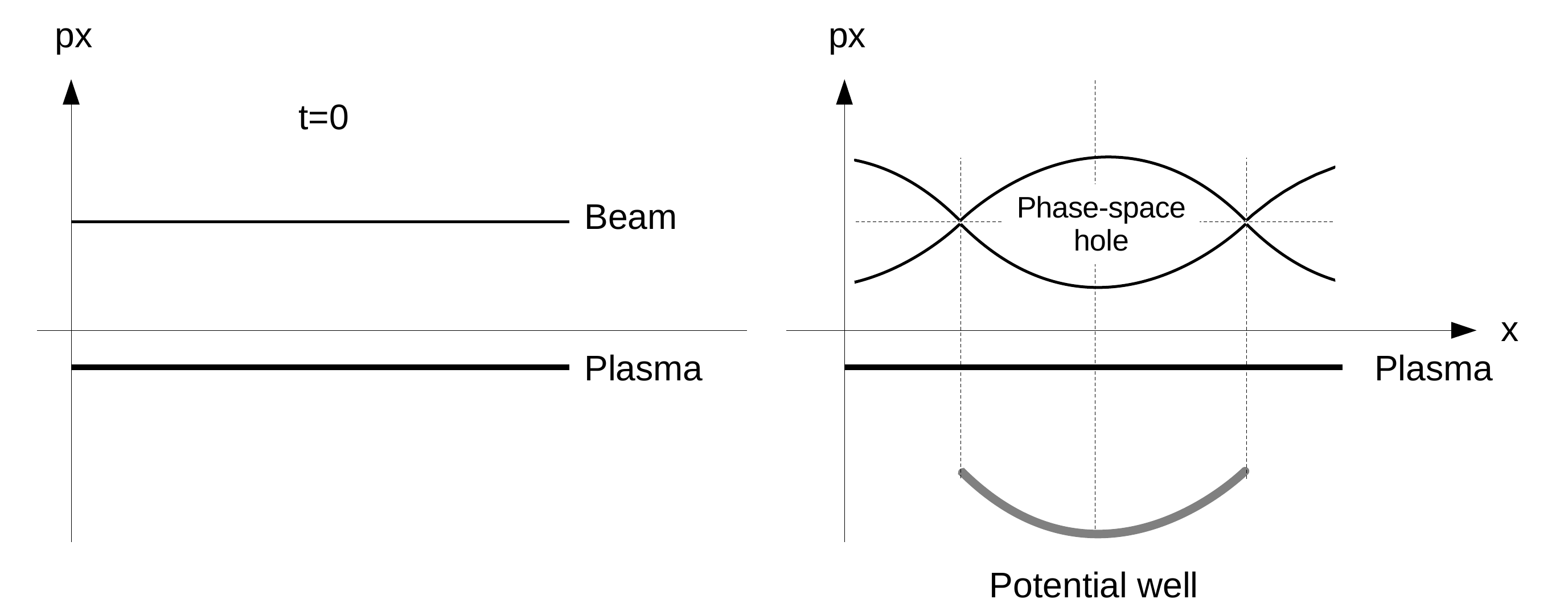}
\end{center}
\caption{Phase-space hole.} \label{fig:holes}
\end{figure}

As the field grows and traps the beam electrons, they start oscillating in the wave potential. PIC people love to plot density graphs in the $(x,v_x)$ phase-space (say $x$ is our dimension) such as the one pictured on Fig. \ref{fig:holes}. At $t=0$, the beam and the plasma are just two lines. In the instability phase, the plasma does not move a lot, but the beam particles start oscillating in the wave, creating ``holes'' in the phase-space.

\subsubsection*{Two-stream, Relativistic}
A naive reasoning gives the right answer. As the oscillatory trapping motion is along the flow, we replace $m\rightarrow\gamma_b^3 m$ in the bouncing frequency (\ref{eq:Ebounce}). This gives the field at saturation\footnote{Let's denote quantities \textit{in the wave frame} with a prime. If the beam dynamic in the wave frame is non-relativistic, we have
\begin{equation}
\omega_b'^2=\frac{q E k'}{m},\nonumber
\end{equation}
where the field $E$ does not need prime as it is parallel to the motion. The bouncing frequency in the lab frame is simply $\omega_b=\omega_b'/\gamma_b$. In addition, $k=\gamma_b k'$ since in the wave frame, $\omega'=0$. We thus find
\begin{equation}
\omega_b^2=\frac{q E k}{\gamma_b^3m},\nonumber
\end{equation}
retrieving the $m\rightarrow\gamma_b^3m$ ``rule''.
},
\begin{equation}
E_s=\frac{\delta^2}{\omega_p}\frac{\gamma_b^3mv_b}{q}.
\end{equation}
The relative energy loss is then computed as in Eq. (\ref{eq:DeltaEe}), replacing the growth-rate by its relativistic value. It reads,
\begin{equation}
\frac{\Delta W_b}{n_b \gamma_b m c^2}\sim \gamma_b\beta^2\left( \frac{n_b}{2n_p}\right)^{1/3}.
\end{equation}
Of course, this quantity has to remain small since linear regime means unperturbed trajectories, that is, $\Delta W_b\ll W_b$.
The energy loss eventually relies on the parameter $S$ with\footnote{See \emph{Review Paper}.}
\begin{equation}\label{eq:S}
S=\gamma_b\beta^2\left( \frac{n_b}{2n_p}\right)^{1/3}.
\end{equation}
If you do the ``clean'' calculation going to the wave frame, like in the footnote, you need the beam dynamic in that frame to be non-relativistic in order to compute easily a bouncing frequency. Doing so requires $S\ll 1$, as explained in the Review. For arbitrary $S$'s, one has
\begin{equation}\label{eq:WS}
\frac{\Delta W_b}{W_b}\sim \frac{S}{(S+1)^{5/2}},
\end{equation}
yielding a maximum energy loss in the linear phase
\begin{equation}\label{eq:WSmax}
\frac{\Delta W_b}{W_b}_{max}=\frac{6}{25}\sqrt{\frac{3}{5}}\sim 0.18~~~\mathrm{for}~~~S=\frac{2}{3}.
\end{equation}

\subsubsection*{Oblique instability}
Poorly known. What is known is that Eq. (\ref{eq:WS}) roughly works until $S\sim 0.45$. For larger $S$'s, $\Delta W_b/W_b$ seems quite insensitive to $S$, and remains around $0.18$ given by Eq. (\ref{eq:WSmax}).

\subsection*{Filamentation instability}
There are 3 different ways to evaluate the field at saturation!
\begin{enumerate}
 \item \textbf{Cyclotron frequency = growth rate.} Filamentation instability grows magnetic field. Such field \textbf{B} affects particles on a time scale given by the cyclotron frequency,
\begin{equation}\label{eq:cyclo}
\omega_c=\frac{q B_1}{\gamma_b m c}.
\end{equation}
Writing again that the linear regime keeps on until $\omega_c=\delta$, we get,
\begin{equation}\label{eq:Bs1}
B_{s1} =   \frac{\gamma_b m c}{q} \delta.
\end{equation}
If one tries to compute the relative energy loss $\frac{B_{s1}^2}{8\pi}/n_b\gamma_bmc^2$, the cold result gives a factor of order $\beta^2$, without \textit{any} more scaling in $\gamma_b$, and almost none in $\alpha=n_b/n_p$ (see Table \ref{tab:1}). The conclusion is that estimating the energy loss requires a finer calculation than this one, and that the result is quite stable in terms of these variables.

\item \textbf{Bouncing frequency = growth rate.} Historically, the field at saturation has rather been evaluated this way. With $\mathbf{v}_{0b}=(0,0,v_b)$ and $\mathbf{k}=(k_x,0,0)$, the growing magnetic field reads $\mathbf{B}_1=(0,B_1\sin k_xx,0)$. At first order, Newton's law projected on the $x$ axis gives,
\begin{equation}
\gamma_b m\frac{d^2x}{dt^2}=qB_1 \frac{v_b}{c} \sin k_x x.
\end{equation}
Particles at $x\sim 0$ oscillate at,
\begin{equation}\label{eq:Bbounce}
\omega_b^2=\frac{q B_1 v_b k_x}{\gamma_b m c}.
\end{equation}
Note that we assumed there's only a \textbf{B} field here. We know that unless the system is strictly symmetric, it's wrong. Here again, we can claim exponential growth keeps on while the motion is almost unperturbed, that is until
\begin{equation}\label{eq:Bs2}
\frac{q B_1 v_b k_x}{\gamma_b m c} =\delta^2~~\Rightarrow~~B_{s2} = \frac{\gamma_b m c}{q }\frac{\delta^2 }{v_b k_x}.
\end{equation}
Comparing Eqs. (\ref{eq:Bs2},\ref{eq:Bs1}) gives,
\begin{equation}
B_{s2} =  B_{s1}\frac{\delta}{v_b k_x},
\end{equation}
so that both estimates give the same result \textit{only} with $k_x=\delta/v_b$. For example, with the symmetric cold case where $\delta=\omega_p\beta\sqrt{2/\gamma_b}$, that implies
\begin{equation}
k_x =  \frac{\omega_p}{c}\sqrt{\frac{2}{\gamma_b}}.
\end{equation}
For non-relativistic setting, the $k_x$ is the typical expected one. For $\gamma_b$, very recent \textit{cold} PIC's \cite{Bret2012} found indeed that it \textit{is} the fastest growing $k_\perp$. Why exactly, as the cold growth sate juste saturates at large $k_\perp$'s?

\item \textbf{Larmor radius = characteristic $1/k_\perp$.} Equating the Larmor radius of an electron in a field \textbf{B} to the characteristic $k_c$ of the insta gives,
\begin{equation}
B_{s3} = \frac{\gamma_b m c}{q}v_0k_c = B_{s1}\frac{v_0k_c}{\delta}.
\end{equation}
\end{enumerate}

The 3 results are summarized in Table \ref{tab:1}, considering the cold symmetric case, and taking $k=\omega_p/c$ for the typical $k_\perp$. Taking $k_\perp\propto \gamma_b^{-1/2}$, the 3 criteria give the same scaling for $\alpha=1$  \cite{Bret2012}.

\begin{table}[t]
\begin{center}
\begin{tabular}{llll}
Criteria &  $B_s$ &  $B_s,$ $mc\omega_p/q$ units & $\Delta W_b/W_b$ \\
\hline\hline
Cyclotron frequency = growth-rate & $\frac{\gamma_b m c}{q} \delta$ & $\beta[\alpha(\alpha+1)\gamma_b]^{1/2}$ & $\beta^2(\alpha+1)/2$ \\
Bouncing frequency = growth-rate & $\frac{\gamma_b m c}{q }\frac{\delta^2 }{v_b k_x}$ & $\beta\alpha(\alpha+1)$ &  $\beta^2\alpha(\alpha+1)^2/2\gamma_b$ \\
Larmor radius = characteristic $k_c^{-1}$ & $\frac{\gamma_b m c}{q}v_0k_c$ & $\beta\gamma_b$ & $\beta^2\gamma_b/2\alpha$ \\
\hline
\end{tabular}\caption{Summary of the 3 ways to evaluate the field at saturation $B_s$ for the filamentation instability, \textit{considering} $k_\perp=\omega_p/c$. Results for the cold case. Taking $k_\perp\propto \gamma_b^{-1/2}$, the 3 criteria give the same scaling for $\alpha=1$ \cite{Bret2012}.}\label{tab:1}
\end{center}
\end{table}

\subsubsection*{Fate of the filaments}
Opposite filaments repel, but like filaments attract. In our 3D world, filaments turn around each other, and like filaments merge. The merging process has been modeled, and successfully simulated with PICs\footnote{Medvedev \textit{et al.}, \textit{The Astrophysical Journal} \textbf{618}, L75 (2005).}.

\subsection*{More realistic settings, successive instabilities}
We've been so far interested in the short term evolution of the system, that is, the end of the linear phase. What's next? I'll just comment \textit{Fig. 40 of the Review Paper}. The initial setup was:

\emph{Beam}: Maxwellian, $n_b=n_p/10$, $\gamma_b=3$, $T_b=50$ keV.

\emph{Plasma}: Maxwellian, $T_p=5$ keV.

\begin{itemize}
 \item $0<\omega_p t<80$: System initially governed by \textit{oblique} modes. \textbf{E} field grows at $0.07\omega_p$. Heating ``kills'' de oblique.
 \item $80<\omega_p t<160$: System switches to a \textit{two-stream} regime. \textbf{E} field grows at $0.016\omega_p$. Heating ``kills'' two-stream.
 \item $200<\omega_p t<600$: Remaining drift feeds \textit{filamentation}. \textbf{B} field grows at $0.005\omega_p$.
\end{itemize}

By the end of the simulation $\omega_p t\sim 600$, the beam had lost about 30\% of its energy, entirely transferred to plasma electrons. Open questions:
\begin{itemize}
 \item Is filamentation the necessary end state of every initial setup?
 \item Does the drift eventually ends (in the frame of the fixed ion background)? That is, is the drift energy eventually converted at 100\% into heat? Sounds reasonable. Is that sure?
\end{itemize}

 Indeed, the interesting question might be \textit{how long} does it take?

%% file: PlasmaTalk13.tex
\begin{center}
\section{Ohm's law and the Biermann battery}
\end{center}

The original 1950 paper is Ref. \cite{Biermann}, ``{\"U}ber den Ursprung der Magnetfelder auf Sternen und im interstellaren Raum'', published in \textit{Zeitschrift Naturforschung Teil A}. Cited more than 200 times, and probably read by no one but the happy few who 1) read German and 2) could access it.

$\blacksquare$ The MHD equations are formed from the fluid  equations for electrons and ions (dropping subindices),
\begin{eqnarray}\label{Fluid13}
\frac{\partial n}{\partial t}+\frac{\partial}{\partial \mathbf{r}}\cdot(n\mathbf{v})&=&0,\nonumber\\
mn\left(\frac{\partial \mathbf{v}}{\partial t}+\mathbf{v}\cdot\frac{\partial \mathbf{v}}{\partial \mathbf{r}}\right) &=&
qn \left(\mathbf{E}+\frac{\mathbf{v}}{c}\times\mathbf{B}\right)-\nabla p + n m \mathbf{g}.
\end{eqnarray}

 The MHD variables are defined from  $n_{e,i}(\mathbf{r},t),\mathbf{v}_{e,i}(\mathbf{r},t)$ as\footnote{
In \textit{Plasma Talk 3}, the MHD velocity $\mathbf{V}$ was defined through $(m_e+m_i)\mathbf{V}=m_e\mathbf{v}_e + m_i\mathbf{v}_i$. Most books \cite{Spitzer13,Goedbloed} present definition (\ref{MHDVar13}) above. It is more rigorous, as it gives a $\rho \mathbf{V}$ MHD term exactly equal to the total momentum. At any rate, the difference between the two quantities is $\propto(n_e-n_i)(\mathbf{v}_e-\mathbf{v}_i)$. This is a second order quantity in the MHD regime, where electrons are expected to closely follow the ions, so that $n_e\sim n_i$ and $\mathbf{v}_e\sim\mathbf{v}_i$.},
\begin{eqnarray}\label{MHDVar13}
\rho(\mathbf{r},t)&=&m_in_i+m_en_e,\nonumber\\
\mathbf{J}(\mathbf{r},t)&=&q n_i\mathbf{v}_i -q n_e \mathbf{v}_e,\nonumber\\
\mathbf{V}(\mathbf{r},t)&=& \frac{1}{\rho(\mathbf{r},t)}(n_e m_e\mathbf{v}_e+m_i n_i\mathbf{v}_i).\nonumber
\end{eqnarray}

Merging the fluid Euler equations for both species gives,
\begin{equation}
\rho\left(\frac{\partial \mathbf{V}}{\partial t}+\mathbf{V}\cdot\frac{\partial \mathbf{V}}{\partial \mathbf{r}}\right) =
\frac{\mathbf{J}}{c}\times\mathbf{B} -\nabla (\overbrace{p_i+p_e}^P)+\rho \mathbf{g}.
\end{equation}

Ohm's law is the equation giving the current $\mathbf{J}$. Where does it come from? What we did in \textit{Plasma Talk 3} was to follow the basic, ``business as usual'' procedure: sit in the frame of the fluid locally at $\mathbf{V}$. There, the electric field is $\mathbf{E}'$. Ohm's law gives the current in the lab frame from $\mathbf{E}'$, as $\mathbf{J}'=\sigma\mathbf{E}'$. Because $\mathbf{E}'=\mathbf{E}+\mathbf{V}\times\mathbf{B}/c$, we get the famous (non-relativistic)
\begin{equation}\label{Ohm13}
\mathbf{J} = \sigma\left(\mathbf{E}+\frac{\mathbf{V}}{c}\times\mathbf{B}\right).
\end{equation}

What about $\mathbf{J}'=\sigma\mathbf{E}'$? It arises from the microscopic picture that under the action of an electric field, particles, mostly electrons, are accelerated in the direction of the field, while collisions with the ions act like a friction force\footnote{Solid state physics call this the ``Drude model'', from Paul Drude, who came up with this idea in 1900.}. Writing something like $m\partial_t\mathbf{v}=q\mathbf{E}'-\nu \mathbf{v}$ where $\nu$ is some collision frequency, and setting $\partial_t =0$, indeed yields $\mathbf{v}=\frac{q}{\nu}\mathbf{E}'$ and then $\mathbf{J}'=qn\frac{q}{\nu}\mathbf{E}'$. Ideal MHD assumes $\nu=0$, giving
\begin{equation}
\mathbf{E}'=\mathbf{E}+\frac{\mathbf{V}}{c}\times\mathbf{B}=\mathbf{0}.
\end{equation}

$\blacksquare$  But we inadvertently assumed many things. For example, $\partial_t =0$ assumes $\mathbf{E}'$ varies slowly enough with time. If variations are too fast, the stationary regime does not have enough time to set in, and some $\omega$ dependency appears. Also, we assumed particles are accelerated only by $\mathbf{E}'$ between two collisions. What if $\mathbf{B}$ is strong enough to curve the trajectories in between? In this case, the resistivity in the direction normal to the field is higher ($\times 1.9$) than along the field (\cite{Spitzer13} p. 28, or \cite{Kulsrud} p. 43).

Since conductivity comes from the electrons\footnote{The $\sigma$ of $\mathbf{J}' = \sigma\mathbf{E}'$ has the mass on the denominator.}, let's write their full Euler equation,
\begin{equation}
m_en_e\left(\frac{\partial \mathbf{v}_e}{\partial t}+\mathbf{v}_e\cdot\frac{\partial \mathbf{v}_e}{\partial \mathbf{r}}\right) =
q_en_e \left(\mathbf{E}+\frac{\mathbf{v}_e}{c}\times\mathbf{B}\right)-\nabla p_e-\nu(\mathbf{v}_e-\mathbf{v}_i).
\end{equation}
Neglecting the left-hand-side for the small electrons inertia, and setting $\nu=0$ for ideal MHD yields,
\begin{equation}
\mathbf{E}+\frac{\mathbf{v}_e}{c}\times\mathbf{B}=\frac{\nabla p_e}{q_en_e}.
\end{equation}
According to Kulsrud \cite{Kulsrud} p. 405, the pressure term above is negligible when there is a $\mathbf{B}$ field. \textbf{WHY?} But for small $\mathbf{B}$'s, or even $\mathbf{B}=0$, you need to keep it. Inserting the electric field above in $\partial_t\mathbf{B}=-c\nabla\times\mathbf{E}$ and setting $\mathbf{v}_e\sim\mathbf{V}$ gives\footnote{We need a little drag between ions and electrons to write $\mathbf{v}_e\sim\mathbf{V}$.},
\begin{equation}\label{battery}
\frac{\partial \mathbf{B}}{\partial t}=\nabla\times(\mathbf{V}\times\mathbf{B})+c\frac{\nabla n_e \times \nabla p_e}{q n_e^2}.
\end{equation}
The second term is our \textit{Biermann Battery}. There's no $\mathbf{B}$ in there, so that it can make it from nothing. Still, the microscopic derivation shows we need ionization, just to be able to create electronic currents moved by the electronic pressure.

\begin{figure}[t]
\begin{center}
\includegraphics[width=0.5\textwidth]{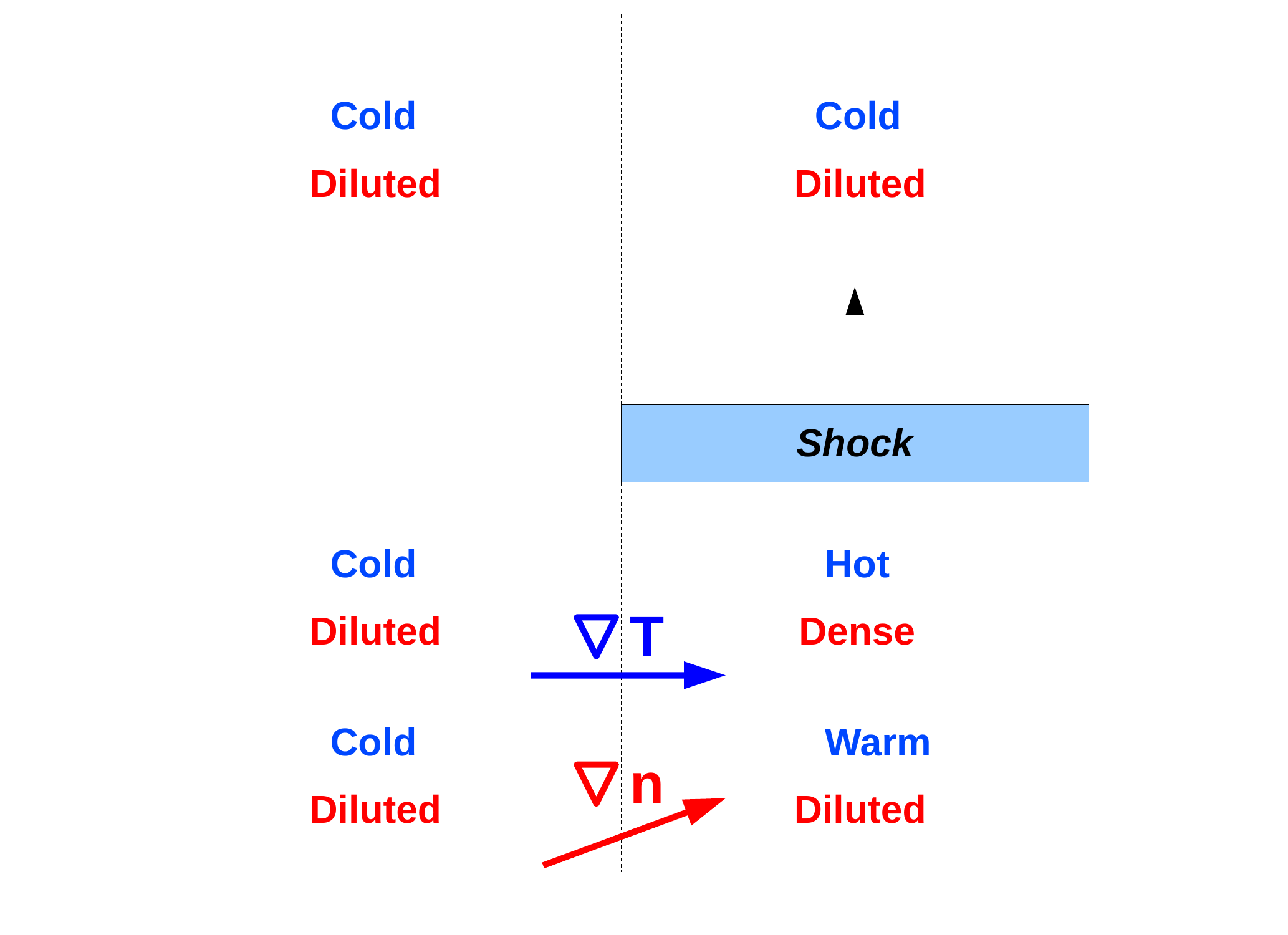}
\end{center}
\caption{How a finite size shock can generate non-parallel density and temperature (hence pressure) gradients. From \cite{Kulsrud}, p. 406.} \label{fig:bier}
\end{figure}

$\blacksquare$  But, an equation of state usually gives $p_e(n_e)$. The gradients for $p_e$ and $n_e$ are thus likely to be parallel, so that the cross product vanishes. How do we break this? Several possibilities,
\begin{enumerate}
 \item Try a system in rotation around $z$, with $\Omega(z)$. The pressure gradient reads $\nabla P \propto \Omega^2(z)\rho(r,\theta)$. It has a $z$ component, while $\nabla \rho(r,\theta)$ has not. Those who've read the paper say Biermann considered this option.
 \item Try the scheme of Figure \ref{fig:bier}. Suppose a finite size shock travels through a cold, diluted upstream. The downstream just behind the shock is dense and hot. Far downstream though, the plasma expands but remains warm, so that it is now warm \textit{and} diluted. The figure shows well how this can generate non-parallel density and temperature (hence pressure) gradients. A \textit{non}-spherical shock also does the job for similar reasons.
\end{enumerate}

Everything is easily adapted to a system \textit{partially} ionized. If $n_n$ is the density of the neutral, define $\chi=n_e/(n_i+n_n)$. Eq. (\ref{battery}) can be adapted dividing the Battery term by $1+\chi$ (\cite{Kulsrud}, p. 406). See \cite{Mestel1983}.

Kulsrud proposed this mechanism to produce Cosmic Fields from scratch in 1997 \cite{Kulsrud1997}. Yet, he co-authored in 1992 another paper \cite{Tajima} advocating spontaneous plasmas fluctuations.

Also Khanna \cite{Khanna1998}, showed that a rotating BH in a plasma will always
generate toroidal and poloidal magnetic fields.

Ref. \cite{xu2008} studied the Biermann Battery effects in Cosmological MHD Simulations of Population III Star Formation. In its own terms, ``We find that the Population III stellar cores formed including this effect are both qualitatively and quantitatively similar to those from hydrodynamics-only (non-MHD) cosmological simulations''. No dynamical effects.

The Biermann battery mechanism was successfully tested in the lab in 2012 \cite{BiermannNature}.